\documentclass[tighten,times,twocolumn]{aastex63}

\graphicspath{{./}}

\usepackage{amsmath}
\usepackage{mathtools}

\renewcommand{\vec}[1]{{\boldsymbol{\mathbf{#1}}}}

\newcommand{\appropto}{\mathrel{\vcenter{
  \offinterlineskip\halign{\hfil$##$\cr
    \propto\cr\noalign{\kern2pt}\sim\cr\noalign{\kern-2pt}}}}}

\usepackage{hyperref}
\hypersetup{ bookmarks = true }
\usepackage{xspace}
\usepackage{graphicx}
\usepackage[enable]{easy-todo}

\usepackage{comment}

\usepackage{savesym}
\savesymbol{tablenum}
\usepackage{siunitx}
\restoresymbol{SIX}{tablenum}
\DeclareSIUnit\degr{deg}
\DeclareSIUnit\parsec{pc}
\DeclareSIUnit\dBm{dBm}
\DeclareSIUnit\jansky{Jy}
\DeclareSIUnit\beam{beam}

\definecolor{ucol}{rgb}{.6,0,0}
\definecolor{webgreen}{rgb}{0,.5,0}
\definecolor{halfgray}{gray}{0.55}

\usepackage[capitalize]{cleveref}
\newcommand{\secref}[1]{Section~\ref{#1}}

\newcommand{\tcm}{21$\,$cm\xspace}  

\newcommand{\Nfeed}{\ensuremath{N_\mathrm{feed}}}

\newcommand{\sw}[1]{\textsc{#1}\xspace}

\newcommand{\reviewercomment}[1]{#1}

\shorttitle{CHIME Overview}

\begin{document}

\title{An Overview of  CHIME,  the Canadian Hydrogen Intensity Mapping Experiment}
\begin{abstract}
The Canadian Hydrogen Intensity Mapping Experiment (CHIME) is a drift
scan radio telescope operating across the \SIrange{400}{800}{\mega \hertz} band. CHIME is located at the Dominion Radio Astrophysical Observatory near Penticton, BC Canada. The
instrument is designed to map neutral hydrogen over the redshift
range 0.8 to 2.5 to constrain the expansion history of the Universe.
This goal drives the design features of the instrument. CHIME consists of four parallel cylindrical reflectors, oriented north-south, each $\SI{100}{m} \times \SI{20}{m}$ and outfitted with a 256~element dual-polarization linear feed array. CHIME observes a two degree wide stripe covering the entire meridian at any given moment, observing 3/4 of the sky every day due to Earth rotation. An FX correlator utilizes FPGAs and GPUs to digitize and correlate the signals, with different correlation products generated for cosmological, fast radio burst, pulsar, VLBI, and \tcm absorber backends. For the cosmology backend, the $\Nfeed^2$ correlation matrix is formed for 1024 frequency channels across the band every \SI{31}{\milli \second}. A data receiver system applies calibration and flagging and, for our primary cosmological data product, stacks redundant baselines and integrates for \SI{10}{\second}. We present an overview of the instrument, its performance metrics based on the first three years of science data, and we describe the current progress in characterizing CHIME's primary beam response. We also present maps of the sky derived from CHIME data; we are using versions of these maps for a cosmological stacking analysis as well as for investigation of Galactic foregrounds.

\end{abstract}


\shortauthors{The CHIME Collaboration}

\correspondingauthor{Simon Foreman, Juan Mena-Parra}
\email{sforeman@perimeterinstitute.ca, jdmena@mit.edu}

\newcommand{\UBC}{Department of Physics and Astronomy, University of British Columbia, Vancouver, BC, Canada}
\newcommand{\MITP} {Department of Physics, Massachusetts Institute of Technology, Cambridge, MA, USA}
\newcommand{\MITK} {MIT Kavli Institute for Astrophysics and Space Research, Massachusetts Institute of Technology, Cambridge, MA, USA}
\newcommand{\TRU}{Department of Physical Sciences, Thompson Rivers University, Kamloops, BC, Canada}
\newcommand{\PI}{Perimeter Institute for Theoretical Physics, Waterloo, ON, Canada}
\newcommand{\DRAO}{Dominion Radio Astrophysical Observatory, Herzberg Astronomy \& Astrophysics Research Centre, National Research Council Canada, Penticton, BC, Canada}
\newcommand{\UBCO}{Department of Computer Science, Math, Physics, and Statistics, University of British Columbia-Okanagan, Kelowna, BC, Canada}
\newcommand{\McGill}{Department of Physics, McGill University, Montreal, QC, Canada}
\newcommand{\UofTastro}{David A.\ Dunlap Department of Astronomy \& Astrophysics, University of Toronto, Toronto, ON, Canada}
\newcommand{\UofTphys}{Department of Physics, University of Toronto, Toronto, ON, Canada}
\newcommand{\WVU} {Department of Computer Science and Electrical Engineering, West Virginia University, Morgantown WV, USA}
\newcommand{\WVUA} {Department of Physics and Astronomy, West Virginia University, Morgantown, WV, USA}
\newcommand{\WVUGWAC} {Center for Gravitational Waves and Cosmology, West Virginia University, Morgantown, WV, USA}
\newcommand{\Yale}{Department of Physics, Yale University, New Haven, CT, USA}
\newcommand{\YaleA}{Department of Astronomy, Yale University, New Haven, CT, USA}
\newcommand{\Dunlap}{Dunlap Institute for Astronomy and Astrophysics, University of Toronto, Toronto, ON, Canada}
\newcommand{\RRI}{Raman Research Institute, Sadashivanagar,   Bengaluru, India}
\newcommand{\ASIAA}{Institute of Astronomy and Astrophysics, Academia Sinica, Taipei, Taiwan}
\newcommand{\CITA}{Canadian Institute for Theoretical Astrophysics, Toronto, ON, Canada}
\newcommand{\CIFAR}{Canadian Institute for Advanced Research, 180 Dundas St West, Toronto, ON, Canada }
\newcommand{\WVUphysastro} {Department of Physics and Astronomy, West Virginia University, Morgantown, WV, USA}

\collaboration{100}{The CHIME Collaboration:}
\author[0000-0001-6523-9029]{Mandana Amiri}
\affiliation{\UBC}
\author[0000-0003-3772-2798]{Kevin Bandura}
\affiliation{\WVU}
\affiliation{\WVUGWAC}
\author{Anja Boskovic}
\affiliation{\UBC}
\author[0000-0003-0173-6274]{Tianyue Chen}
\affiliation{\MITK}
\author[0000-0001-6509-8430]{Jean-Fran\c{c}ois Cliche}
\affiliation{\McGill}
\author[0000-0001-8123-7322]{Meiling Deng}
\affiliation{\DRAO}
\affiliation{\PI}
\affiliation{\UBC}
\author[0000-0003-2381-9804]{Nolan Denman}
\affiliation{\UofTastro}
\author[0000-0001-7166-6422]{Matt Dobbs}
\affiliation{\McGill}
\author[0000-0002-6899-1176]{Mateus Fandino}
\affiliation{\UBC}
\affiliation{\TRU}
\author[0000-0002-0190-2271]{Simon Foreman}
\affiliation{\PI}
\affiliation{\DRAO}
\author[0000-0002-1760-0868]{Mark Halpern}
\affiliation{\UBC}
\author[0000-0002-8513-5603]{David Hanna}
\affiliation{\McGill}
\author[0000-0001-7301-5666]{Alex S. Hill}
\affiliation{\UBCO}
\affiliation{\DRAO}
\author[0000-0002-4241-8320]{Gary Hinshaw}
\affiliation{\UBC}
\author[0000-0003-4887-8114]{Carolin H\"ofer}
\affiliation{\UBC}
\author[0000-0002-3354-3859]{Joseph Kania}
\affiliation{\WVUphysastro}
\author[0000-0003-3405-7470]{Peter Klages}
\affiliation{\UofTastro}
\author[0000-0003-1455-2546]{T.L. Landecker}
\affiliation{\DRAO}
\author[0000-0001-8064-6116]{Joshua MacEachern}
\affiliation{\UBC}
\author[0000-0002-4279-6946]{Kiyoshi Masui}
\affiliation{\MITK}
\affiliation{\MITP}
\author[0000-0002-0772-9326]{Juan Mena-Parra}
\affiliation{\MITK}
\author[0000-0001-8292-0051]{Nikola Milutinovic}
\affiliation{\UBC}
\author[0000-0002-2626-5985]{Arash Mirhosseini}
\affiliation{\UBC}
\author[0000-0002-7333-5552]{Laura Newburgh}
\affiliation{\Yale}
\author{Rick Nitsche}
\affiliation{\UBC}
\author[0000-0002-2465-8937]{Anna Ordog}
\affiliation{\UBCO}
\affiliation{\DRAO}
\author[0000-0003-2155-9578]{Ue-Li Pen}
\affiliation{\CITA}
\affiliation{\Dunlap}
\affiliation{\ASIAA}
\affiliation{\PI}
\author[0000-0002-9516-3245]{Tristan Pinsonneault-Marotte}
\affiliation{\UBC}
\author[0000-0002-5283-933X]{Ava Polzin}
\affiliation{\YaleA}
\author[0000-0001-6967-7253]{Alex Reda}
\affiliation{\Yale}
\author[0000-0003-3463-7918]{Andre Renard}
\affiliation{\Dunlap}
\author[0000-0002-4543-4588]{J. Richard Shaw}
\affiliation{\UBC}
\author[0000-0003-2631-6217]{Seth R. Siegel}
\affiliation{\McGill}
\author[0000-0001-7755-902X]{Saurabh Singh}
\affiliation{\McGill}
\affiliation{\RRI}
\author[0000-0002-6873-2094]{Rick Smegal}
\affiliation{\UBC}
\author[0000-0002-7436-2325]{Ian Tretyakov}
\affiliation{\UofTphys}
\author{Kwinten Van Gassen}
\affiliation{\UBC}
\author[0000-0003-4535-9378]{Keith Vanderlinde}
\affiliation{\UofTastro}
\affiliation{\Dunlap}
\author[0000-0002-1491-3738]{Haochen Wang}
\affiliation{\MITK}
\affiliation{\MITP}
\author[0000-0002-6669-3159]{Donald V. Wiebe}
\affiliation{\UBC}
\author{James S. Willis}
\affiliation{\Dunlap}
\author[0000-0001-7314-9496]{{\rm and}  Dallas Wulf}
\affiliation{\McGill}

\section{Introduction}
\label{sec:intro}

The emergence of cosmic acceleration -- the increasingly rapid expansion of the Universe since redshift $\sim$1.5 -- has signalled that either a gravitationally repulsive dark energy dominates the energy density of the Universe today, or that Einstein's General Relativity does not correctly describe gravity on cosmological scales.  The impact of this discovery on fundamental physics and astrophysics is revolutionary, and decoding the physics of cosmic acceleration requires new, higher-quality measurements of the expansion rate of the Universe as a function of time.

Nature has provided a standard ruler with which to measure the expansion history of the Universe: the baryon acoustic oscillation (BAO) scale \citep{Seo:2003,Seo:2007}.  Acoustic waves propagated through the primordial plasma in the early Universe for a fixed amount of time -- 379,000 years -- until the plasma cooled and became neutral gas, primarily hydrogen.  The distance these waves travelled has been precisely measured in the Cosmic Microwave Background (CMB) radiation \citep{Hinshaw:2013,PlanckParam:2018}.  These waves imparted slight baryonic over-densities on the BAO scale which are imprinted in the large-scale distribution of matter in the Universe.  By measuring cosmic structure as a function of time (i.e., redshift), we can deduce the apparent size of the BAO scale as a function of cosmic epoch, and hence the expansion history of the Universe.

The signature of BAO was first detected in large scale structure, at redshift $z\approx0.35$ \citep{Eisenstein:2005} and $z\approx0.2$ \citep{Cole:2005}, using galaxies as tracers.  More recently, measurements of the BAO scale at redshifts up to $z \sim 0.8$ have been made by observing the distribution of optically-detected galaxies, using either spectroscopic \citep{Percival:2007,Beutler:2011,Blake:2011,Padmanabhan:2012,Anderson:2012,Ross:2015,Alam:2017,Alam:2021} or photometric \citep{Seo:2012,Abbott:2019,Abbott:2021} catalogs, and at higher redshifts in Lyman-alpha systems (e.g.\ \citealt{Busca:2013,Slosar:2013,duMasdesBourboux:2020}) and quasars \citep{Ata:2018,Neveux:2020}.
All of these efforts produce measurements of the distance-redshift relation that are consistent with the notion that the dark energy is a cosmological constant with an equation of state $p_{\rm DE} = -\rho_{\rm DE}$ ($w=-1$) \citep{Alam:2021}.  However, improved precision in the distance-redshift relation is still possible due to the fact that only a small fraction of the accessible large scale structure has been mapped to date, especially at redshifts greater than 1.  Several efforts are ongoing to map ever-larger volumes of large-scale structure to yield improved precision, particularly by the ground-based experiments DES \citep{DES:2016} and DESI \citep{DESI:2016}, and the space-based telescopes Roman \citep{WFIRST:2019}, Euclid \citep{Euclid:2018}, and SPHEREx \citep{Dore:2014}. 

A complementary way to map the large scale distribution of matter, called hydrogen intensity mapping, has been successfully demonstrated by several analyses \citep{Pen:2009,Chang:2010,Masui:2013,Switzer:2013,Anderson:2018,Wolz:2021}.  The technique uses modest-angular-resolution observations of redshifted \tcm emission from the hyperfine transition of neutral hydrogen to trace the distribution of hydrogen gas, and thus matter, in the Universe.  Hydrogen intensity mapping allows the apparent angular and radial BAO scale to be measured through cosmic history without the expensive and time-consuming step of resolving individual galaxies. 

While the intensity mapping technique was first demonstrated using conventional radio telescopes, a dedicated instrument is needed to make a measurement of cosmic acceleration with the sensitivity required to test dark energy models.  In order to reduce power spectrum uncertainties due to sample variance, we need to map cosmic hydrogen over nearly half the sky, which requires a telescope with a {\em much} higher mapping speed than previously existed.

\begin{figure}
    \includegraphics[width=\linewidth,keepaspectratio]{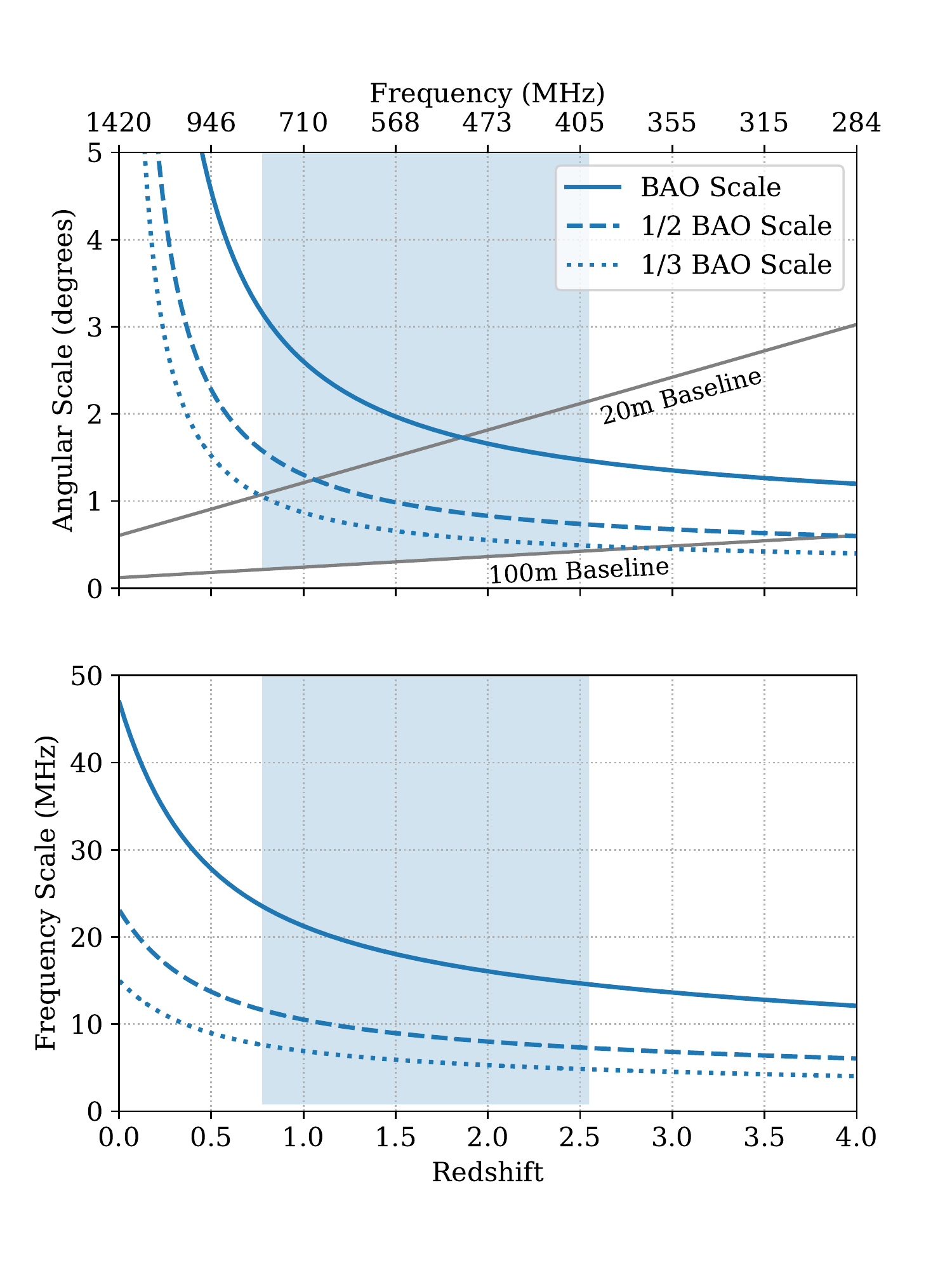}
    \caption{The BAO scale ($r_{\rm BAO}\approx\SI{150}{\mega\parsec}$ comoving) compared to CHIME's angular and frequency resolution.   
     {\em Top:} The blue solid curve shows the angular scale associated with $r_{\rm BAO}$, while the other linestyles show the first few harmonics (corresponding to the peaks of successive BAO ``wiggles" in $k_\perp$ Fourier space, located at multiples of $k_{\rm BAO} \approx 2\pi / r_{\rm BAO}$).  The shaded region shows the range of angular scales accessible to CHIME as a function of frequency, for antenna baselines ranging from \SIrange{0.3}{100}{\metre}. The grey straight lines show the angular resolution associated with feed separations of \SI{20}{\meter} and \SI{100}{\meter}.  {\em Bottom:} the solid curve shows the frequency separation associated  with   the line of sight  BAO diameter for \tcm radiation as a function of redshift. 
     The other linestyles indicate the frequency resolution required to resolve the first two BAO harmonics in $k_\parallel$ Fourier space. 
     CHIME's frequency resolution is \SI{390}{\kilo\hertz}.  
     For all curves, we take $H_0 = \SI[per-mode = reciprocal-positive-first]{70}{\kilo\metre\per\second\per\mega\parsec}$, $\Omega_m = 0.3$, and $\Omega_\Lambda = 0.7$.    In both panels the shaded region denotes  the frequency and redshift coverage of CHIME.
    } \label{fig:bao_scale}
\end{figure}

As described in this paper, the Canadian Hydrogen Intensity Mapping Experiment (CHIME) consists of an array of four \SI{20}{\meter} $\times$ \SI{100}{\meter} cylindrical telescopes, with no moving parts or cryogenic systems, which can observe the northern sky every day over the frequency range \SIrange[range-units = single, range-phrase=-]{400}{800}{\mega\hertz}.  As shown in \cref{fig:bao_scale}, CHIME's angular resolution of $\sim\ang{;40;}$ and frequency resolution of \SI{390}{\kilo\hertz} are well suited to measuring the BAO scale in \tcm emission over the redshift range $0.8 \leq z \leq 2.5$.  This range covers the important epoch in cosmic history when the expansion transitioned from decelerating to accelerating \citep{Riess:2004}.

CHIME's large scale structure map will constitute the largest survey of the Universe ever undertaken.  In addition to facilitating measurements of the BAO scale, CHIME data will constitute a rich dataset for cross-correlating with other probes of large scale structure.  In a companion paper, we present a CHIME detection of cosmological \tcm emission in cross correlation with three separate tracers of large scale structure extracted from the Sloan Digital Sky Survey \citep{stacking2021}.

The main challenge associated with \tcm intensity mapping is the very bright synchrotron foreground emission from the Milky Way and from other nearby galaxies (e.g.\ \citealt{Santos:2005,Liu:2012}).
We are investigating several approaches to foreground filtering and subtraction, that rely in various ways on recognizing the difference between the smooth Galactic spectrum and the chaotic BAO spectrum along each line of sight (e.g.\ \citealt{2015shaw}).
Separately, we note here that CHIME provides a detailed and high signal-to-noise ratio dataset for probing the interstellar medium.

CHIME will map the northern sky in polarization, and we will apply the Faraday synthesis technique \citep{brentjens:2005} to obtain three-dimensional information about magnetized interstellar structures in the Galaxy. This dataset will be without precedent in the Northern hemisphere and will form a component of the Global Magneto-Ionic Medium Survey (GMIMS). GMIMS is the first effort to measure the all-sky three-dimensional structure of the Galactic magnetic field, using telescopes around the world to obtain maps with sensitivity to the range of Faraday depth structures we expect in the diffuse medium \citep{Wolleben:2019,Wolleben:2021}; the CHIME frequency range is a critical component of GMIMS.

CHIME has the same collecting area as the Green Bank telescope and also has a large fractional bandwidth and large instantaneous field of view.
It scans the entire sky visible from Southern Canada at daily cadence with sub ms sampling.  The data from CHIME are passed commensally to separate instruments which search for fast radio bursts (FRBs), monitor
known pulsars visible from the site and search at high spectral resolution for 21-cm line absorption systems.  Additionally, CHIME supports very long baseline interferometry \citep[VLBI,][]{Cassanelli2021} observations with other telescopes.

In \secref{sec:instrument}, we present an overview of the CHIME instrument, including its mechanical design, analog and digital systems, and low-level data processing. In \secref{sec:beams}, we describe recent progress in characterizing CHIME's primary beam response. \secref{sec:Performance} is devoted to various performance metrics based on the first three years of science data, including sources of data loss, gain stability, thermal noise, excision of radio-frequency interference, and preliminary sky maps. We conclude in \secref{sec:Summary}, discussing the outlook for future \tcm measurements and showing an idealized forecast for the precision with which CHIME could measure the cosmic expansion history in the absence of foregrounds or systematics. (The details of this forecast are included in Appendix~\ref{sec:appforecast}.)

\section{Instrument and Low-Level Processing}
\label{sec:instrument}

\begin{figure*}
    \centering
    \includegraphics[width=0.98\linewidth,keepaspectratio]{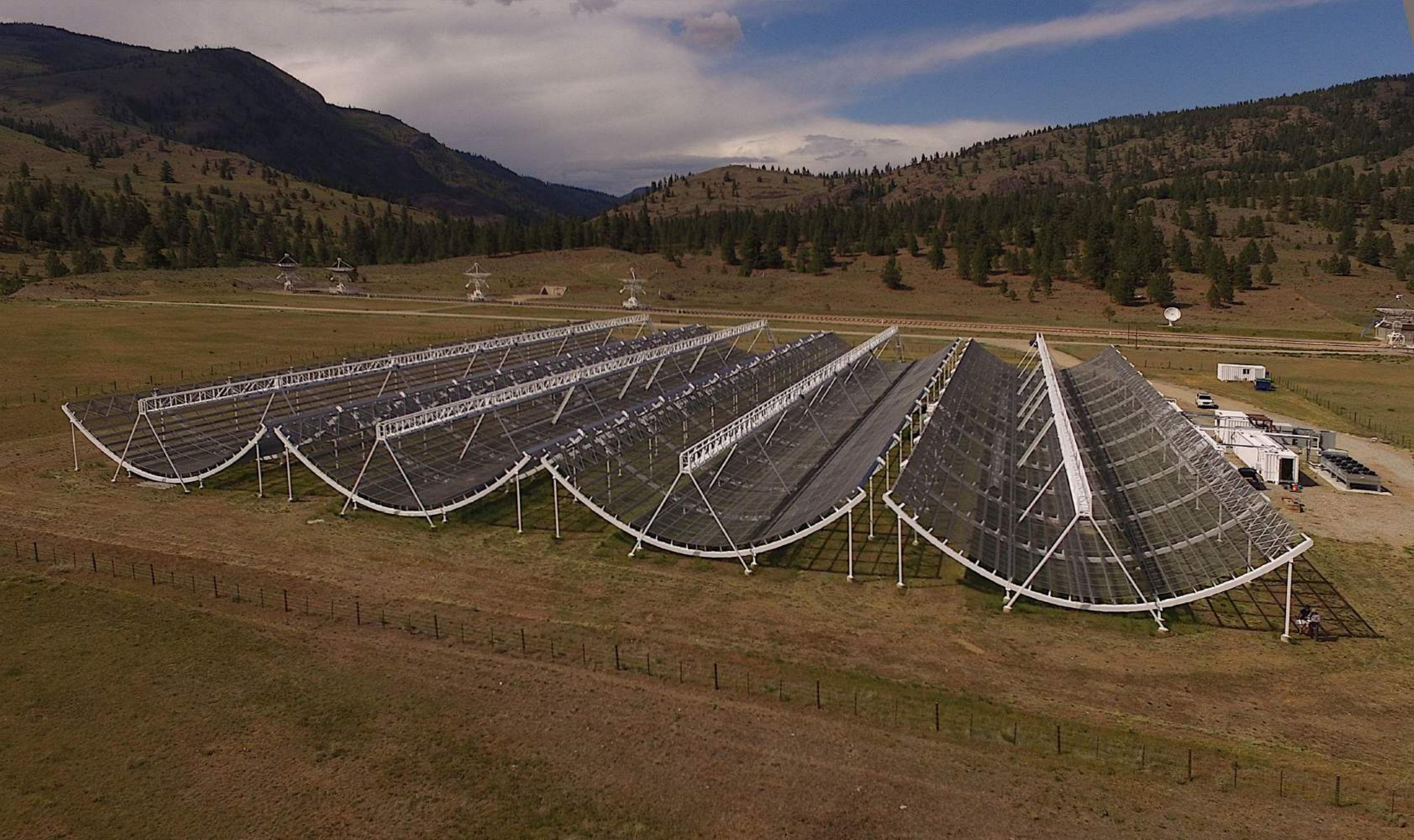}
    \caption{ A photograph of CHIME looking North.  The parabolic reflecting surface of each of the four cylinders is \SI{20}{\metre} aperture, \SI{5}{\metre} focal length and \SI{100}{\metre} long.  The two people standing at the southeast corner (right, foreground) help to show the physical size.  256 dual-polarization antennas are placed along the central \SI{78}{\metre} of the focal line of each cylinder, beneath an $\SI{88}{\metre} \times \SI{0.65}{\metre}$ groundplane.   The focal lines are covered by and suspended from the walkways visible along each cylinder axis.  Signals are amplified at each feed and brought by low-loss coaxial cables to receiver huts located in commercial RF shielded rooms within customized RF-protective shipping containers, one located  between the first and second cylinders, another between the third and fourth.  After band-defining amplification, analog-to-digital conversion, a time-to-frequency transform and half of a `corner-turn', signals are brought from the two receiver huts to additional RF rooms within the white shipping containers seen at the right, where the corner-turn is completed and  a spatial transform and other processing are performed.  The grey and black structure at the far right is an ambient air heat exchanger associated with the water-cooling system for the X-engine in the adjacent RF rooms.  Behind that, also grey, is a \SI{0.5}{\mega\watt} power substation to power the instrument.  CHIME is located at the Dominion Radio Astrophysical Observatory which is protected by law and the adjacent hills from terrestrial radio interference. In the background one can see five dishes of the DRAO Synthesis telescope and a solar radio monitor. }
    \label{fig:chime_photo}
\end{figure*}

CHIME  is a transit radio telescope.  It consists of linear arrays of feeds along the focus of each of four cylindrical parabolic reflectors.  The optical system has no moving parts, and CHIME scans the sky as the Earth turns. 
A photograph of the telescope and surrounding site is shown in \cref{fig:chime_photo}.

\begin{figure}
    \centering
    \includegraphics[width=0.85\linewidth,keepaspectratio]{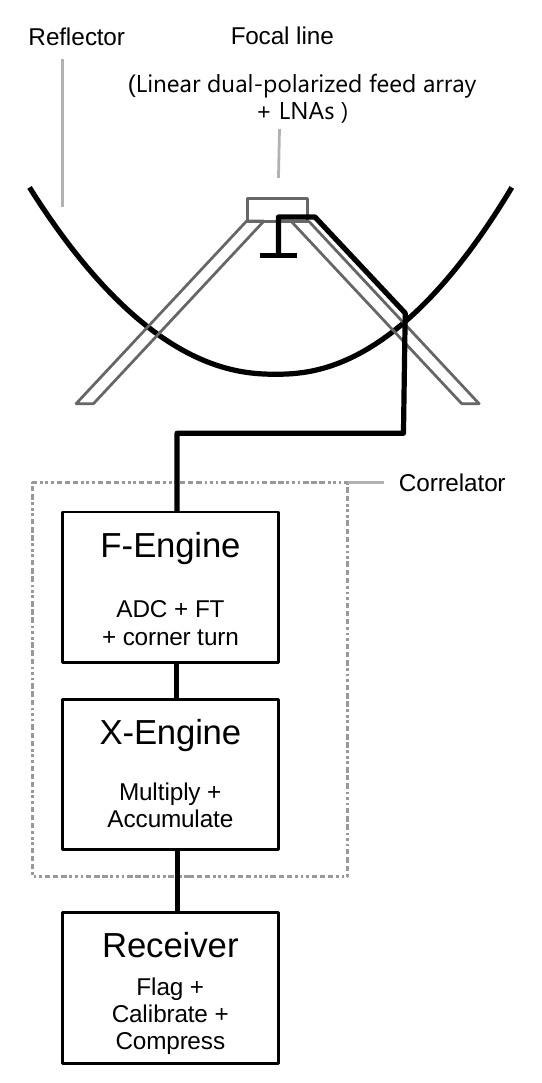}
    \caption{
    Schematic diagram of data flow through CHIME.  Signals focused onto a linear array of  broad-band dual-polarization antennas, are amplified (each polarization separately) using room temperature receivers with a noise performance below \SI{30}{\kelvin} that amplify and filter the signals to \SIrange{400}{800}{\mega\hertz}. The correlator is an FX design, where the F-engine digitizes and channelizes the signals from the 2048 analog receivers and also implements the majority
    of a \textit{corner-turn} network that rearranges the channelized data for spatial correlation.   The X-engine completes the corner turn and  performs the cross-multiplications and averaging to compute the $\Nfeed^2$ spatial correlation matrix separately at each frequency. The X-engine also performs additional  real-time data processing operations  to beamform and to increase spectral resolution for the pulsar, FRB, and absorber back end instruments. }
    \label{fig:diagram_overview}
\end{figure}

In this section, we walk through the design of the instrument,  showing how its main features have been   designed coherently to meet the performance requirements established in \secref{sec:intro}.  The signal flow is captured schematically in \cref{fig:diagram_overview}
and we will follow this same path in our description: from reflectors which define the field of view, through feeds and analogue electronics, to an FX correlator, and the digital back end we call the data receiver.  

   As described in the introduction, the frequency coverage of CHIME is chosen to interrogate the epoch when dark energy first emerged in the dynamics of the Universe.   A wide observing bandwidth increases the total cosmic signal power and allows interrogation of a wide range in redshift. Limiting the frequency range to cover a factor of two  eases the challenges in antenna design and allows digital sampling in the second Nyquist zone, which permits slower sampling and a substantial savings in the cost of electronics.    CHIME takes advantage of the historic drop in the cost of low noise amplifiers and digital electronics to fill the aperture of its cylindrical reflectors  with radio feeds in one dimension.  In this geometry, every feed scans the full North-South meridian synchronously and simultaneously, and the instrument scans the full overhead sky every day with no moving parts, reducing systematic errors.

\subsection{Site}
\label{sec:site}

CHIME is built at the Dominion Radio Astrophysical Observatory (DRAO), near Penticton, B.C., Canada. DRAO is operated as a national facility for radio astronomy by the National Research Council Canada.  Working at the DRAO has provided the CHIME team with very welcome connections  to a community of experienced radio astronomers and engineers.  

The site is in the White Lake Basin, within the traditional and unceded territory of the Syilx/Okanagan people. Prior to construction we walked the land with elders, and during initial excavation Okanagan Nation observers were present. The site offers flat land protected from radio frequency interference (RFI) by Federal, Provincial, and local regulation and by surrounding mountains.  The climate is semi-arid, with low snowfall levels (relative to other places in Canada), important for a stationary telescope. The DRAO's John A. Galt Telescope, a 26-m steerable single-dish telescope with an equatorial mount, is located \SI{230}{\m} east of the centre of CHIME, and \SI{20}{\m} North. We use the Galt Telescope for holographic beam mapping. The DRAO supports CHIME with roads, AC power,  machine shop access, well-equipped electronics laboratories, office space, and staff accommodation.

\begin{figure}
    \centering
    \includegraphics[width=0.95\linewidth,keepaspectratio]{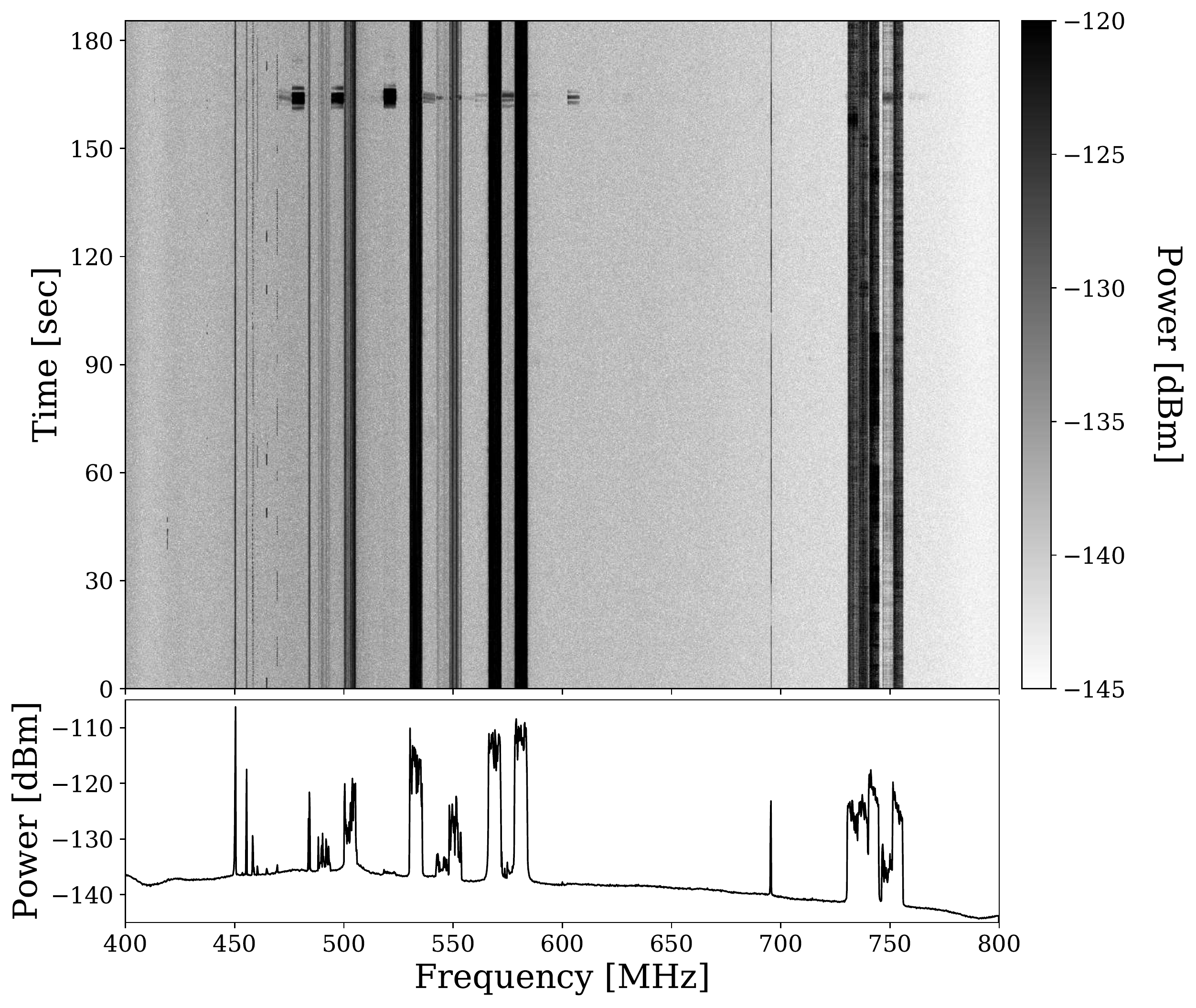}
    \caption{Example of the radio frequency interference (RFI) measured by CHIME antennas. \textit{Top}: Dynamic spectrum obtained from the squared magnitude of the visibility between two E-W polarized  antennas located near the southern end of two adjacent cylinders.  The frequency resolution is \SI{195}{\kilo\hertz}, obtained by direct transform of the raw data rather than through the CHIME PFB pipeline,
    and the time resolution is \SI{100}{\milli\second},  averaged from the \SI{1}{\milli\second} cadence at which this dataset was collected.  The greyscale denotes power in units of \si{\dBm} at a feed in the focal line.  Data were collected on 
    May 4, 2019
    at 16:00 PDT. An intermittent RFI event is visible at \SIrange[range-units = single, range-phrase=-]{460}{600}{\mega\hertz} at around \SI{165}{\second}.
    \textit{Bottom:}  The median value over time of the image shown in the top panel. 
    Notable features include LTE bands at \SIrange[range-units = single, range-phrase=-]{730}{755}{\mega\hertz}, several \SI{6}{\mega\hertz} TV station bands between \SIrange[range-units = single, range-phrase=-]{480}{580}{\mega\hertz}, and UHF repeaters around \SI{450}{\mega\hertz}. The narrow line at \SI{690}{\mega\hertz} is the local oscillator of a nearby synthesis telescope. }
    \label{fig:drao_rfi}
\end{figure}

The mountains around the observatory shield the site from RFI from nearby cities, but a significant portion of the CHIME frequency band is still contaminated by satellites, airplanes, wireless communication, and TV broadcasting bands. This includes LTE bands in the \SIrange{730}{755}{\mega\hertz} range, TV station bands between \SIrange{480}{580}{\mega\hertz}, and UHF repeaters around \SI{450}{\mega\hertz}. These features are clearly visible in the spectrum shown in \cref{fig:drao_rfi}. Besides cell-phone and TV-station bands that are static in nature, there are many sources of intermittent RFI events such as direct transmission from satellites and airplanes, as well as scattering of distant ground-based sources. One such event is visible in \cref{fig:drao_rfi}  from \SIrange{460}{600}{\mega\hertz} at around \SI{165}{\second}.
These scattering events typically appear as \SI{6}{\mega\hertz} wide bursts which last for a few seconds, and are caused by the reflection of distant broadcast TV bands from meteor ionisation trails or aircraft. 

\subsection{Mechanical and Optical Design}
\label{sec:Structure}

\begin{figure}
    \centering
   \includegraphics[width=1.\linewidth,keepaspectratio]{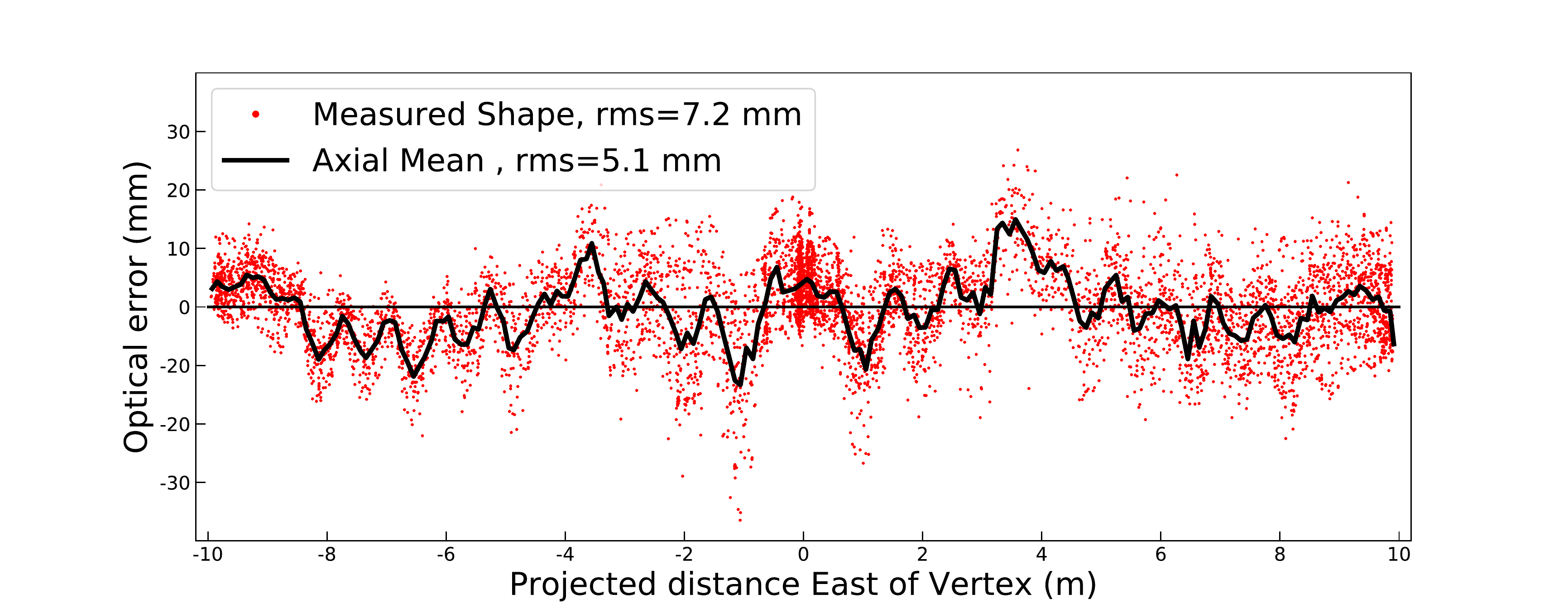}
    \caption{Measured surface error of CHIME Cylinder A compared to a best-fit parabola plotted against cylinder $X$, the horizontal distance East of the vertex. Points on the surface are measured with a surveyor's total station tracking a retro-reflector on a small wheeled cart as it moves over the reflector surface.  The survey accuracy is nominally \SI{3}{\milli\metre} in \SI{100}{\metre} which has not been subtracted from the scatter seen here. The quantity plotted is half the optical delay error from the sky to the reflector to the focus, equivalent to simple surface error for a flat mirror. There are two main terms in the shape visible here.  The mesh which forms the surface appears to sag approximately \SI{1}{\centi\metre} in each of the \SI{1}{\meter} gaps between the supporting purlins compared to the desired parabolic shape.  Additionally one can see that the rolled parabolic truss is formed of three segments which also depart from the desired shape by near to \SI{1}{\centi\metre}.  The net surface deviation is \SI{7.2}{\milli\metre} RMS, or $\lambda / 50$ at CHIME's shortest wavelength.  These deviations are clearly coherent over the entire structure of a cylinder.  Each of the four cylinders looks similar to this one example in all the key features.  }
    \label{fig:surface_error}
\end{figure}

The  design of CHIME is focused on enabling the measurement of BAO across the redshift
range where dark energy begins to impact the dynamics of the Universe. The spectral response,
reflector geometry and RF feeds are designed together to form an instrument
tuned to perform this measurement in a way that allows control and
characterization of systematic errors.  Total estimated cost was also a strong
design driver.

Measuring BAO in the redshift range  from 0.8 to 2.5 covers the region of
interest for probing
dark energy, and fills in a redshift gap which is sparsely covered by optical
measurements.   At these wavelengths, sufficient angular resolution
to resolve  BAO features in the power spectrum of the sky is easily achieved by a \SI{100}{\meter}
baseline (see \cref{fig:bao_scale}). 

An East-West  array of cylindrical, \SI{100}{\meter}-long reflectors each coupled to a linear feed
array along its focus meets these needs.   Such a system scans a North-South
stripe of the sky interferometrically and observes 
most of the 3/4  of the celestial
sphere visible from our site every day as the Earth turns.  Given that each
feed in this system requires a feed response of $\pm$ \SI{1}{\radian} along the
cylinder axis, choosing a reflector shape to be an $f / 0.25$ parabola allows the
use of feeds with approximately symmetric angular response patterns.  At this
f-ratio, the focus is level with the edges of the reflector, protecting the
feed array from terrestrial radiation.

The required  East-West separation of feed arrays can be achieved by varying the
number of cylinders and the aperture of each.  Deploying four \SI{20}{\meter}-aperture
reflectors was chosen as a reasonable compromise of costs of the reflectors and
costs of the electronics to collect and process the signals while still
providing massively redundant measurements of the most important ($u,v$) baselines.
This redundancy simultaneously provides lower system noise and protection from minor
variations of the response of individual elements of the instrument.

 We describe the layout of the telescope in a 3D Cartesian system with $+Z$ pointing to the zenith, $+X$ to the East and $+Y$ North.  Thus, the linear feed arrays are oriented along the $Y$ axis with $X$ and $Y$ polarization directions.  When we describe the angular response of the telescope we use the orthographic projected angles $x$ and $y$ defined in section \secref{sec:beam_data}.

A steerable telescope can be turned to low elevation angles to shed snow, but this is
not possible with the CHIME reflectors. Therefore, the reflector surface is formed with
wire mesh to allow snow to fall though. Larger gaps in the mesh shed snow with more
assurance but also allow thermal radiation from the ground to leak through to the
focus, raising the system temperature. Heavy wire gauge lowers the RF leakage. Using
tools from \citet{Mumford4066356}, 
we evaluated RF leakage across the CHIME band of
commercially available sheets of heavy-duty mesh, settling on \SI{19}{\milli\meter}
spacing woven mesh made of \SI{2.2}{\milli\meter} diameter galvanized steel. This
material is easily available in large flat  sheets. The leakage through these sheets add from
\SIrange{1}{2}{\kelvin} to the system temperature across the CHIME band. 

The central \SI{78}{\meter} of each focal line is instrumented with feeds and low noise amplifiers (LNAs). The \SI{100}{\meter}-long
reflectors intercept the beams of the end feeds out to a zenith angle of
\SI{65}{\degree}. These end feeds do see more RFI and more thermal loading than
typical feeds, and this is accounted for in our analysis pipeline (see \secref{sec:operations}).

The reflector structure itself was designed in collaboration with 
Empire Dynamic Systems, Coquitlam
BC, a civil engineering firm with substantial experience building astronomical
facilities  using standard steel fabrication techniques.  Each \SI{8}{\meter}-long section
of the reflector is formed from three panels. These are rolled steel beams
connected by \SI{8}{\meter} long purlins running parallel to the axis, assembled on site and
lifted into place.  The mesh reflector surface is bolted to the purlins once the
structure of an entire cylinder is complete.  The structure is supported on
steel legs which stand on  cement footings placed deep enough that the base is below the anticipated frost depth.

The surface accuracy, shown in \cref{fig:surface_error}, corresponds to
between $\lambda/50$ and $\lambda/100$ across the CHIME band.  The surface errors
are dominated by two terms: a consistent imperfect shape formed by  the purlins
welded to the curved steel frames and by   almost \SI{1}{\centi\meter} of sag of the mesh in each
of the \SI{1}{\meter} gaps between purlins.  These perturbations are coherent for the full length of each cylinder in
the North-South direction and were measured by tracking a
retro-reflector across the full surface using a surveyor's total station.

The ground plane of the linear feed array, at the focus of the cylinder, is just wide
enough that it can shield the narrowest 
building code-compliant
walkway placed above it. Removable panels of the walkway facilitate access to amplifiers and cables.
Access stairs at the North end of every focal line are in line with the optic
axis and the same width as the ground plane.

Observations of bright point sources acquired with CHIME exhibit an unexpected phase error that scales linearly with east-west baseline distance, frequency, and the sine of the source's zenith angle.  This can be explained by a clockwise rotation (looking down from the sky) of the telescope structure by $0.071 \pm \SI{0.004}{\degree}$ with respect to the true astronomical north-south direction.  Alternatively, it can be explained by a linear offset in the north-south positions of the feeds from one cylinder to the next of $-2.73 \pm \SI{0.15}{\centi\meter}$ per cylinder (from west to east).  The quoted values were measured by minimizing the phase of visibilities when beamformed to the location of \SI{24}{} bright point sources ranging in declination from \SIrange{5}{65}{\degree}.  We are currently unable to distinguish between these two explanations due to confusion between this effect and the phase of the beam response as a function of hour angle.  We assume an overall rotation of the telescope when constructing the baseline distances that are used in our analyses.

\subsection{Analog System}
\label{sec:AnalogSystem}

\newcommand{\vect}[1]{\vec{#1}}
\begin{figure}
    \centering
    \includegraphics[width=0.98\linewidth,keepaspectratio]{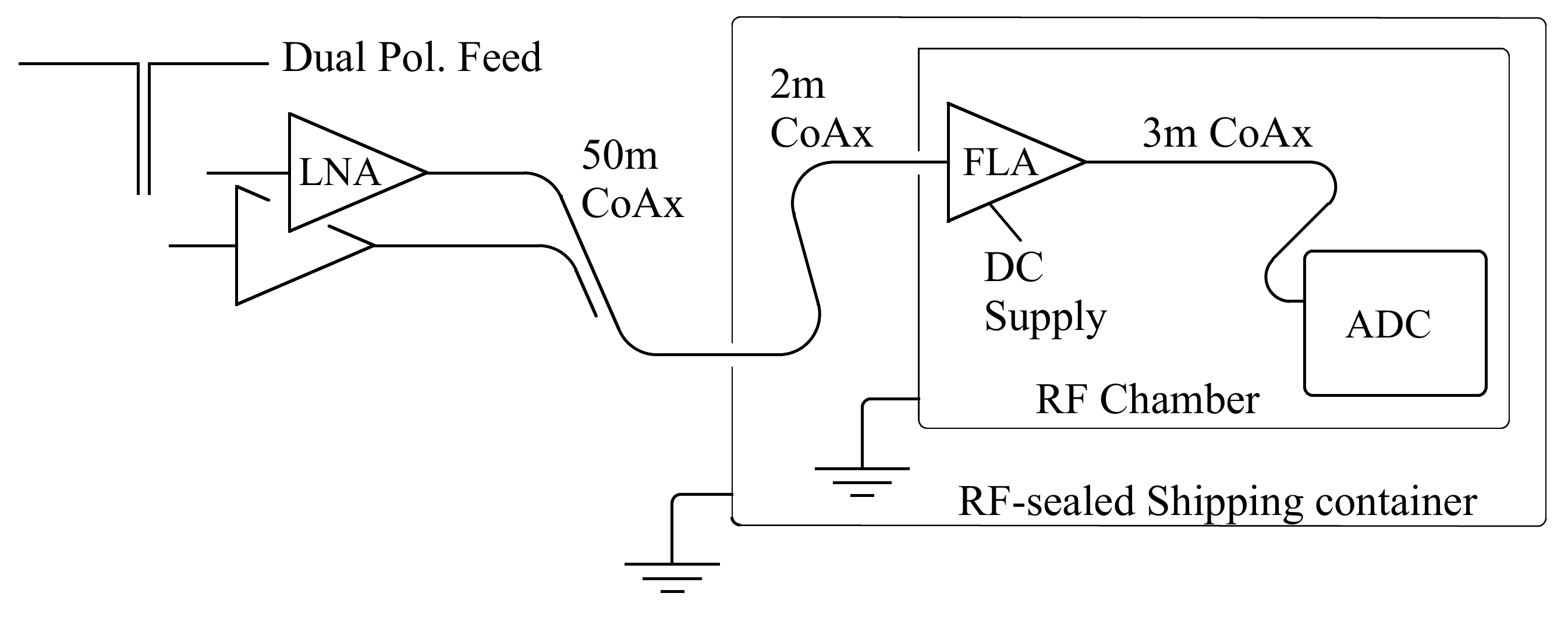}
    \caption{Block diagram of one channel of the analog front-end. A  signal  from  one  port of a
dual polarization antenna  is connected to an LNA, and carried
via a $\sim$\SI{1}{\metre} cable and a \SI{50}{\metre} low-loss coaxial cable into the double shielded Receiver hut. FLAs, mounted on
the inside surface of the inner RF chamber wall define the instrument
passband, provide additional gain, and transmit the signal to the ADCs. A DC Power Supply in the RF chamber is used to power the FLAs, which in turn, provide DC power to the LNAs over the coaxial cables using the built-in bias-tees in the LNAs and FLAs. Any FLA/LNA chain can be turned on or off remotely.
The antennas, amplifiers and \SI{50}{\metre} cables are all labelled with bar-codes, which are scanned upon assembly, allowing their interconnections to be documented in a database.   S-parameters have been measured for every individual component over the full CHIME Band.
}
    \label{fig:diagram_analog}
\end{figure}

\begin{figure}
   \centering
   \begin{tabular}{ll}
   \includegraphics[width=0.45\linewidth,keepaspectratio]{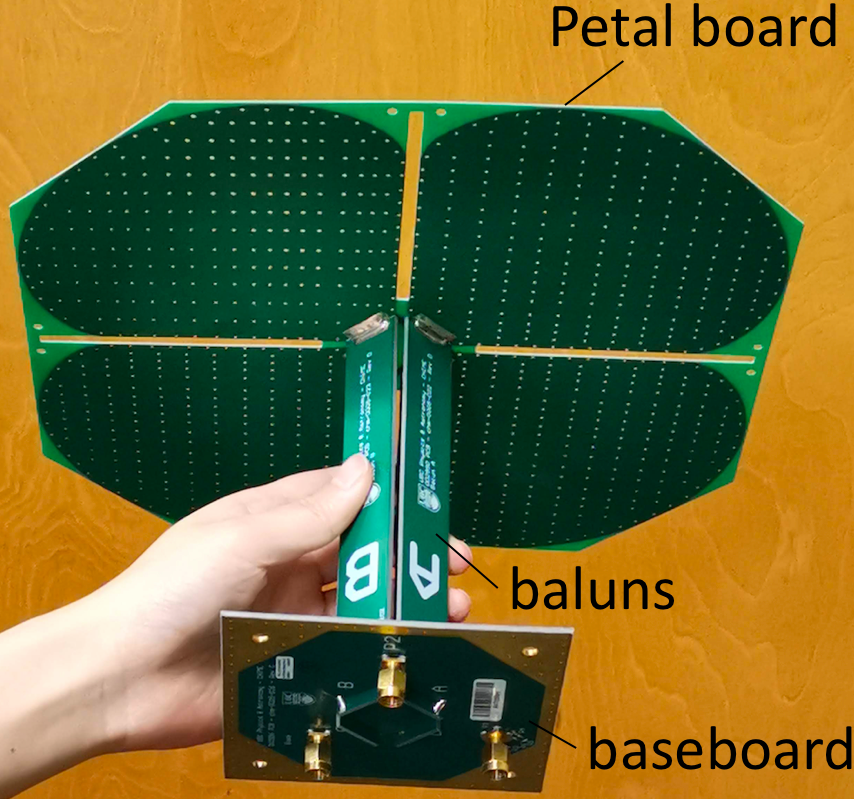} &
   \includegraphics[width=0.45\linewidth,keepaspectratio]{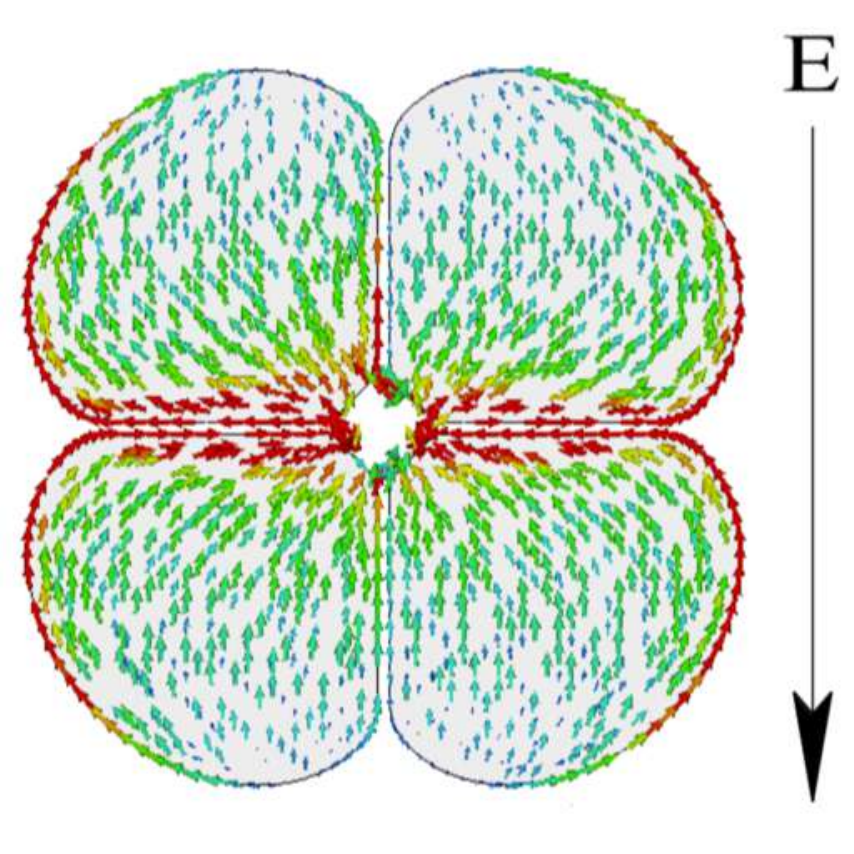} \\
   \multicolumn{2}{c}{\includegraphics[width=0.95\linewidth,keepaspectratio]{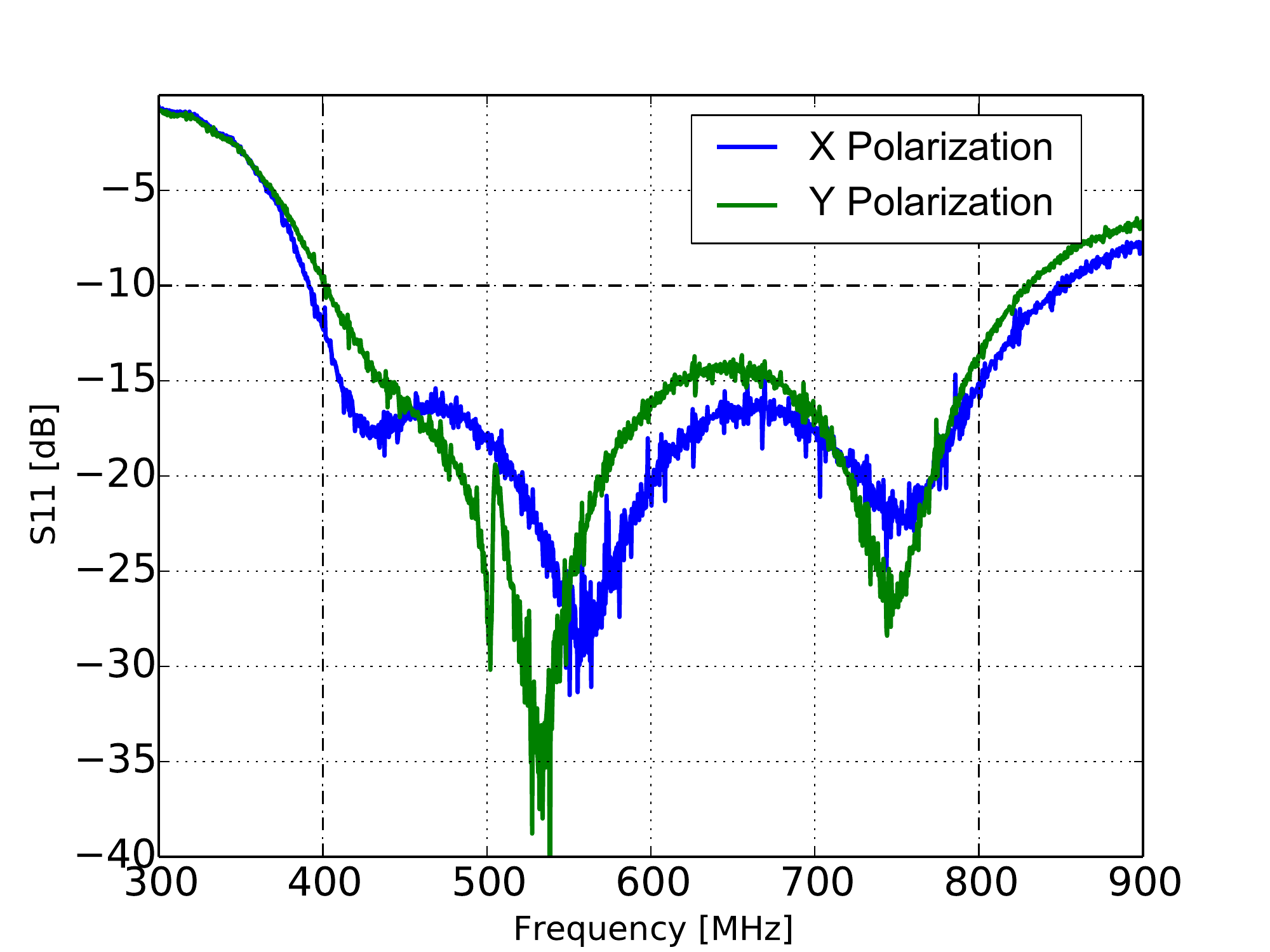}}
   \end{tabular}
   \caption{{\em Top Left}: A photograph of a CHIME cloverleaf antenna element. {\em Top Right}: The simulated current pattern on the petal-shaped radiating board of the cloverleaf antenna at \SI{600}{\mega\hertz}. Feeds are constructed using commercial printed circuit board (PCB) materials and techniques, resulting in precise and economic antennas.
   {\em Bottom}: Measured S11 of the two polarizations of the cloverleaf antenna. The design substantially exceeds the goal to have a return loss of more than 10 dB over the full CHIME band, illustrated by the horizontal dashed line.
   }
     \label{fig:cloverleaf_antenna}
\end{figure}

 The analog signal path consists of 256 dual-polarized \textit{cloverleaf} antennas \citep{Deng_2020,Deng_2014} in a linear feed-array along the focus of each cylinder, with each linear polarization coupled to a low-noise amplifier (LNA), coaxial cables, a band-defining filter and  amplifier (FLA) and the input to an analog-to-digital converter (ADC).  A single channel is shown in \cref{fig:diagram_analog}.  The system components have been designed together
 to optimize overall performance for interferometric measurement of the BAO.
 With 256 dual-polarized antennas per cylinder and four cylinders, there are 1024 antennas and 2048 analog signal chains.

\begin{figure*}
    \centering
   \includegraphics[trim={2cm 0 2cm 0},width=1.\linewidth,keepaspectratio]{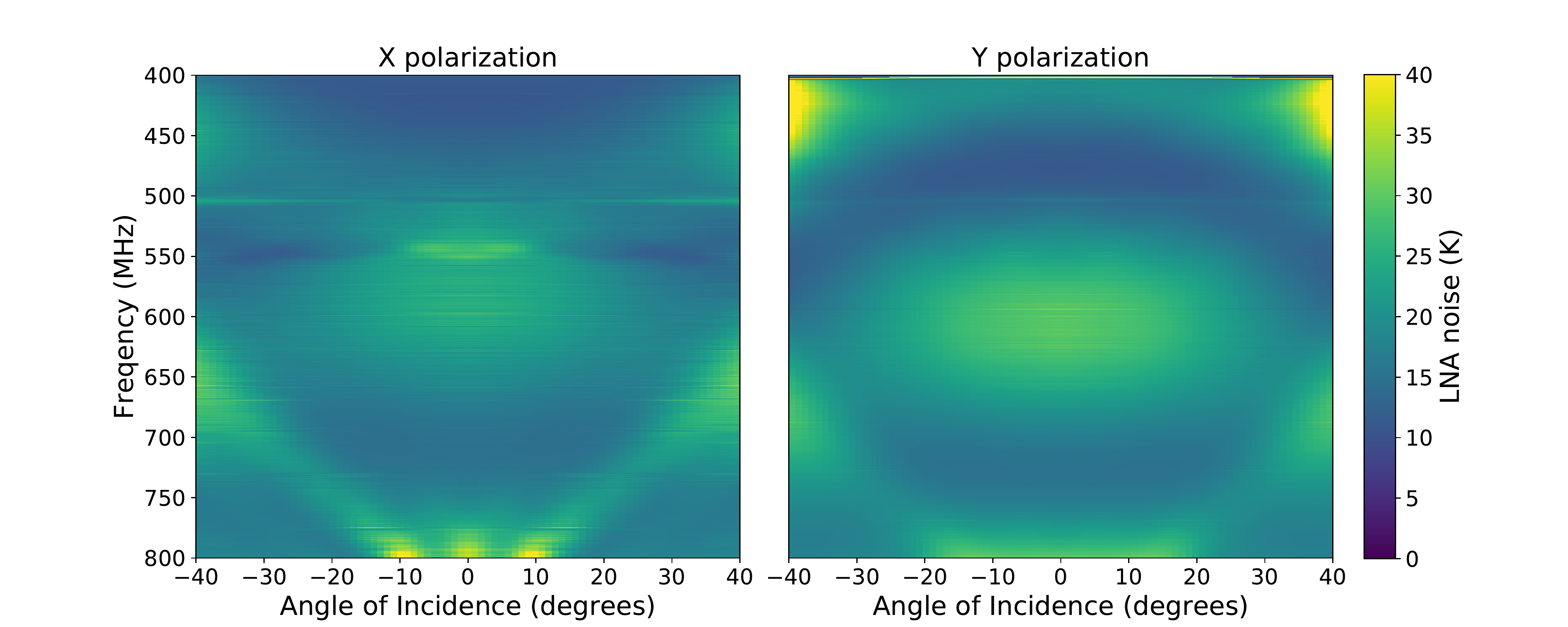}
    \caption{Modeled LNA noise temperature for the central element of a linear feed array.
    Using measured feed-to-feed coupling parameters ($S_{21},~S_{31},~S_{41},~\dots$), the \textit{effective} impedance for the central feed in a linear array has been calculated as a function of frequency and incident angle.  The noise is calculated using this impedance and a high-fidelity model of our LNA performance.  Because of the stronger coupling for $X$ polarization, particularly in the vicinity of the feature near \SI{550}{\mega\hertz},  15 elements are used in the  $X$-impedance model, and 13 for $Y$. The sharp feature at \SI{500}{\mega\hertz} in both polarizations is a property of isolated CHIME antennas.
    }
    \label{fig:scan_angle}
\end{figure*}

 \begin{figure*}
    \centering
   \includegraphics[trim={3cm 0 3cm 0},width=0.75\linewidth,keepaspectratio]{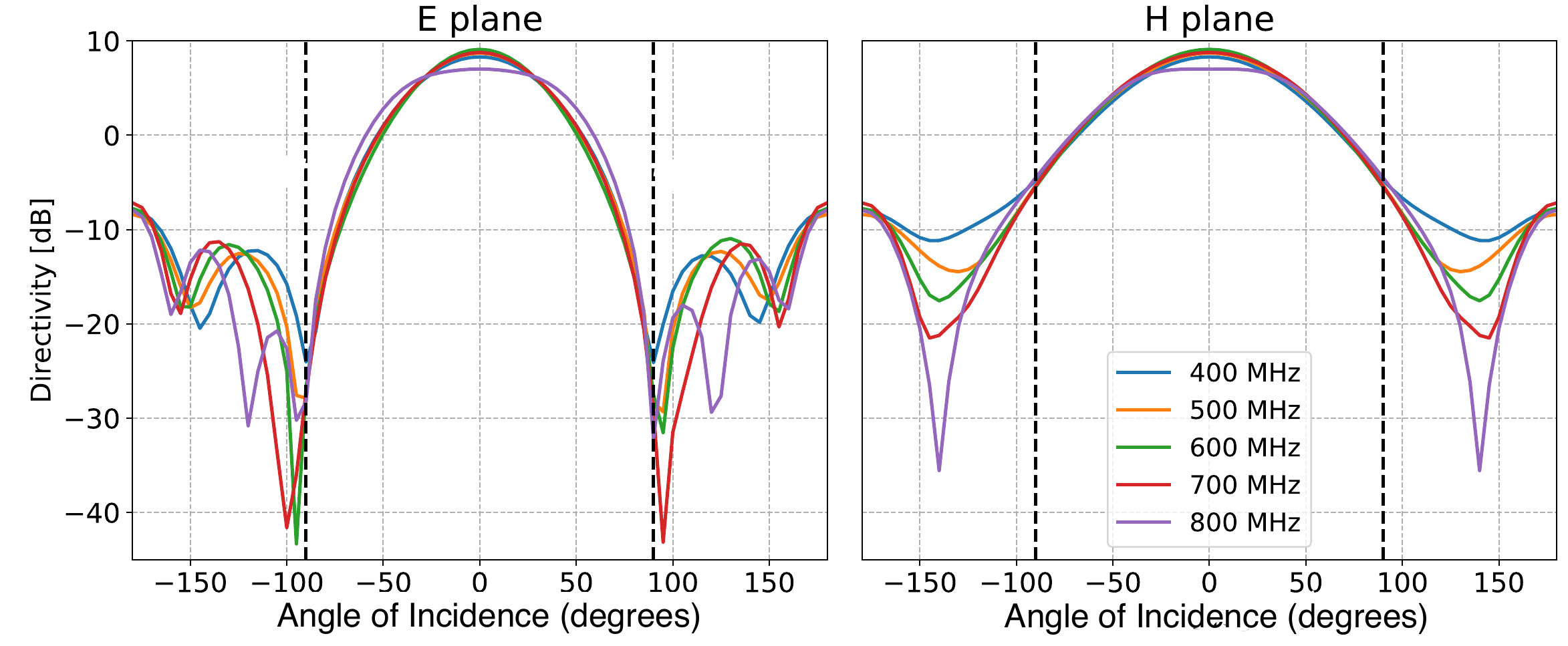}
    \caption{ Sections of the modelled angular response of a CHIME feed in the E and H planes for several frequencies across the CHIME band.  Although the dual-feed  antenna is symmetric with respect to its $X$ and $Y$ axes, each beam is slightly elliptical between its E and H planes, and therefore $Y$-polarized and $X$-polarized beams illuminate the reflector differently. The vertical dashed lines in the E and H panels are at $\pm 90^\circ$, corresponding to the edges of the reflector for $X$ and $Y$ polarized radiation.}
    \label{fig:feed_response}
\end{figure*}

 Each cloverleaf antenna, together with its image antenna in the ground plane, has an effective focus nominally located at the ground plane, independent of frequency.  The radiating board, whose current pattern is shown in \cref{fig:cloverleaf_antenna}, is designed to have a smooth petal shape in order to be free of resonances and match to the CHIME LNA over the octave bandwidth from \SIrange{400}{800}{\mega\hertz}. \citet{Deng_2020} described this optimization. For each linear polarization, pairs of balanced signals from the four petals are combined via a tuned set of microstrip transmission lines (a \textit{balun}) to form a single-ended signal at the input to the LNA on the base of the antenna.
 The petals are printed on the top and bottom surface thin (0.031") FR4 PCB material and liberally connected with vias, while  the stem and base are printed on low-loss Arlon DiClad 880 (Dk=2.2)  material using ordinary printed circuit techniques \citep{Leung-thesis}.

Feeds are \SI{305}{\milli\metre} apart along the focal line (the telescope $Y$ axis), and communicate  with one another with coupling coefficients that depend on separation, polarization, signal frequency and  angle of incidence. Coupling between feeds separated by as much as five times the basic interval is not negligible. The baluns are designed to produce an  \textit{effective} impedance of each element of the linear antenna array, including these coupling terms, which is noise-optimal for our LNA. Balun designs are therefore different for $X$ and $Y$-polarized elements because  inter-feed coupling is stronger for the $X$ ($\overrightarrow{ E} \perp$ to separation)  polarization  than for $Y$ ($\overrightarrow{ E} ~\parallel$ to separation).  The calculated noise temperature for the central element of a linear array  is shown in \cref{fig:scan_angle} as a function of frequency and incident angle.

\cref{fig:feed_response} shows models of the  angular response of an individual feed, modelled using CST Studio \citep{CST}, for several frequencies across the CHIME band.  As desired for feeds facing an \textit{f}/0.25 cylindrical reflector, the beam shape is broad and the beam width is largely independent of frequency over the CHIME band.  Notice that the E-plane and H-plane beam widths are slightly different from each other.  Therefore the $X$-polarized and $Y$-polarized channels have slightly different illumination patterns on the reflector, and slightly different far-field angular response patterns. The consequences of this variation will be discussed in \secref{sec:beams}.

\begin{figure}
   \centering
   \includegraphics[width=1.\linewidth,keepaspectratio]{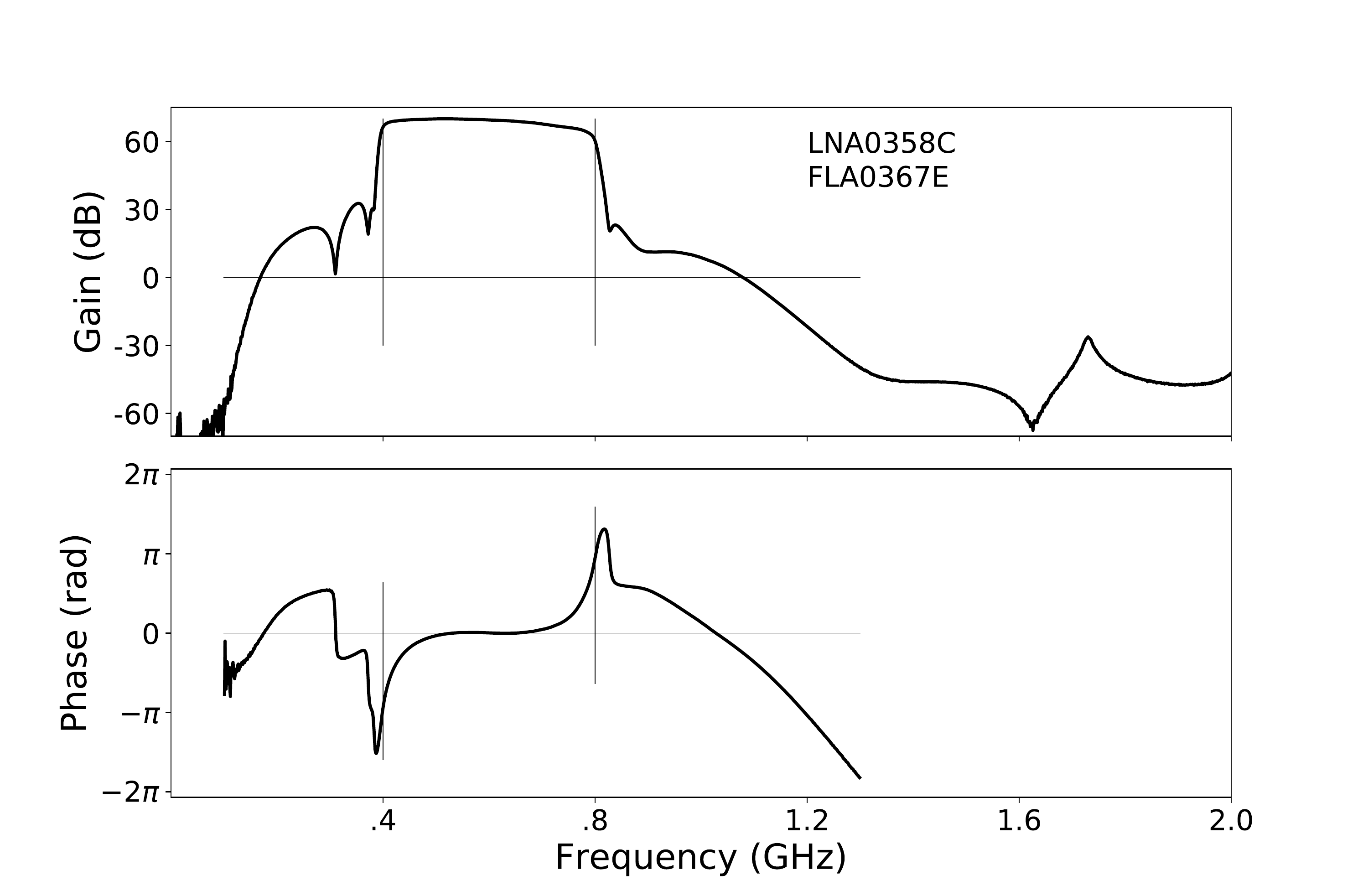}
   \caption{
   {\em Top} : Gain of the analog chain from the LNA at the CHIME feed to the ADC input. The vertical dashed lines show the edges of the second Nyquist band for the CHIME ADC sampling cadence of \SI{800}{\mega\hertz}, corresponding to the CHIME bandwidth of digital signals.  The chief elements in this analog chain are a low noise amplifier with a peak gain of \SI{42}{\dB} and a gentle roll off above $f= \SI{1}{\giga\hertz}$, a filter amplifier with a peak gain of \SI{38}{\dB} and a well defined passband provided by a custom filter from Mini-circuits (BPF-600-2+), \SI{50}{\metre} of low loss LMR-400 type coaxial cable and \SI{5}{\metre} of higher loss cable located within the receiver huts.  {\em Bottom}: The sum of measured phase shifts of all components of the analog chain plotted against frequency.  A single delay term is subtracted to show a flat phase curve at the centre of the band. Phase shifts associated with the very steep edges of the CHIME band-defining filters are evident.}
   \label{fig:analog_gain}
\end{figure}

The amplification and phase response of the remaining analog chain are plotted in \cref{fig:analog_gain}.  The very sharp band edges at \SI{400}{\mega\hertz} and \SI{800}{\mega\hertz} are designed to allow half-Nyquist sampling of the signal.  The response is achieved with a custom bandpass filter built for CHIME by Mini-circuits\footnote{\url{https://www.minicircuits.com/}}, model  BPF-600-2+, and installed following the first gain stage of the second stage amplifiers (FLA).  One sees in \cref{fig:scan_angle} that the LNA noise across the CHIME band is roughly \SI{20}{\kelvin}.  The gains of the LNA and FLA are chosen so that all other noise contributions are minor. The FLA contributes \SI{0.6}{\kelvin}  at the very top end of the CHIME band.  Cable losses and ADC input noise are less than this.

\begin{figure}
    \centering
    \includegraphics[width=1.0\linewidth,keepaspectratio]{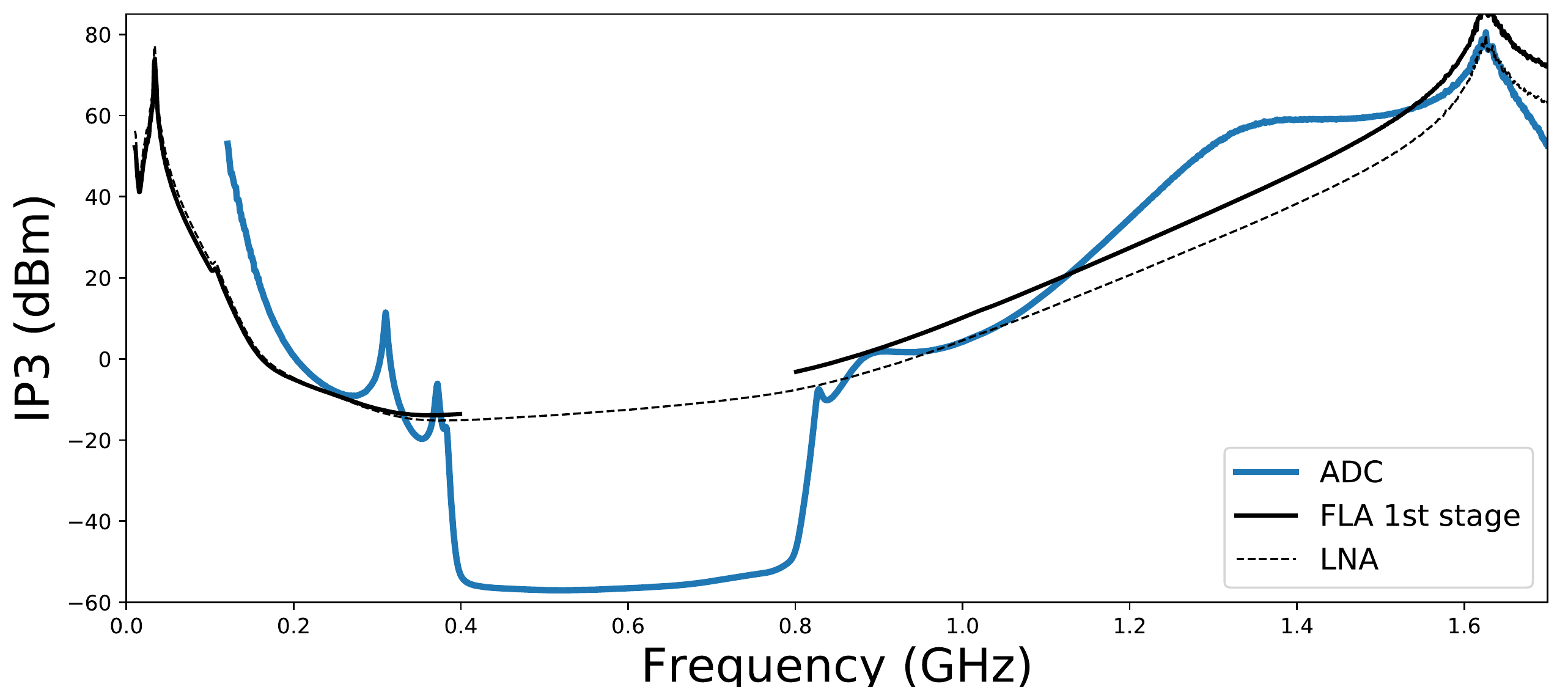}
    \caption{Analog chain linearity parameters, referred to the LNA input, are plotted against frequency.  The non-linearity parameter IP3 is \SI{13}{\dBm} at the input of the CHIME ADCs, where amplified RF power is highest.  The output coefficients, OP3 for the FLA and LNA are measured to be \SI{35}{\dBm} and \SI{30}{\dBm} respectively,  nearly independent of frequency.  These coefficients are more useful referred to a common point and so we have referred them all to the equivalent coefficients at the input of the LNA, taking account of gains, bandpasses and cable losses in front of each element.  We have nearly achieved our design goal that the system limit is set by the ADC at all frequencies.  In normal operation RFI signals at CHIME do not reach these levels either in band or nearby out of band.}
    \label{fig:analog_ip3}
\end{figure}

The non-linear response coefficients for the CHIME analog chain are plotted in \cref{fig:analog_ip3}, with all coefficients referred to the LNA input.  By design, the system third-order intercept point (IP3) within the CHIME band is dominated by that of the ADC.  The LNA and the first stage of the FLA are not protected  by the bandpass filter so in principle strong out-of-band RFI could produce in-band harmonics from non-linear response of the front end.
Extreme care has been taken with the non-linearity of the front end electronics to avoid this.  RFI at the CHIME site does not reach the levels that would produce a non-linear response in our electronics.

 It is worth a few remarks about the technical details of deploying 4,000 amplifiers and a similar number of cables over a \SI{100}{\metre} square.  The LNA and FLA are built into folded steel boxes which are soldered shut.  A small slab of RF absorber is glued inside the FLA boxes to suppress oscillations of the final stage to which earlier generations of our amplifier were prone.    Aluminum segments of the focal line which we call cassettes, consisting of four antennas, eight LNAs and associated \SI{1}{\meter} long SMA-to-N type cables are assembled indoors and carried to the focal line where they are mounted
 in place and bolted to each other.  Thus, the inter-feed spacing is set by digital machining. The \SI{50}{\metre} low-loss N-type coaxial cables connecting the LNAs to the FLAs at the receiver hut are cut to be the same length to within 0.1\%, and the optical delay of each cable has been measured separately.   Excess cable length for the antennas nearest the hut is stored in cable trays running the length of each cylinder in a geometry we call an optical trombone.   A full set of S-parameters is measured at the factory for each cable and serial numbers are for each recorded on bar codes.  This  is the practice for all components of the analog chains.  During system assembly,   pair-wise connectivity of all analog components is recorded using  a hand scanner and an interactive script operating on a mobile device.

 The FLAs sit within
a radio-frequency shielded room   with their input connectors protruding through a bulkhead in the wall. DC power is supplied to the LNA from the FLA over the coaxial cable.  The amplifiers of each individual signal chain can be powered off by remote command if desired. The RF room provides
 \SI{100}{\deci\bel} of attenuation and houses  the ADC and F-engine. Once installed, physical access to any antenna or LNA is available by lifting the floorboards of a walkway along each focal line.  This system is  less waterproof than we wish, and in heavy rains water can get to the baseboards of the antennas, causing temporary unacceptable  performance.  The focal line structure, consisting of an elevated enclosed dry volume mildly heated by the LNAs, is a nearly ideal bird habitat; consequently, we have found it is very important that there are no holes as large as \SI{2}{\centi\metre} diameter anywhere in the structure since these would would allow starlings to enter.

\subsection{FX Correlator}
\label{sec:FX-engine}

CHIME employs an FX correlator in which the time-domain signal from each feed is transformed to form a frequency spectrum in a part called the F-engine.  At each frequency, data from every feed are collected at a single designated computation node and a spatial transform is made of these signals to form visibilities. This spatial transform is performed in a part of the instrument called an X-engine.  These two processes are described below.  The F-engine consists  of eight 16-card electronics crates housed in two separate RF-shielded rooms located in modified, cooled,  20-foot shipping containers between pairs of cylinders.  These two containers are connected by optical fibre to the X-engine, which is housed in a pair of RF-rooms enclosed in 40-foot shipping containers, adjacent to the telescope.  The X-engine is built from 256 GPU nodes and is water cooled.

\subsubsection{F-Engine}
\label{sec:Fengine}

The F-engine is implemented using the ICE \citep{2016JAI.....541005B}
platform. ICE uses a field programmable gate array (FPGA) and is a general-purpose
astrophysics hardware and software framework that is customized to
implement the data acquisition, frequency channelization, and corner-turn
networking operations of the CHIME correlator.

\begin{figure*}
    \centering
    \includegraphics[width=0.97\textwidth,keepaspectratio]{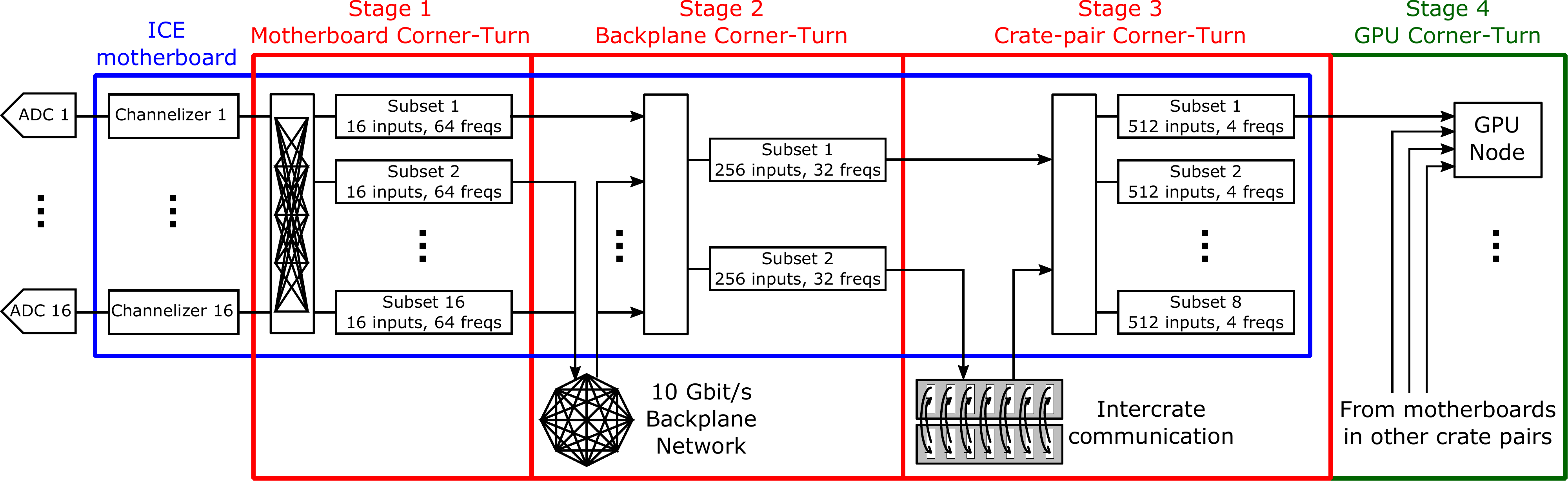}
    \caption{Data flow through the F-engine. A
    total of 128 ICE motherboards are required to process 2048 sky signals.
    These motherboards are installed in eight crates, with each crate handling the signals for one polarization  from every antenna on one cylinder.   Each motherboard digitizes 16 analog signals into 8 bits at 800~MSPS. The data stream from each digitized signal is fed to a
    FFT/PFB that splits the \SI{400}{\mega\hertz} bandwidth into 1024
    frequency channels. A four-stage corner-turn network
    re-arranges the data to allow    spatial   cross-multiplication and integration at each frequency in the X-engine. In stage one, each motherboard
    creates 16 new data streams, each one having 64 frequency channels from
    each of the 16 input signals. 
    In stage two, motherboards within a crate exchange data through a
    high-speed backplane network such that each board holds the data for 64 unique
    frequency channels  from all of the 256 inputs processed by that crate. In stage three,
    each motherboard sends the data from half of its frequency channels to a
    sister motherboard in an adjacent crate. With this inter-crate data
    exchange, each board within a crate pair contains the data for a subset
    of 32 unique frequency channels and 512 inputs. Stage four is completed
    within the X-engine GPU nodes. Each ICE motherboard re-orders the data
    into eight \textit{subsets}, each containing 4 frequency channels for 512 inputs.
    Each subset is sent to a different GPU node. Each of the 256 GPU nodes
    receives data from four different motherboards such that it ends up with
    the information from all the 1024  polarized antennas for four unique frequency
    channels. 
    }
    \label{fig:F_engine}
\end{figure*}

A schematic diagram of the data flow through the F-engine is shown in
\cref{fig:F_engine}. The core of the system is built around ICE motherboards which handle signal processing and networking using Xilinx Kintex-7 FPGAs. Each motherboard supports two custom ADC daughter boards. FPGA firmware and software are customized for the CHIME application. Each ICE motherboard digitizes 16 analog signals into 8~bits at 800~million samples per second (MSPS)   Thus, the
\SIrange{400}{800}{\mega\hertz} sky signals are directly sampled in the  second Nyquist zone.

The data stream from each digitized signal is fed to the FPGA, which implements a
polyphase filter bank (PFB) efficiently using a fast Fourier transform \citep{2008PASP..120.1207P}. Data are processed in frames of 2048 samples, separately for each stream. 
A PFB is more compact in frequency than a simple FFT would be, greatly aiding RFI excision by localizing any disturbance. 
At the cadence of individual  data frames, the PFB applies a sinc-Hamming  window to 4 consecutive data
frames, and outputs a single frame of 1024 complex values, one value per \SI{390}{\kilo\hertz} wide frequency
channel, in 18+18 bit real and imaginary format. 
After the PFB, the data are rounded  to
1024 4+4~bit complex values per frame.
Adjustable scaling factors (complex gains) are applied to each frequency channel before
this step in order to optimize the data compression
\citep{2018JAI.....750008M}.

After the frequency channelization, each ICE motherboard holds the
data for 1024 frequency channels of signals from 16 analog inputs. However, in the X-engine for each frequency,   data from every  input must  be presented to one processor in order to compute  the
cross-multiplications and averaging required to form the visibilities. A
total of \SI{6.6}{\tera\bit\per\second} of data needs to be re-arranged and
transmitted to the X-engine, an operation performed in a four-stage
corner-turn network \citep{2016JAI.....541004B}. The first stage is performed in each ICE motherboard, where the frequency-domain data from each input are split into 16
subsets, each containing 1/16 of the frequency channels from all 16 inputs.

Each group of sixteen ICE motherboards is packaged in a crate, and all the
boards within a crate are interconnected through a custom backplane that
implements a passive high speed full-mesh network. CHIME uses a total of
eight crates or 128 ICE motherboards. The second corner-turn stage is a
data exchange between the boards in the crate, after which each board has all
the data from 256 inputs  for 64 of the frequency channels.

The third stage is a data exchange between pairs of ICE motherboards located in adjacent crates using
high-speed serial links. 
After this third stage,
the data from 512 inputs are split into 256 subsets distributed
through the ICE motherboards of the two crates, and each subset
contains four unique frequency channels. Each crate
pair contains all the data for one quarter of the CHIME array, both polarizations from
one cylinder.

The fourth stage of the
corner-turn network takes place inside the GPU nodes of the X-engine.
Each ICE motherboard sends its data
stream to eight different GPU nodes through two
active \SI{100}{\meter} multi-mode optical fiber QSFP+ to 4$\times$SFP+ cables.
Each GPU node receives one frequency subset from one ICE motherboard
in each crate pair and recombines
the data to compute the correlation matrix for data from all  2048 inputs in four unique frequency channels.

The four F-engine crate pairs are housed in independent racks distributed
between two separate RF-shielded rooms installed within 20~ft modified, RF-shielded shipping containers, known as Receiver Huts. Each receiver hut serves two cylinders and is placed
between them at their midpoint. This arrangement minimizes the total length of coaxial cables running from the focal line of the cylinders to the receiver huts.

A GPS-disciplined, oven-controlled crystal oscillator provides the \SI{10}{\mega\hertz}
clock for the F-engine system. The GPS receiver also generates the
IRIG-B timecode signal used to insert time-stamps in the data. A copy of the clock and
absolute time signals is sent to each of the F-engine crates. From there, the signals
are distributed to each ICE motherboard and digitizer daughter board through
a low-jitter distribution network.  A broadband noise source system, which will be
described in \secref{sec:OfflinePipeline}, is used to monitor and correct
for drift between copies of the clock provided to each digitizer daughter board.

The F- and X-engines communicate over  256 optical fibers. Each fiber
cable contains four strands that connect one ICE motherboard to four different GPU nodes. These are
carried within a waterproof cable tray that goes underneath the cylinders and above the
huts. Also within the cable tray are the coaxial cables that distribute a
clock and absolute time signals to the F-engine huts.
The mapping of which RF frequencies are sent to which nodes in the X-engine is adjustable.  This allows, for example, sending the data from frequency channels heavily corrupted by RFI to nodes which are temporarily down for repair, preserving useful bandwidth.

\subsubsection{X-engine}
\label{sec:Xengine}

\newcommand{\kotekan}{\sw{kotekan}}

\begin{figure}
    \centering
    \includegraphics[width=0.46\textwidth,keepaspectratio]{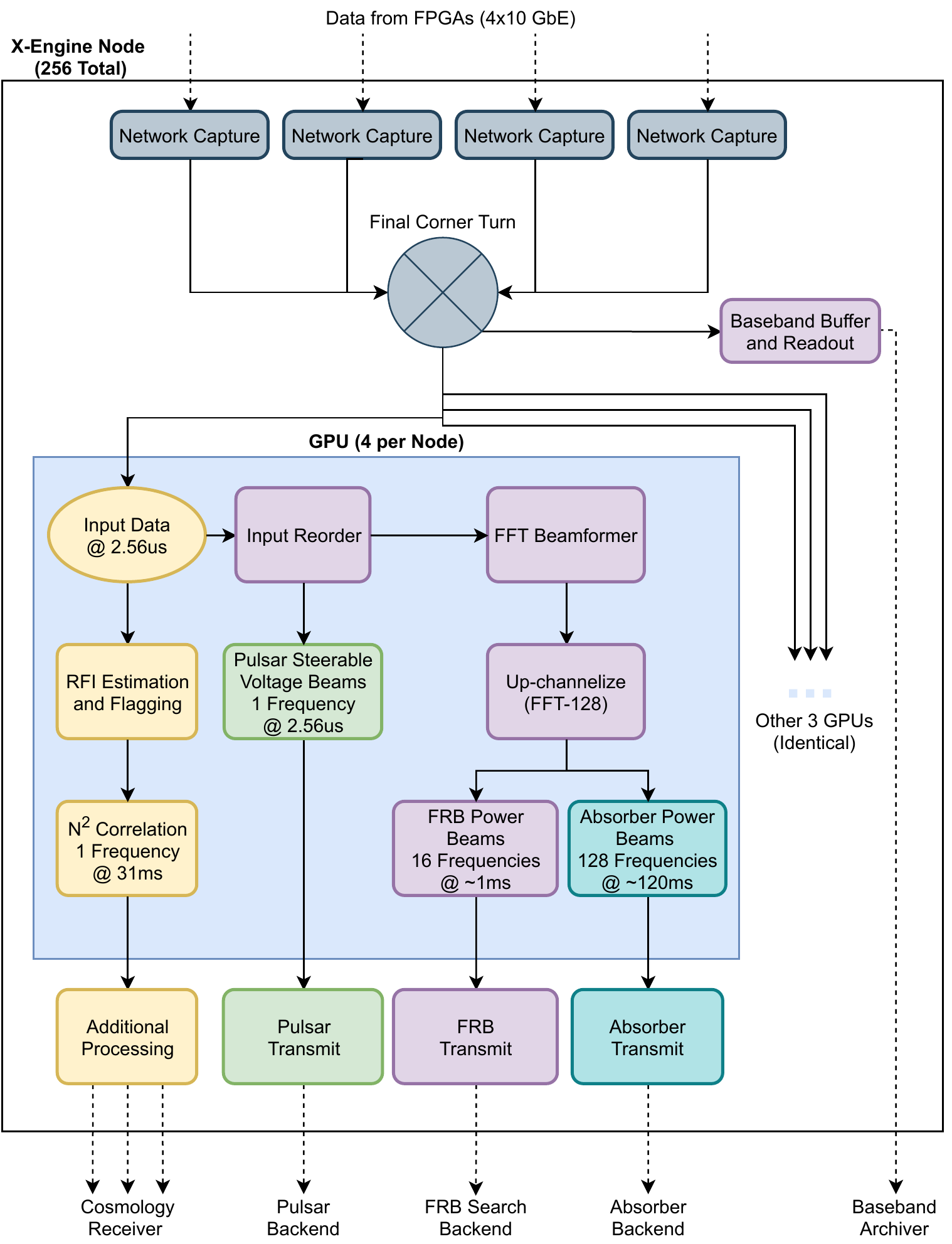}
    \caption{Processes performed by each X-engine node. Data arrive from the
    F-engine, the final corner-turn is performed by the CPU in the X-engine, and  signals from all 2048
    feeds within one single  frequency channel are transferred to one of the four GPUs. On each GPU the
    spectral kurtosis is computed as an estimation of RFI, and
    contaminated samples are removed. After flagging the data for RFI at $\sim$\SI{0.6}{\milli\second} cadence, the
    data are correlated to produce an $\Nfeed^2$ visibility matrix and summed over \SI{31}{\milli\second}.
    Each \SI{31}{\milli\second} correlation product is copied off the GPU and tested again, this time for
    long-duration RFI, which is either removed or processed further (see \cref{fig:receiver}).
    The  data are also branched off to two distinct  beamforming engines, a tracking
    voltage beamformer with 12 steerable beams and an FFT spatial beamformer which
    generates 1024 power beams at increased frequency resolution. Those power
    beams are further split into two combinations of frequency and temporal resolution. The
    tracking voltage beamformer is used primarily for the CHIME/Pulsar backend, and the
    FFT beamformer is used for both the CHIME/FRB search backend and a \tcm narrow-band absorber search backend.
    A buffer of the most recent 33 seconds of  data from the F-engine is
    updated in RAM. When triggered by the FRB search engine, the raw voltage data in this buffer, corresponding to one event,  is transmitted to an archive.
    }
    \label{fig:X_engine}
\end{figure}

The CHIME X-Engine  performs spatial correlations  and other real-time signal processing operations,
using  256 nodes, each with 4 GPU chips.
Details of the nodes and support infrastructure can be found
in \citet{Denman2020}.  These nodes run a soft
real-time pipeline built using the \kotekan framework \citep{kotekan,Renard2020},
which handles the X-engine, RFI flagging, and multiple real-time beamforming operations.  The processes performed by each node are shown in \cref{fig:X_engine}.

Data arrive at each of the the nodes from the F-engine on four \SI{10}{\giga\bit\per\second}
fibre SFP+ links. Each link conveys data from 512 feeds from four frequency bins
from each of the four F-engine crate pairs. Packet capture is handled in \kotekan
using the \sw{DPDK}\footnote{The Data Plane Development Kit. See
\url{https://dpdk.org}} library to reduce UDP packet capture overhead normally
associated with using Linux sockets. Once in the system, the packet data from each
link is split into 4 different staging memory frames, one for each frequency, which
completes the final corner-turn. Following packet capture there are 4 frames each
with data from one frequency channel and from  all 2048 feeds, for 49,152 time samples. These frames are
transferred to the GPU chips, resulting in each GPU chip processing data for exactly one of the frequency channels.

Once the data frames are on the GPU, a number of operations are applied to the data using
\sw{OpenCL} and hand optimized GPU kernels. The primary operation is the creation of the visibility matrix by the correlation kernel, using about $75\%$ of the processing time. For each frequency channel, the complex data from each feed are
multiplied by the complex conjugate of the corresponding signal from
each other feed to
create the visibility matrix. This is a Hermitian matrix, and  only the
upper triangle is directly computed.

 This calculation dominates the computational cost in CHIME, and we worked hard  to optimize it.
 Data are processed independently in blocks of 32$\times$32 feeds, distributed across 64 collaborating computational instances (``work items'' in a ``work group'').
 These work items employ Cannon's algorithm \citep{Cannon_alg}, collectively loading 8 sequential timesteps for all 32+32 inputs under consideration,
 and sharing these over high-speed local interconnects.
 Unsigned 4-bit values can be packed into 32-bit registers, allowing efficient multiplication and in-situ accumulation \citep{Klages2015}.
 Ultimately, 6 of the 8 arithmetic operations required for a complex multiply-accumulate (cMAC) operation are performed in a single GPU instruction. The remaining two are paired with another cMAC, for a total of 3 instructions per pair of cMAC computations.
 These intermediate products are accumulated in active registers, with top bits periodically peeled off and accumulated to high-speed local memory to prevent overflow.
 Products are summed in time over 12288 input time samples, before being unpacked and read out, to produce visibility products with a temporal resolution of roughly 31ms.
 To maximize throughput, this kernel was directly implemented in AMD's assembly-level Instruction Set Architecture (ISA), and the resulting high performance both left space for additional processing kernels (e.g. beamforming, RFI), and also allowed for a substantial reduction in observatory power envelope via low-power operation of the GPUs.

To excise RFI-contaminated data  prior to the correlation operations, a spectral kurtosis value is
computed over all inputs and 256 successive time samples (total $\sim \SI{0.66}{\milli\second}$)
\citep{Taylor2018}.  Each \SI{0.66}{\milli\second} block of data with a kurtosis value deviating from the
expected value by a configurable threshold is given 0 weight. The amount of data which are excised or
otherwise lost (for example to lost network packets) is accounted for in the
metadata and normalized later in the pipeline. These kurtosis values are
extracted from the GPU and used in a second-stage RFI test which can drop entire
\SI{31}{\milli\second} samples after they leave the GPU based on the statistics of
the 48 $\times$ \SI{0.66}{\milli\second} spectral kurtosis samples within. This second stage
is designed to excise RFI events that are lower in power but longer in duration than those found in the first stage.  This second stage excision is turned off during Solar transit.

The \SI{31}{\milli\second} visibility frames which are not excised are  processed in the  CPU associated with each node
and  transmitted to a receiver system running another configuration of \kotekan which does further processing. See \cref{fig:receiver} and \secref{sec:ReceiverSystem} for more detail.

In addition to the correlation, RFI estimation and flagging, the GPUs
perform two kinds of beamforming operations. The first type is a tracking
voltage beamformer, which takes
right ascension and declination coordinates and generates a set of
dynamic phases that are applied to input voltage data and summed over all
feeds to generate a single coherent beam used to observe celestial sources
while in the CHIME field of view. Currently CHIME forms 12 of these beams
simultaneously. The data from these formed beams are scaled to 4+4-bit complex data
at full \SI{2.56}{\micro\second} time resolution and transmitted over the 1 Gigabit Ethernet
(GbE) links on the nodes. The data streams  from 10 of these beams
are sent to 10 CHIME/Pulsar processing nodes.
The remaining two beams are used for other operations such as VLBI and calibration.

The second type of beamforming operation is an FFT-based spatial imaging beamformer
\citep{Ng2017} which generates  1024 power beams in fixed terrestrial coordinates for use in the FRB engine and the high-resolution absorber search.  A spatial FFT is performed   for the data from each cylinder to
generate 512 beams for each polarization.  Of these,  256 are selected to achieve roughly achromatic pointing. A 4-way transform is  computed
across all these beams in rows between cylinders.  This combination produces 1024 beams at each frequency, and for each of these 128 successive temporal samples are Fourier transformed
to extract higher frequency
resolution.  For the high frequency-resolution absorber search,  the data are squared  at full 128 sub-frequency spectral resolution
($\sim \SI{3}{\kilo\hertz}$), and integrated to $\sim
\SI{120}{\milli\second}$ time resolution. After leaving the GPU these high resolution data are
 integrated again to \SI{10}{\second}, and  stored on a
backend running a special configuration of \kotekan,  to enable a search
for \tcm narrow-line absorbers.
For the FRB search engine, the data are squared, summed over polarizations, and  summed over 16
frequency bins and 384 time samples to produce 1024 power-beams with 16
sub-frequency bins ($\sim \SI{24}{\kilo\hertz}$) per original CHIME channel at $\sim
$\SI{1}{\milli\second} time resolution. This tuning of sampling time and frequency resolution is made to match the data to the conflicting goals in the FRB engine of resolving short pulses and performing de-dispersion in a discretely sampled spectrum. These data are  sent from each
GPU to the FRB search backend in custom UDP packets over a 1
GbE link to be searched in real time for FRBs.

\subsection{Real-Time Processing}
\label{sec:RealtimePipeline}

\label{sec:ReceiverSystem}

\begin{figure*}
    \centering
    \includegraphics[width=0.95\textwidth,keepaspectratio]{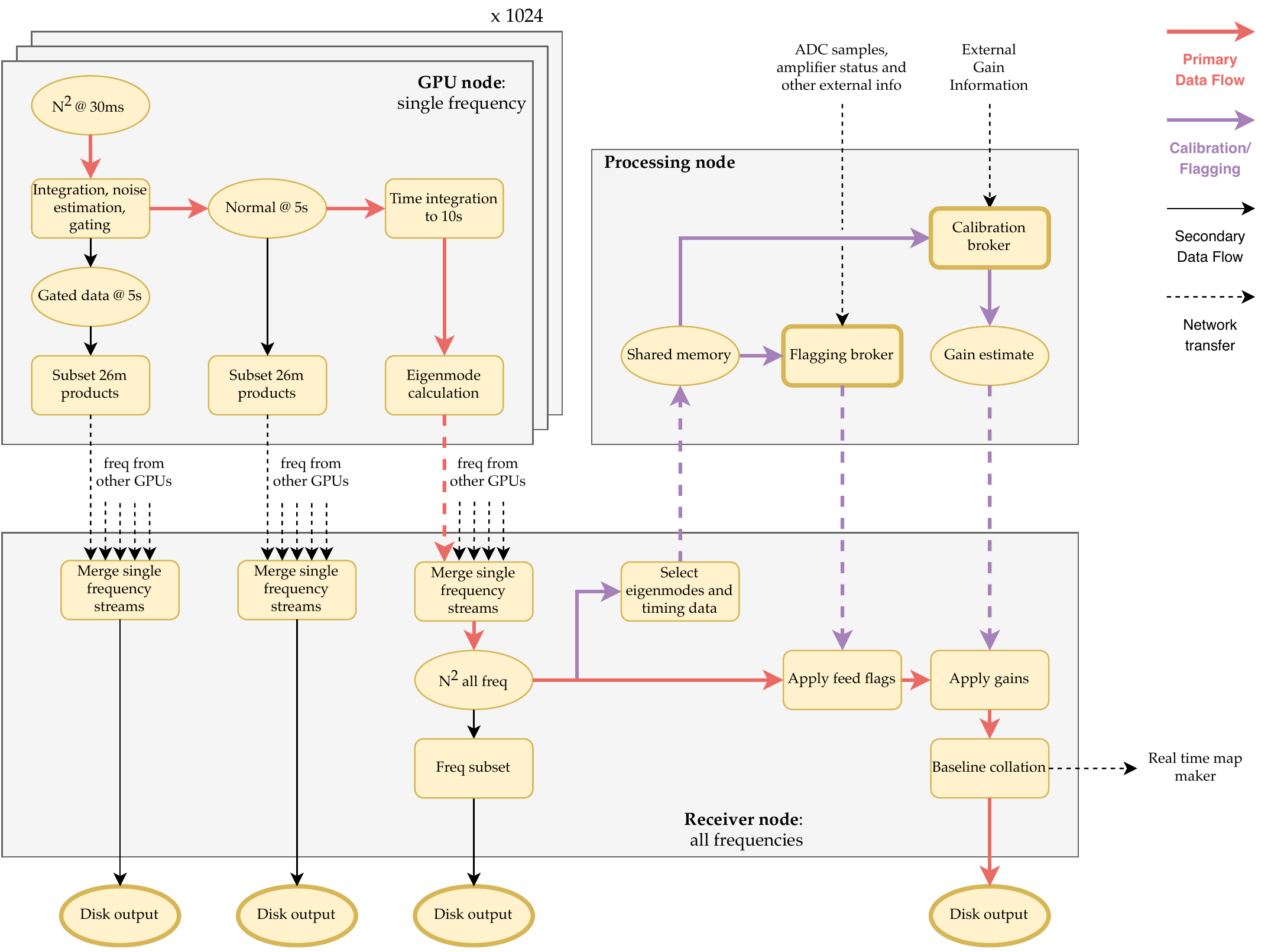}
    \caption{
        A diagram of the data flow through the CHIME post-correlation receiver system. The receiver system processes a full $\Nfeed^2$ correlation matrix for each of the 1024 frequency channels every $\sim$\SI{31}{\milli\second} from the X-engine (see \secref{sec:Xengine}). Initially this stream is processed within the CPUs of each GPU node to accumulate the data up to a \SI{10}{\second} cadence, estimate its noise and, if desired,  perform gating for pulsar observations. Cross correlations with the Galt \SI{26}{\metre} Telescope used for bright source and pulsar holography can be extracted at \SI{5}{\second} cadence to prevent fringe smearing. For use in calibration, we solve for the highest four eigenvalues and vectors of each $\Nfeed^2$ frame. The data for all of this is sent over the network to a single receiver node for further processing, including flagging, calibration and baseline stacking before being written to disk. Flags for bad correlator inputs are derived by a broker process running on a separate node that assimilates various sources of data quality information into a mask for each correlator input. Similarly, gain solutions for calibration are derived by a broker that uses eigenvector and noise source timing data from the correlation products as well as environmental data to produce gains that are applied in real time to the $\Nfeed^2$ data.
        }
    \label{fig:receiver}
\end{figure*}

The ensemble of the 1024 GPUs generate $\Nfeed^2$ correlation products at a \SI{31}{\milli\second}
cadence for each of  1024 frequencies. This amounts to a raw data rate of
$\sim$\SI{4.6}{\tera\bit\per\second}. It is not feasible to write out and store such a fire-hose of
data. The receiver system is tasked with aggregating and processing the data
stream in preparation for archiving. In the process it produces ancillary data
products that are tapped for system and data quality monitoring.
\cref{fig:receiver} provides a schematic representation of the receiver system.
The various stages are distributed across multiple computers (aka nodes). The first of them
occur on the GPU nodes themselves (executed on the CPU) before being transmitted
over the network to the single receiver node, where the remainder of the pipeline
occurs. Another computer, the processing node, hosts parallel processing tasks
that are not time-critical for subsets of data. Notably this includes deriving
the calibration solutions that are fed back into the main receiver node pipeline.
The final data products are sent over the network to an archive node. Aside from
a few exceptions, all of these stages are built on the \sw{kotekan} framework.

\paragraph{Accumulation and gating}
In order to reduce the data rate, the first stage following the GPU co-adds  RFI-cleaned
\SI{31}{\milli\second} frames for \SI{5}{\second}. A later stage
co-adds samples further to the final \SI{10}{\second} cadence, but optionally the
subset of the data  comprised of correlation products with the Galt
\SI{26}{\meter} telescope are kept at the finer time resolution to avoid
smearing due to the faster fringing of the $\sim$\SI{230}{\meter} baseline between
Galt and CHIME. This is the last chance for any operations on the fast-cadence
data. The variance over the \SI{31}{\milli\second} samples is calculated to
estimate the noise level in the accumulated frame and passed along with it.
Gated accumulation is also supported, where samples are weighted and binned
into \textit{on} and \textit{off} gates and the difference of the two is returned at the end of
the integration window. Gating can be initiated, or its parameters updated on the
fly without interrupting data acquisition. Currently, gating is used for
simultaneous observations with the Galt telescope of slow ($P
>$~\SI{300}{\milli\second}) pulsars for beam holography (\secref{sec:beam_data}).

\paragraph{Eigendecomposition}
The four leading eigenvalues/vectors of the $\Nfeed^2$ visibility matrix are
estimated for every time sample and passed on down the pipeline. It is necessary
to perform this step in the X-engine in order to distribute the computational load over
the 256 CPUs located there. The eigenvectors represent the response of every individual
array element to the dominant modes on the sky at that moment, making them a
valuable tool for real-time calibration. Importantly, it is not
possible to perform this decomposition after the redundant baseline collation
step, and  the full $\Nfeed^2$ visibility matrix is only stored for  a small number of frequencies, so
these eigen-data  are important for offline analysis as well. Since noise-coupling
between nearby feeds is significant and will outweigh the sky modes, the
diagonal values of up to 30 feed separations are excised from the matrix prior to the
decomposition. To avoid biasing the result, an iterative scheme is employed to
progressively complete the masked region.

\paragraph{Calibration broker}
A daily complex gain calibration for every sky signal is derived from the transit of a
bright astronomical point source. The calibration broker is a service running on the processing
node that produces  gain solutions by fitting the eigenvector data immediately
following the transit of a chosen point source. The eigenvectors are continuously provided to the
broker via a shared memory ring buffer and the broker can access a timestream spanning
the transit by reading the buffer file approximately \SI{20}{\minute} after transit.
During transit, the bright source is the dominant contribution to the sky signal and
the visibility matrix can be approximated as an outer product of the input gain
vector (a rank-1 approximation), identified as the leading eigenvector. A
complication is that the 2048 sky signals include two polarizations, so there are in fact
two near-orthogonal components to the matrix. There is no guarantee that these two
vectors neatly divide the inputs by polarization as is required to interpret the
eigenvectors as gain solutions. An additional orthogonalisation with respect to the
two-dimensional space of polarizations must be performed by the broker to isolate
them. The intrinsic flux density of the source across the band is corrected for using the
measurements of \cite{2017Perley}. Frequencies affected by RFI are flagged by
comparing the ratio of the eigenvalue on- and off-source, and those with anomalous
gain amplitudes are also flagged. Gains for the flagged frequencies are recovered by
interpolating between the gain solutions for adjacent  good frequencies. The four brightest
sources are processed in this way at every transit, but only one is used for
calibration. The choice of which source is used changes throughout the year to avoid calibrators near the Sun,
and any differences in the primary beam patterns are corrected using the average ratio of past gains from the source, to past gains from Cygnus A (Cyg A).  The calibration procedure therefore normalizes the primary beam
pattern at each frequency to unity on meridian at the declination of Cyg A.
See \secref{sec:OfflinePipeline} for additional corrections applied later in the pipeline.

\paragraph{Flagging broker}
The role of the flagging broker is to perform real-time identification of correlator
inputs that should be excluded from further analysis.  It runs on the processing
node and provides regular updates to the relevant stages of the receiver pipeline.
It uses a variety of data products and housekeeping metrics to repeatedly evaluate
10 different tests, with each test designed to identify malfunctioning or otherwise
anomalous correlator inputs.  Below we list its data sources and briefly summarize
the corresponding tests.  Note that there can be multiple tests derived from a single data source.
\begin{itemize}
    \item \textit{Layout database:} Reject inputs that are not currently connected to an antenna
    or that have been flagged manually by a user.
    \item \textit{Power server:} Reject inputs whose amplifiers are not currently powered.
    \item \textit{ADC data:} Reject inputs whose raw ADC data has an outlier RMS, histogram, or spectrum.
    \item \textit{RFI broker:} Reject inputs determined to have highly non-gaussian statistics based on a
    monitoring stage internal to the X-engine.
    \item \textit{Calibration broker:} Reject inputs for which the complex gain calibration failed,
    or whose gain amplitudes exhibit large, broadband changes relative to its median over the past 30 days.
    \item \textit{Autocorrelation data:} Reject inputs that have outlier noise or whose autocorrelation
    shows large, broadband changes relative to past values.
\end{itemize}
If a correlator input fails any one of the tests all baselines formed from that input
will be given zero weight when averaging over redundant baselines.

\paragraph{Gain/flag application and redundant baseline collation}
The $\Nfeed^2$ and \SI{26}{\meter} streams from the GPU nodes are
merged into all-frequency streams as they arrive at the receiver node. The
\SI{26}{\meter} streams undergo no further processing, as do a subset of four
frequencies from the $\Nfeed^2$ visibility calculation that are output at this stage to preserve some of the
full array information. Keeping $\Nfeed^2$ terms for all frequencies,   amounting to a data rate of over
\SI{200}{\tera\byte\per\day}, is not feasible
because of storage constraints. A lossy compression is effected by averaging redundant
baselines within each cylinder pair together. Baselines are not combined between the six  cylinder pairs
to maintain the possibility of correcting for any non-redundancy between the
cylinders or between the signal paths that are routed to separate receiver huts. The
daily rate of data archived is thus reduced to $\sim$\SI{1}{TB}. Prior to
collating visibilities along redundant baselines, the gain calibration and flags generated by their
respective brokers are applied to the data. This compression method is lossy due to any non-redundancy that might arise from support structures, edge effects, and imperfections in the reflectors, or non-uniformity in the feed responses, as well as imperfect calibration.

\paragraph{Real-time map}
A subset of 64 frequencies is tapped from the main pipeline following the
baseline collation stage and transmitted to the processing node, where a
separate pipeline beamforms the visibilities to generate a real-time data stream we call a
\textit{ringmap}. The ringmap is a representation of the data as a timestream of
formed beams, visualising the sky as it drifts through the field of view
of the cylindrical reflectors (see \secref{sec:maps} for details). The maps for
those frequencies are buffered over a period of \SI{24}{\hour} and can be
displayed using a data monitoring web viewer. They are useful for assessing
recent data quality at a glance and we study them every day.

\paragraph{Output datasets}
The branching points in the pipeline lead to three main data products. The
\texttt{stack} dataset is output by the baseline collation stage and contains
the total of CHIME's sensitivity, with all non-flagged baselines contributing
over the entire band. The \texttt{N2} dataset holds complete uncompressed
visibility matrices for  four frequencies. It is useful for instrument
characterisation and understanding the effects of baseline collation. The gated
and ungated \texttt{26m} datasets contain only the cross-products with inputs from the Galt telescope, and  at a \SI{5}{\second} cadence, twice that of the other datasets. These are
produced only during simultaneous observations of point sources for beam
holography (see \secref{sec:beam_data}).

\paragraph{Compression and archiving}
The final module of the real-time pipeline is an archiving service that packages the
data into a structured archive format, applies another stage of compression, and
registers files with the archive database. It takes advantage of the relatively slow
rate of change of the measured sky gradually drifting through the field of view by
ordering the data with time as the fastest varying index and compressing the
redundant information between nearby time samples. All the data are truncated at a
specified fraction of the measured noise level to excise the high variability in the (noise
dominated) least significant bits and thus further improve the effectiveness of the
compression. The \sw{bitshuffle} algorithm \citep{2015bitshuffle} compresses these
data on a bitwise basis, resulting in a typical size reduction of $\sim 2$--$5$ times for
\texttt{stack} files. Data are stored on site for up to six months and indefinitely at
archives located at Compute Canada centres. Archive files are tracked in an SQL
database and all file operations are mediated by a software daemon that validates the
integrity of the data and ensures storage redundancy.

\subsection{System Monitoring}
\label{sec:SystemMonitoring}

\newcommand{\prometheus}{\sw{Prometheus}}
\newcommand{\grafana}{\sw{Grafana}}
\newcommand{\slack}{\sw{Slack}}
\newcommand{\dias}{\sw{dias}}
\newcommand{\theremin}{\sw{theremin}}
\newcommand{\alertmanager}{\sw{Alertmanager}}

CHIME is a complex instrument with 2048 analog signal chains processed by
nearly 400 separate computers spread over six physical locations on site. To keep
the experiment running 24 hours and 7 days a week, it is important to identify and rectify inevitable failures in a timely manner. In this section, we explain how the CHIME operations are
monitored in almost real time to assess instrumental and experimental health.

\paragraph{Instrument health monitoring}
The instrumental health can be monitored by verifying that various hardware and software subsystems are running, data are written to disk, equipment huts are thermally stable, and there is no failure that is an emergency and needs immediate attention, e.g. coolant leak or fire. An array of auxiliary sensors are deployed across CHIME to probe various environmental parameters. These include temperature sensors across one cylindrical reflector; ambient temperature, humidity, smoke and leak sensors in equipment huts; and a weather station with wind and rain-accumulation sensors. Data from all these sensors is streamed in real-time into a central database. In addition, metrics are collected from various hardware and software components, including but not limited to power supplies, operating systems (OS), network statistics from switches. Almost every software and firmware component also generates its own set of internal health metrics.

CHIME uses \prometheus \citep{Prometheus} and \grafana \citep{Grafana} for managing and monitoring the housekeeping data in real-time. \prometheus is an open-source monitoring system and time-series database. The data collected by \prometheus can be  displayed through web-based dashboards in the  \grafana environment. \prometheus allows defining rules for alert conditions and expressing them as a \prometheus query that can invoke an alert to an external service. Alerts are handled by \alertmanager \citep{Prometheus} that sends out notifications through \slack and email to targeted team members when thresholds set on various metrics are violated.

This combination of \prometheus and \grafana environment provides the ability to monitor the operation remotely. As there is only one Telescope Operator on site during working hours, and no one otherwise, the CHIME team provides nearly 24-hour remote monitoring of the operation by taking on shifts on a rotating basis after regular work hours. The person on duty responds to alerts in-situ only if they are critical and causing interruption in the data acquisition. As an example, temperature control in equipment huts is quite sophisticated. As both X-engine and F-engine hardware are cooled by liquid coolant, the greatest attention is paid to detecting any potential leaks in the plumbing. If leaks are detected, valves automatically cut the supply of coolant into huts to minimize any potential damage to the system. Similarly, if smoke or flood sensors are persistently tripped the power is automatically shut to receiver huts. This way the system automatically reacts to catastrophic events ensuring the safety of subsystems.

A subset of housekeeping data stored in \prometheus is exported and written to an HDF5 file on a daily basis. These files are then archived to be used during offline data analysis.

\paragraph{Experimental health and data-integrity monitoring}

Considering the amount of data that CHIME generates, it is challenging to check the data quality and integrity in  real time. The focus of this operation is to highlight only those  data-quality issues that can be addressed and improved by acting swiftly and adjusting certain configurable hardware or software parameters. The timeframe for these assessments can be seconds (e.g.\ RMS of sky signal); minutes (e.g.\ spectra waterfall, correlation triangle); or, a day (e.g.\ calibration quality, downward trend in noise integration). Data quality and integrity are monitored though a mix of manual checks and a  set of automated quick data analysis on a daily basis by the remote operator(s).

\paragraph{ \dias }
\footnote{\url{https://github.com/chime-experiment/dias}} is a software framework for data integrity analysis and generation of daily plots. It runs as a service that schedules the execution of data \textit{analyzers}. This framework replaces slow on-demand script execution with an automated pre-generation of a set of data products, which are not archived and are  only available for a few months. A lightweight package for generating web-based plots, \theremin, is developed in house and used to view these data products.

Using \dias and \theremin we are able to monitor the quality of the data itself in near real-time. This includes estimates of
the RFI environment, the full array sensitivity derived from sub-integration variances, bright source spectra, and real-time sky maps
derived directly from the saved CHIME data products. This allows the CHIME team to get rapid feedback on the end to end performance of the instrument and to make timely adjustments if needed.

\subsection{Offline Processing}
\label{sec:OfflinePipeline}

Post processing of the CHIME data is done via a Python-based, YAML-configurable, offline pipeline. The basic infra-structure is available in \sw{caput} \citep{caput} and most of the non-CHIME-specific functionality is available in \sw{draco} \citep{draco}. CHIME specific parts of the pipeline are found in \sw{ch\_pipeline} \citep{ch_pipeline}. The pipeline structure is flexible, being used not only for the main data product pipeline, but also for a variety of functions such as instrument simulations, holography and cross-correlation analysis, foreground removal, and power spectrum estimation.

The main data pipeline for CHIME runs on Compute-Canada's Cedar\footnote{\url{https://docs.computecanada.ca/wiki/Cedar}} cluster where one of our science data archives is located. The data are processed in units of Local Sidereal Days (LSD)\footnote{CHIME uses sidereal days referenced to a November 15 2013 start.}. The first step of the pipeline is to locate and load all files pertaining to a particular LSD into memory. A number of calibration and transformation operations are  performed in the order presented below.

A timing correction is applied to each file to account for differences in timing between the two receiver huts\footnote{The timing differences and the corrections applied are described in detail in section \ref{sec:phase}.}.
The final step of redundant-baseline stacking is performed, in which redundant baselines corresponding to different pairs of cylinders are stacked together (this step is delayed to this point to allow for the timing calibration to occur). At this point, an offline stage of RFI masking is applied to the data, complementary to the real-time RFI excision that takes place in the receiver pipeline (see \secref{sec:ReceiverSystem}). This stage derives a figure-of-merit for sensitivity estimates based on the radiometer equation applied to cross-polarization data. This figure-of-merit is  fed to a sum-threshold algorithm \citep{2010MNRAS.405..155O} in frequency-time space which outputs a single mask for all baseline stacks. This stage also includes a specific search for intermittent RFI with the \SI{6}{\mega\hertz}-wide bands, characteristic of TV stations.

To allow for later stacking of multiple sidereal days, the data are resampled to go from the original time-of-day basis  to right ascension. This regridding is done via an inverse Lanczos interpolation which takes the data from the native resolution of around \SI{10}{\second} to approximately \SI{5}{\arcmin} in right ascension. The regridded data corresponding to a full sidereal day are combined into a \textit{sidereal stream}, the final data product which is written to disk for analysis and long-term archiving. A few additional products are saved alongside each sidereal stream visibility data. These include ringmaps (see \secref{sec:maps}), delay power spectra,
and bright point source spectra, as well as the sensitivity figure-of-merit and the RFI mask derived from them.

An independent second stage pipeline exists to combine many sidereal streams into higher sensitivity full sidereal day products called \textit{sidereal stacks}. Initially, all sidereal streams in a specified time range are selected. These are specified to be times of mostly uninterrupted observation in which the telescope was operating in a stable mode. For instance, we require all the data that goes in a sidereal stack to have been calibrated on the same source (see \secref{sec:ReceiverSystem} for calibration details).

Before stacking multiple days, an extra step of cleaning is applied to each sidereal stream to remove all day-time data as well as any times flagged as potentially corrupted by a range of environmental indicators (rain, excessive site RFI, bad calibration due to instrument restart, etc.). The data are  combined into aggressively cleaned, sun-free, sidereal stacks which are the main science-ready data products of the CHIME data pipeline.   Corrections for  thermally induced phase shifts as described in  \secref{sec:stability} can  be applied at this point.

\section{Beams}
\label{sec:beams}

The biggest challenge for detecting extragalactic \tcm emission is filtering out the much brighter foreground emission, dominated by diffuse Galactic emission and extra-galactic radio sources \citep{2011PhRvD..83j3006L}. To do so, it is crucial to have precise knowledge of the instrumental beam response. Estimates by \cite{2015shaw} indicate that this response must be characterized to roughly a part in $10^{4}$ in power units, and this has motivated the pursuit of a number of parallel strategies for beam measurement and modelling, as well as efforts to quantify the required precision in more detail. In this section, we first describe how CHIME's instrument design determines the general features of the beam response, and then present the current status of our ongoing work to characterize this response.

\subsection{General Features of the CHIME Beams}
\label{sec:beam_intro}

We define the ``base" beam to be the illumination on the sky (amplitude, phase, and polarization) that results when a single feed broadcasts with all other feeds along the focal line shorted \citep{Deng_2014}.  Although CHIME never operates as a transmitter, this is a useful construct for understanding the beam properties.  In the absence of multi-path effects, discussed below and in \secref{sec:beam_model}, this base beam produces a nearly elliptical illumination of the sky: $\sim$120  degrees long in the unfocused North-South (N-S) direction, along the cylinder axis, and a few degrees wide with frequency-dependent diffraction side-lobes in the East-West (E-W) direction, perpendicular to CHIME's cylinder axis.

Multi-path and other coupling effects  alter this simple description by as much as 50\% at some frequencies.  The physical origin of the multi-path interference is radiation interacting with the focal-line assembly, which consists of the linear feed array and a common ground plane.  In this environment, a signal broadcast by a feed will reflect off the cylinder and a large fraction of that signal will go directly to the sky, but a small portion strikes the focal plane assembly, where some is absorbed by a neighbouring feed and the rest is reflected and/or re-radiated by the assembly, eventually reaching the sky.  The details of this latter interaction are complex and are still actively being characterized.  Nonetheless, the ``primary" beam is the illumination on the sky one gets when these effects are accounted for.  The ``synthesized" beam is the illumination produced by coherently combining the signal from multiple feeds, each with their own (nearly identical) primary beams.  In this section we focus on characterizing CHIME's primary beam.

Since  multi-path propagation is occurring within a \SI{5}{\meter} cavity (CHIME's focal length), new interference fringes arise roughly every \SI{30}{\mega\hertz} in frequency, as seen  below.  In the remaining sections we present the datasets used to calibrate CHIME's primary beam, and discuss approaches to modelling the full response, informed by these data.

\subsection{Datasets for Beam Calibration}
\label{sec:beam_data}

\begin{figure*}
    \centering
    \includegraphics[width=1.0\linewidth,keepaspectratio]{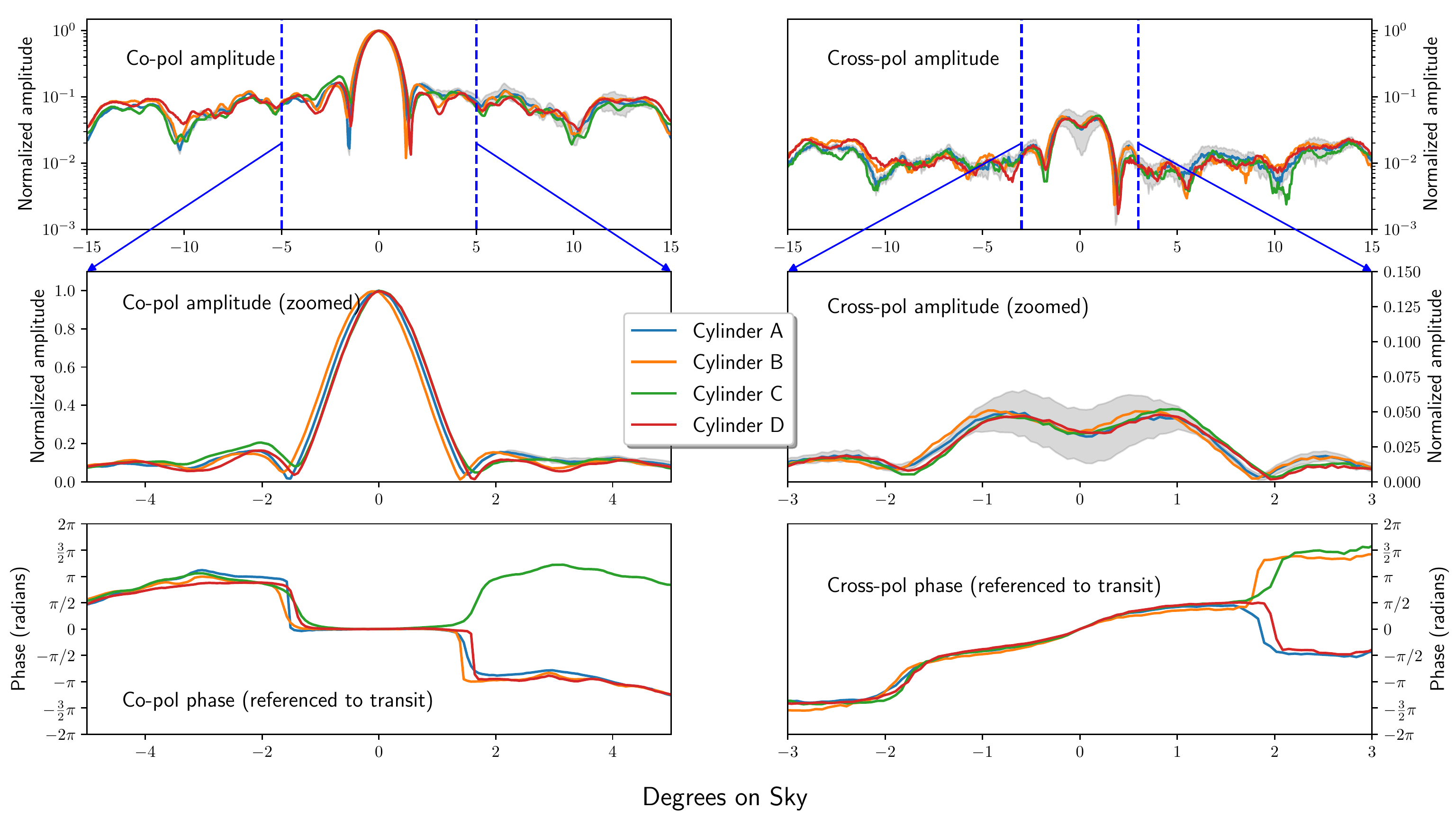}
    \caption{The CHIME $X$-polarized  beam response at \SI{717}{\mega\hertz} from the holographic measurements of a Cyg A transit, taken on 2018 Sep 28. (See \secref{sec:beam_data} for a discussion of CHIME's holographic measurement methodology.) Each panel shows the median response taken over all feeds within a cylinder, normalized such that the co-polar response is $1 + 0j$ at transit.  \textit{Upper left}: The median co-polar amplitude of all feeds per cylinder, normalized by the Gaussian-fit peak height, over the full extent of the observation, converted to degrees on sky, HA$\cdot\cos\left(\delta_{\text{CygA}}\right)$; \textit{Upper right}: The median cross-polar amplitude from the product of a CHIME feed with the opposite polarization on the Galt receiver. The data have been scaled by the same factor as applied to the co-polar response, so the curves give an indication of the level of cross-polarization in the beam; \textit{Middle left}: Same as the upper left panel, but zoomed to a smaller hour angle range and plotted on a linear scale; \textit{Middle right} Same as the upper right panel, but zoomed to a smaller hour angle range and plotted on a linear scale; \textit{Lower left}: The median co-polar phase as a function of scaled hour angle, taken over all feeds in a cylinder (the median was evaluated for the real and imaginary parts separately before evaluating the phase); \textit{Lower right}: Same as the lower left for the cross-polar phase.  The phase difference between cylinders, after accounting for phase wrap, is only large near the first zero crossing of the field.  The gray bands in the amplitude plots indicate the standard deviation over all the Cyg A holography tracks of Cylinder A's median feed response.  (Cylinder A is representative)}
    \label{fig:beam_holo}
\end{figure*}

Ideally, the CHIME primary beam calibration would be based on direct measurements of the telescope's response to a bright (relative to the sky confusion), polarized point source along every direction in the far field, at every frequency. However, a sufficiently complete population of such sources is not available; instead, we make use of several direct measurements, each of which provides beam information in a different regime. Importantly, these regimes often overlap, which allows for multiple cross-checks on the results. Thus far, the most useful information has been obtained from three datasets: holography of bright point sources, which allows beam amplitude and phase measurements for each feed along a limited number of one-dimensional tracks through the beam; transits of bright point sources, which trace the feed-averaged beam response on meridian; and transits of the Sun, which provide similar information to holography (without the phase information) but with near-continuous sampling over a specific range of declination.

When plotting 2-dimensional beam measurements over a large angular extent, we use an orthographic projection with its origin at zenith.  This projection has the advantage of not distorting the apparent beam width at different elevations.  Moreover, the projected coordinates $x$ and $y$ in the tangent plane remain parallel to East and North, respectively. For the unit vector pointing to hour angle $\mbox{\textsc{ha}}$ and declination $\delta$, the corresponding angular coordinates are given by
\begin{equation}
    x = - \cos{\delta} \sin{\mbox{\textsc{ha}}}
\end{equation}
and
\begin{equation}
    y = \cos{\ell} \sin{\delta} - \sin{\ell} \cos{\delta} \cos{\mbox{\textsc{ha}}}
\end{equation}
where $\ell$ is the latitude of the observer ($+49.3^{\circ}$ for CHIME).

\subsubsection{Holography}
\label{sec:holo}

Holography is an established technique for making accurate measurements of the amplitude and phase of antenna beams at radio frequencies (e.g.~\citealt{1976ITAP...24..295B,1977MNRAS.178..539S,Baars2007}).   We use this technique by tracking a celestial source with a nearby moving telescope while the source transits through the stationary CHIME beam. The correlation between the signals from each stationary feed and the tracking reference telescope traces the response of CHIME along the path of the source. For CHIME holography, the John A. Galt $\SI{26}{\meter}$ telescope, located \SI{230}{m} East of CHIME,  is used as the tracking system. For these observations, a \SIrange{400}{800}{\mega\hertz} dual-polarization modified CHIME receiver is mounted on the Galt telescope \citep{Berger:2016}.  The resulting cross correlations yield CHIME's co-polar and cross-polar far-field beam response (amplitude and phase) per feed, per frequency, along a track in hour angle at the declination of each observed source.

The data collected to date comprise 1888 tracks of 24 celestial sources since holographic observations began in October 2017, typically spanning $\pm$\ang{40} or more in hour angle and \ang{-21} to \ang{+65} in declination (\ang{-70} to \ang{+16} in zenith angle). The data are fringe-stopped (phase shifted to account for Earth rotation)
and binned to a celestial grid, with the resulting average and variance per bin stored on disk.  Data from successive observations  of  a given source can be combined,  reducing measurement  noise.  A sample holographic measurement of Cyg A is presented in \cref{fig:beam_holo}, which shows the amplitude and phase of the co- and cross-polar beams in each CHIME cylinder.

\begin{figure}
    \centering
    \includegraphics[width=1\linewidth,keepaspectratio,trim={0 0 20pt 0}]{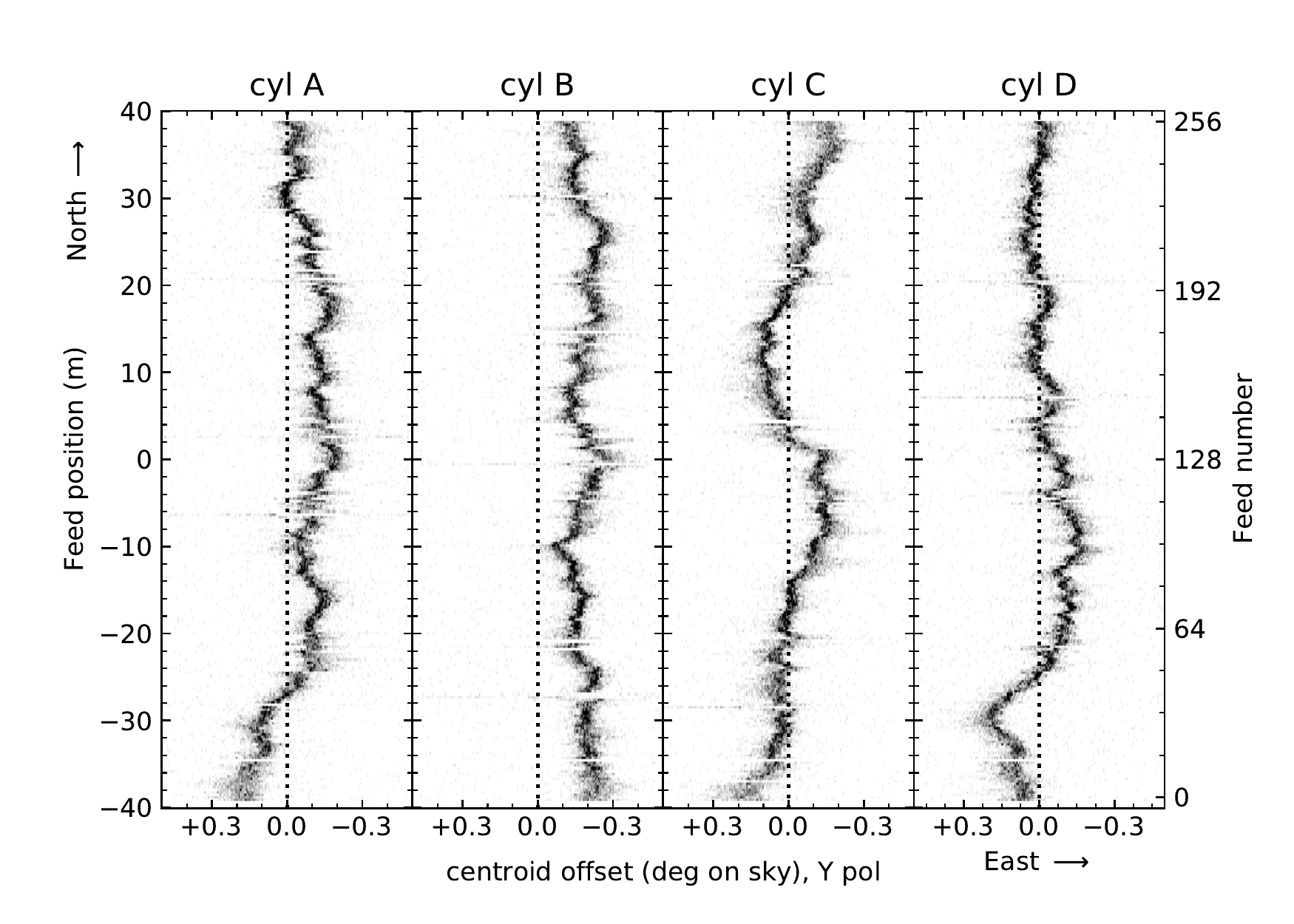}
    \caption{Per-feed measurements of the CHIME E-W beam centroid obtained from a Gaussian fit to the holographic measurements of a Cyg A track.  For each Cylinder A-D (left to right panels), the best-fit centroid is shown as a function of feed position along the cylinder.  Multiple points per feed show results for each non-flagged frequency that was processed for that feed.  The spread with frequency arises from a small but statistically significant oscillation in the centroid with a periodicity of \SI{30}{\mega\hertz}, indicating a small E-W asymmetry in the signal multi-path.  The dominant effect, however, is the position-dependent variation that arises from imperfections in the cylinder surface and primarily from a few mm of  E-W position offsets of  feeds on the focal line.  The Y polarization is shown; the X polarization shows a similar trend with a slightly larger frequency variation.}
    \label{fig:centroid}
\end{figure}

\begin{figure}
    \centering
    {
    \includegraphics[width=1\linewidth, trim={0 0 1cm 0}, keepaspectratio]{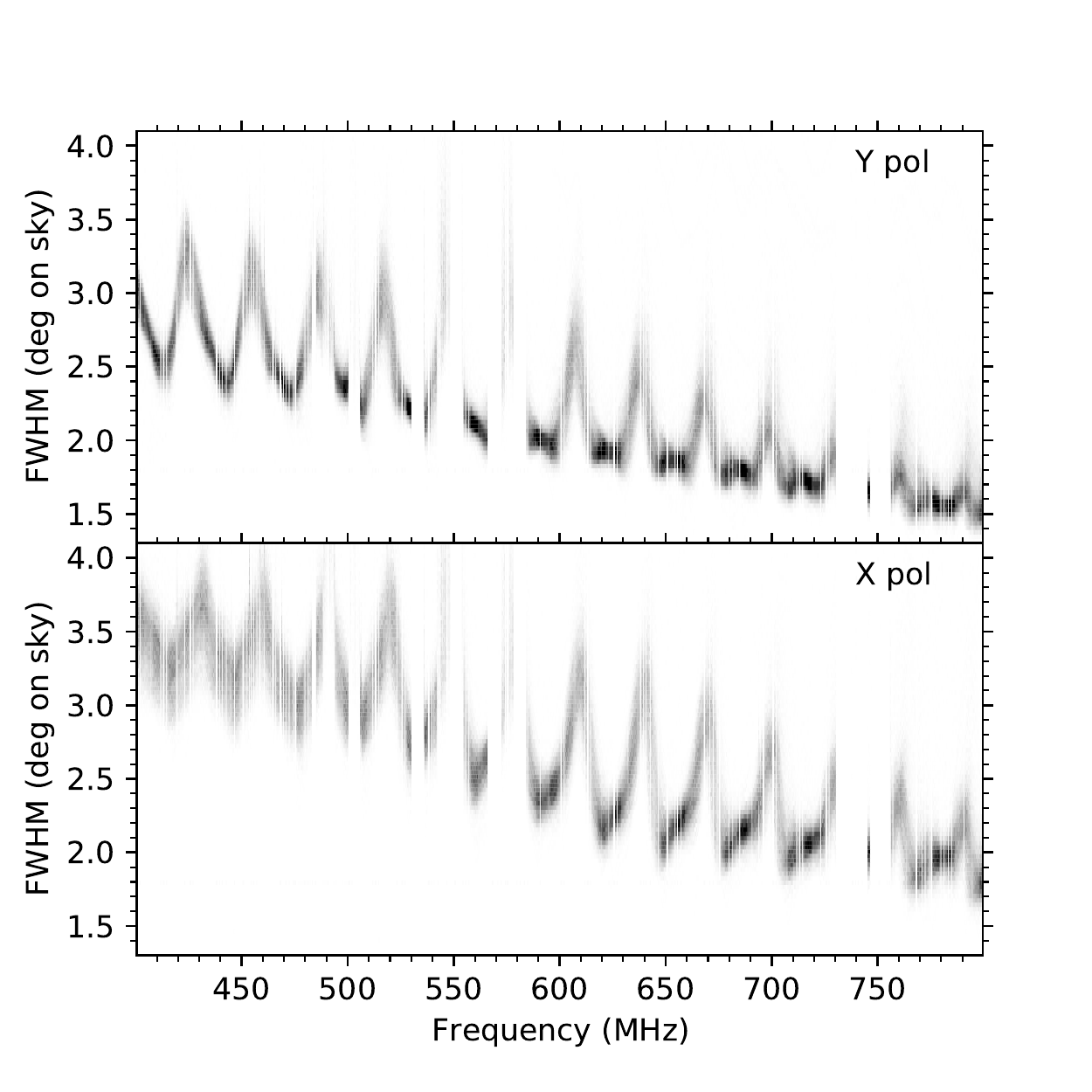}} \\
    \caption{Measurements of the CHIME $X-Z$ plane (East-West)  beam FWHM obtained from a Gaussian fit to the holographic measurements of a Cyg A track, plotted as a function of frequency.  Multiple points per frequency show results for each non-flagged feed that was processed for that frequency.  The top and bottom panels show results for the Y and X polarizations, respectively.  The dominant variation in the FWHM arises from signal multi-path which introduces a \SI{30}{\mega\hertz} periodicity in the beam response.  Characterizing this multi-path is the dominant ongoing effort in the CHIME beam calibration program.  
    }
    \label{fig:width}
\end{figure}

For each frequency and co-polar correlation product in the holography data, we fit the sum of a Gaussian profile and a constant offset to the amplitude response as a function of hour angle.  The resulting centroid and Gaussian full-width half-max (FWHM) parameters are shown in Figs.~\ref{fig:centroid} and~\ref{fig:width}, respectively, for all feeds and frequencies.

The centroid parameter shows a small but significant dependence on focal line position which is correlated for nearby feeds (\cref{fig:centroid}).  This suggests that the centroid offsets are due to physical displacements of the focal lines and/or cylinder structures from their design positions.  Note that, given the \SI{5}{\meter} focal length of CHIME, a \ang{0.2} centroid offset requires an effective position offset of \SI{1.7}{\centi\meter} between the E-W feed position and the symmetry plane of the cylinder.  In cylinders A, C, and D, the median centroid offset (taken over feed number) is close to zero, whereas in Cylinder B, all feeds are offset to the east (i.e., towards negative hour angle), implying that the focal line as a whole is offset by $\sim$\SI{1}{\centi\meter} from the symmetry plane of Cylinder B's parabolic figure. 
\reviewercomment{Multiple points for each feed on a given cylinder in \cref{fig:centroid} show measurements for that feed at different frequencies, and the spread of these points represents a small frequency dependence in the E-W centroid}.  This variation has a periodicity of $\sim$\SI{30}{MHz} which arises from an E-W asymmetry in CHIME's signal multi-path.  Multi-path effects are discussed in \secref{sec:beam_model}.

\reviewercomment{\cref{fig:width} shows the FWHM parameter as a function of frequency for both polarizations,
with multiple points per frequency representing measurements
for all the non-flagged feeds for that frequency.
As expected given the dipole illumination pattern of the feed, the FWHM is roughly twice as large at \SI{400}{\mega\hertz} as at \SI{800}{\mega\hertz} and $\sim$20\% higher in the X polarization than in the Y polarization. Multi-path effects cause the $\sim$\SI{30}{\mega\hertz} ripple in the FWHM 
for both polarizations.
There is a larger spread in the FWHM measurements for the X-polarization, especially
at low frequencies. 
This difference between polarizations remains after including flags for feeds near 
structural elements like support struts, 
so the exact cause of the larger spread in X polarization, and its impact on the cosmology data analysis, remains under investigation.}

\begin{figure*}
    \centering
    \includegraphics[width=1.1\linewidth,keepaspectratio, trim={3cm 0 0 0}]{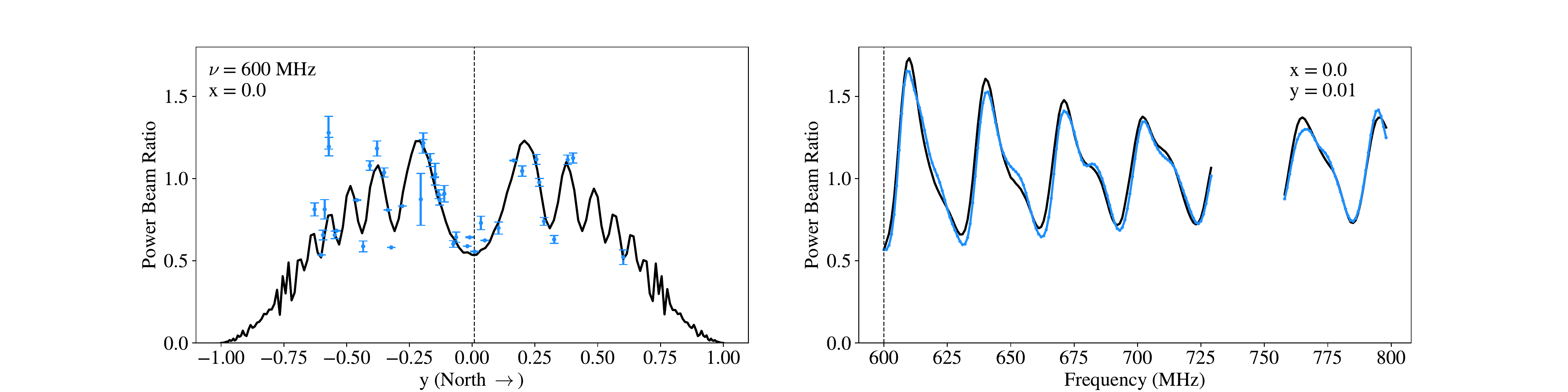}
    \caption{The on-meridian power beam response as determined from 37 bright celestial sources (blue points and curve) and from a coupling model (black curves, \secref{sec:coupling_model}). \textit{Left} - The response as a function of orthographic $y$ at \SI{600}{\mega\hertz}. \textit{Right} - The response as a function of frequency for a source, 3C147, within $0.5^{\circ}$ of zenith. The vertical dashed lines show the zenith angle and frequency used for constructing 1D beams in right and left panels respectively.  The blue points and curve are obtained from the beam-formed response to 37 point sources at transit, divided by their expected flux from the literature.  The uncertainties are dominated by uncertainties in the literature flux densities and are highly correlated across the band.  The current best-fit coupling model (\secref{sec:coupling_model}), fit to all 37 sources in the range \SIrange{600}{800}{\mega\hertz}, is shown in black.  This relatively simple model clearly captures the main features of CHIME's on-meridian response.
    }
    \label{fig:beam_ns}
\end{figure*}

\subsubsection{Celestial Sources Near Transit}
\label{sec:ptsrc}

There are 37 bright point sources in CHIME's declination range with flux greater than \SI{10}{\jansky} at \SI{600}{\mega\hertz}, which is significantly above our estimated confusion noise of $\sim$\SI{0.1}{\jansky}.  These sources span zenith angles of $\SI{37}{\degree}$ north of zenith to $\SI{38.9}{\degree}$ south of zenith.  We measure the spectra of these sources at transit by phasing the CHIME array to the declination of the source and recording the observed spectrum as a separate dataset.  Given our Cyg A calibration strategy, the ratio of the observed spectrum to its spectrum reported in the literature gives the ratio of CHIME's on-meridian beam response at the zenith angle of the source to its on-meridian response at the zenith angle of Cyg A.  Examples of these data are shown in \cref{fig:beam_ns}, along with a preliminary fit to a ``coupling model'' described in \secref{sec:coupling_model}.

This technique can be extended to a much larger number of fainter sources if we restrict attention to inter-cylinder baselines which have a large east-west baseline component and therefore lower confusion noise from diffuse synchrotron emission.  For the cross correlation of CHIME data with large scale structure traced by the eBOSS survey \citep{stacking2021}, we used this technique to produce a model of CHIME's main lobe response from the north to south horizon.  A detailed description of the procedure and the model is given there, so we provide only a brief summary here.

Inter-cylinder baselines with a large east-west component are largely insensitive to diffuse sky signals, such as Galactic synchrotron emission.  Thus, one can approximate the emission measured by these baselines as solely composed of radio point sources (ignoring the subdominant cosmological signal).  We construct a model of this sky using catalogs of source spectra measured by the VLA Low-frequency Sky Survey (VLSS; \citealt{cohen2007}), the Westerbork Northern Sky Survey (WENSS; \citealt{Rengelink1997A&AS..124..259R}), the NRAO VLA Sky Survey (NVSS; \citealt{Condon1998AJ....115.1693C}), and the Green Bank survey (GB6; \citealt{Gregory1996ApJS..103..427G}).
This sky model is put into a simulation pipeline that produces mock (noise-free) visibilities which have no CHIME beam convolution applied.  Then, as described in Appendix A of \cite{stacking2021}, we form beams on the sky using both the simulated and measured visibilities and regress the two data sets to infer the primary beam response in the data.  The resulting beams are filtered to remove small-scale features that likely originate from flux errors in the catalog.

\begin{figure*}
    \centering
    \includegraphics[width=0.98\linewidth,keepaspectratio]{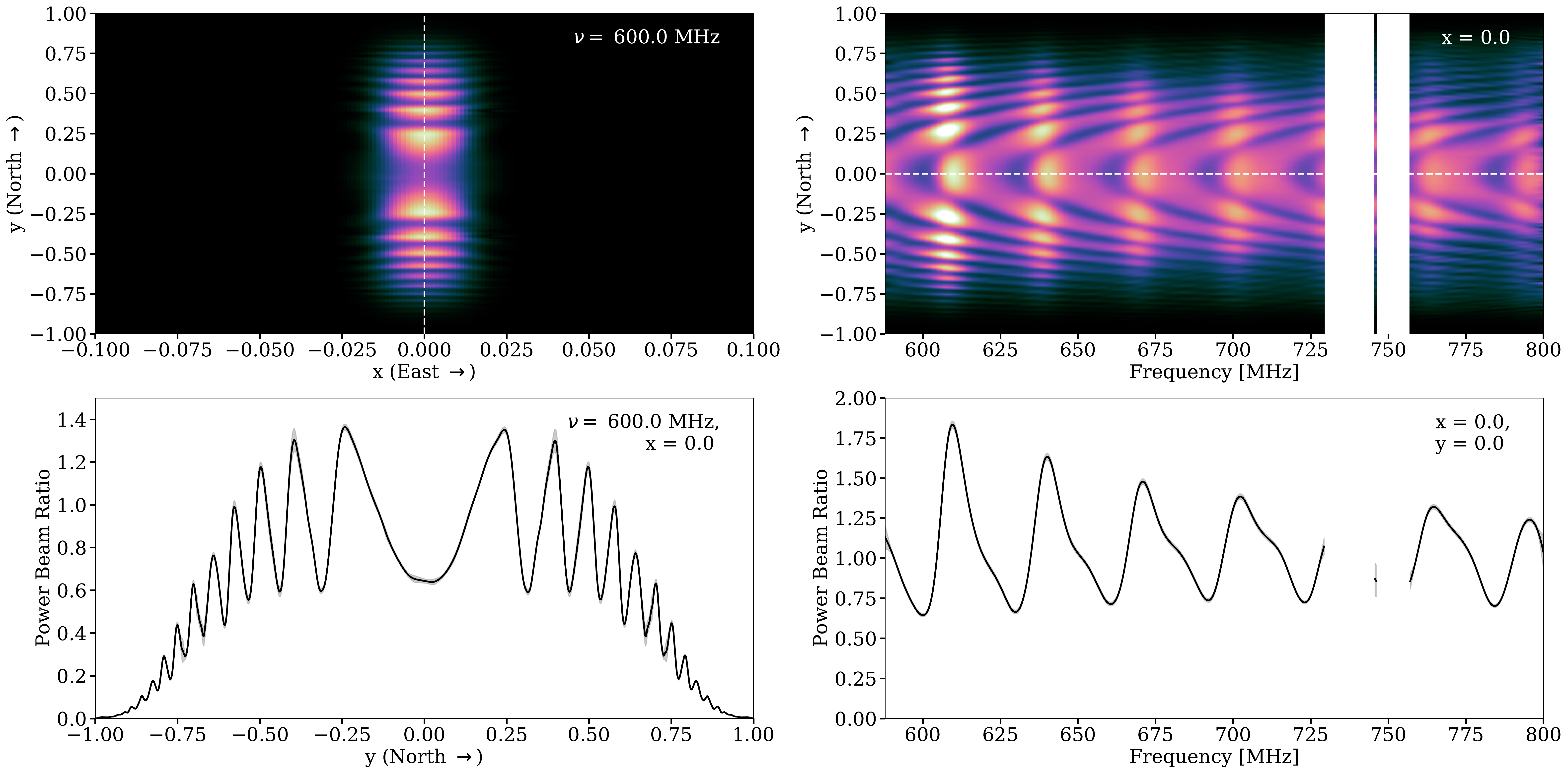}
    \caption{Model for the near-meridian primary beam of the Y polarized array, obtained by fitting the visibilities measured with long E-W baselines to a model for the radio emission from extragalactic point sources.  Top left: beam model at \SI{600}{\mega\hertz} as a function of orthographic angular coordinates $x$ and $y$.  Top right: beam model on meridian as a function of frequency and $y$.  The bottom panels show 1D slices through the beam at the location indicated by the white dashed line in the panels above.  The gray contours in the bottom panel provide an estimate of the uncertainty (\SI{68}{\percent} confidence interval).  The color scale in the top panel spans the range shown on the y-axis in the bottom panel.  The beam model is in power units in all cases and has been normalized to 1.0 on meridian at the declination of Cyg A (\SI{40.734}{\degree}) at each frequency in order to match how the data are normalized by the calibration procedure. The X polarisation response exhibits the same general features, but is slightly wider in both the $x$ and $y$ directions, and also has a lower response at zenith because the dipole is oriented perpendicular to the axis of the cylinder.}
    \label{fig:beam_ptsrc}
\end{figure*}

At present, the model is only derived for hour angles less than roughly \ang{2}, but in principle it can be extended to cover the dominant east-west sidelobes.  \cref{fig:beam_ptsrc} shows the beam response obtained from this method for the Y polarization at 600 MHz.  Our interpretation of the main features of this beam is given in \secref{sec:beam_intro} and \ref{sec:beam_model}.

\subsubsection{Solar Response}

\begin{figure}
    \centering{
    \includegraphics[width=1.0\linewidth,keepaspectratio]{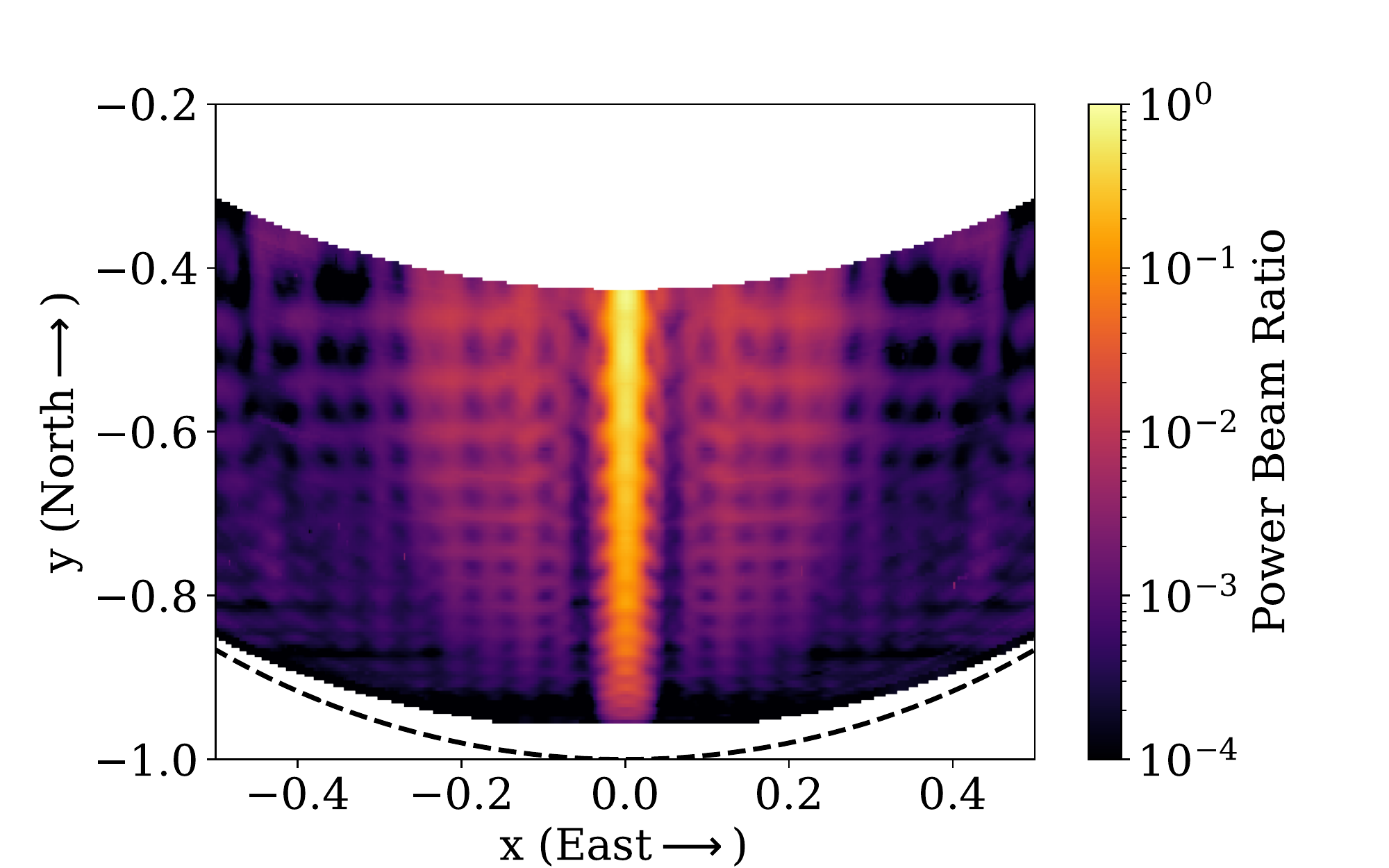}} \\
    \caption{Orthographic projection of the average CHIME beam response in the Y polarization at \SI{679}{\MHz}, from beam-formed measurements of the Sun taken between  31 May 2019 and 11 July 2020. The North-South extent of the data is set by the $\pm\ang{23.5}$ declination range traversed by the Sun over this time interval.  The black dashed line marks the Southern horizon.}
    \label{fig:beam_solar}
\end{figure}

The Sun provides a complementary dataset to astrophysical point sources for beam mapping. Every six months, the Sun moves between $\pm$\ang{23.5} declination, providing quasi-continuous spatial sampling over this declination range.  Additionally, the brightness of the Sun ($>$\SI{100}{\kilo\jansky}) permits un-confused hour angle coverage comparable to the holographic measurements.  The flux of the Sun varies with time, but this can be calibrated at every declination that has a sufficiently bright astrophysical source.  Variability between such calibrations limits the accuracy of these data, as does the finite angular size of the Sun, but even this qualitative information is invaluable for guiding beam modelling efforts.  Data collected in the Fall of 2019 are shown in \cref{fig:beam_solar}.  A more detailed description of CHIME's solar data processing is \reviewercomment{presented} in \cite{solar2022}.

\subsection{Beam Modelling}
\label{sec:beam_model}

Ultimately, we seek to use the datasets described above to construct a single comprehensive beam model. The biggest challenge in this endeavor is accurately accounting for the multi-path and coupling effects that modulate the simple elliptical base beam. In the following, a few complementary approaches to this problem are described: a data-driven approach, where we attempt to extrapolate the datasets described above to the $2\pi$ sr above the horizon; and a semi-analytic approach where we model the coupling between separate feeds with a physically-motivated parameterization. Note that this work is ongoing, and further details are deferred to forthcoming papers.

The models described below are intended to describe a typical feed's beam response.  The response of individual feeds will deviate from this owing, for example, to perturbations in the cylindrical reflector shape (e.g.\ \cref{fig:surface_error}), and/or to feed position and orientation offsets (e.g.\ \cref{fig:centroid}).  Additionally, the presence of structural elements in the vicinity of some feeds, e.g.\ support struts, can scatter radiation and alter the beam response of those feeds \citep{1991RaSc...26..363L}.  Given that CHIME measures numerous redundant visibilities (i.e.\ correlation products with the same baseline), feed-to-feed variations will average down in the stacked data. The extent to which these variations must be accounted for when filtering foregrounds remains to be quantified.

\subsubsection{Data-Driven Extrapolation}

  We exploit the fact that CHIME's beam response is nearly separable in orthographic $(x,y)$ angular coordinates, and use singular value decomposition (SVD) of the solar data to derive a set of beam modes which can be continued to regions not covered by the solar data.  The extrapolations can be guided by additional data, e.g., the holography data (\S\ref{sec:holo}) and/or the celestial source data (\S\ref{sec:ptsrc}); and/or by theory, e.g. the coupling model (\S\ref{sec:coupling_model}).We have been developing a few approaches to this extrapolation problem which we outline below.  However, we have yet to settle on a single approach, so we defer the details to a forthcoming paper.

\begin{figure}
    \centering
    \includegraphics[width=1.0\linewidth,keepaspectratio]{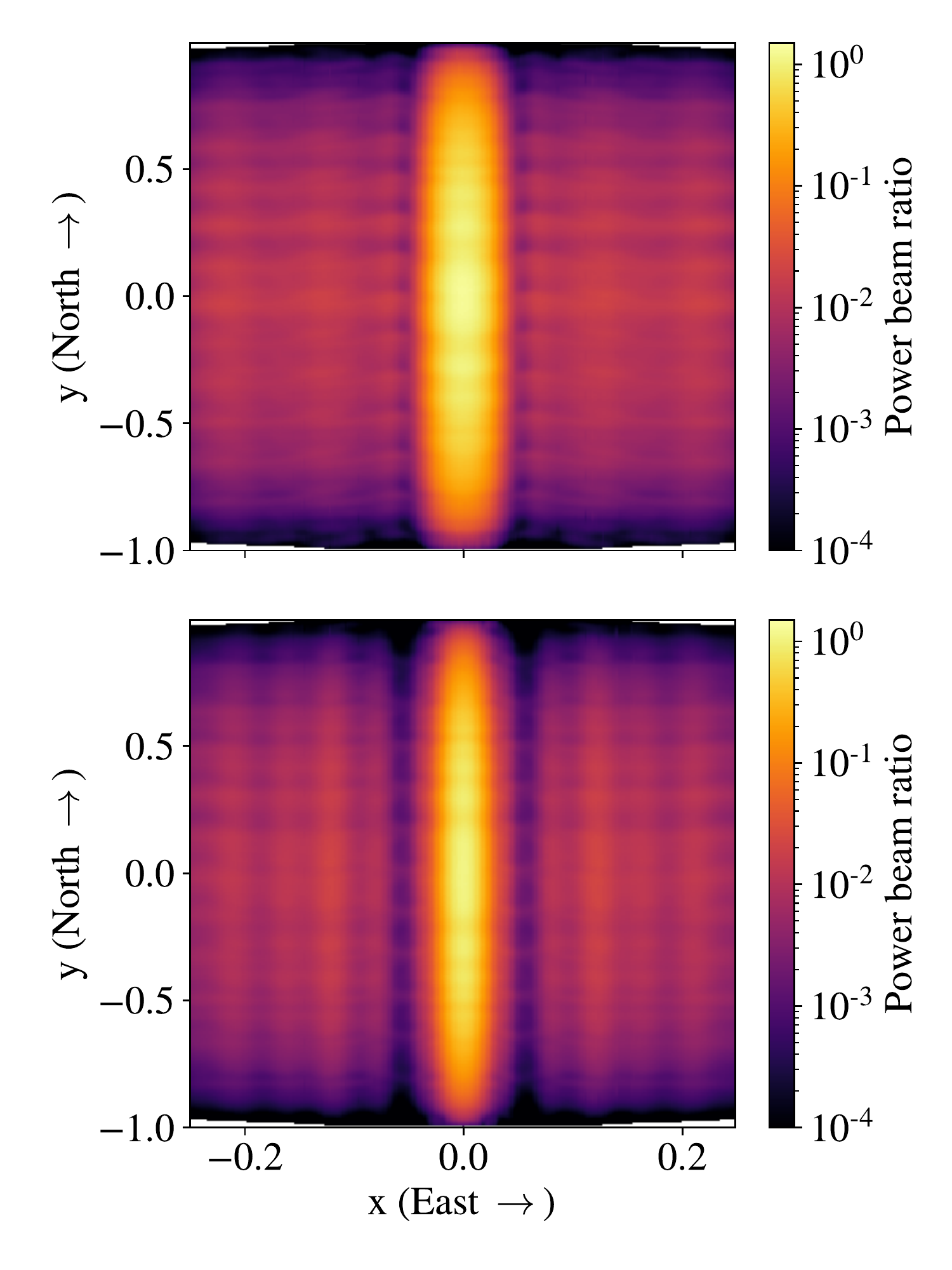}
    \caption{Orthographic projection of the modelled CHIME beam response in X (upper panel) and Y polarization (lower panel) feed, generated at \SI{678}{\mega\hertz} using the data-driven model described in \secref{sec:beam_model}.  It is modeled using basis functions derived from solar data measurements, which are fit to independent measurements of the beam.
    }
    \label{fig:beam_SVD}
\end{figure}

In one approach we form a set of basis functions at a target frequency, derived from the solar data in a small frequency range centred on the target frequency.  We use the coupling model to extrapolate these functions to \SI{2\pi}{\steradian} and fit them to a combination of the holography and celestial source data described above.  The viability of this model rests on the fact that $\sim$99\% of the variance in the solar data can be described by a linear combination of 3 modes which are separable in $(x,y)$ coordinates.  However, our ability to accurately extrapolate these modes to the rest of the sky relies on a model that has known limitations.  Further, our ability to assess the quality of the model is limited by the available holography and source data, which have limited sky coverage. \cref{fig:beam_SVD} shows a current estimate of the \SI{2\pi}{\steradian} beam response at \SI{678}{\mega\hertz}.

In a second approach, we exploit the fact that the sidelobe signal in the solar data, as a function of orthographic $x$ -- once re-scaled by frequency, i.e., $x' \equiv x\cdot(\nu/\SI{600}{\mega\hertz})$ -- is well described by a linear combination three functions of $x'$ over the entire range of $(y,\nu)$ measured by the solar data.  We fit these three modes to the near-meridian celestial source data depicted in \cref{fig:beam_ptsrc}, at each $y$ and $\nu$ separately.  The result is a \SI{2\pi}{\steradian} model which is visually similar to \cref{fig:beam_SVD}.  A detailed description and comparison is deferred to a forthcoming paper.

\subsubsection{Coupling Model}
\label{sec:coupling_model}

This approach is a phenomenological one inspired by physical optics: we form a parameterized model of the base beam and of multi-path effects, and fit those parameters to the data described in \secref{sec:beam_data}.  In its simplest form -- called the coupling model -- the multi-path is attributed entirely to crosstalk between pairs of feeds along the focal line.  In the time domain, we may express this as a superposition of base beam profiles, delayed by specific amounts in time,
\begin{equation}
\tilde{A}_i({\bf n},t) = A_i({\bf n},t) + \sum_j \alpha_{ij} A_j({\bf n},t+\tau_{ij}),
\end{equation}
where $A_i({\bf n},t)$ is the electric field produced by feed $i$ (thought of here as a transmitter) in the absence of neighbouring feeds, $\bf n$ is a directional unit vector, $A_j({\bf n},t+\tau_{ij})$ is the electric field produced by neighbouring feed $j$, delayed by a time $\tau_{ij}$, and $\alpha_{ij}$ is a coupling coefficient that describes the strength of the coupling.  In the frequency domain, the time delay transforms to a phase factor.  In the model's simplest form, we assume that all feeds produce the same pattern, $A({\bf n})$, and that there are two coupling paths between any pair of feeds: a ``direct'' path via signals propagating parallel to the ground plane with delay $\tau_{ij} = |\Delta y_{ij}|/c$, where $\Delta y_{ij}$ is the north-south separation between feeds $i$ and $j$, and a ``1-bounce'' path via signals reflecting once off the cylinder as they travel from feed~$i$ to feed~$j$, with a delay set by analogous geometric arguments. The model is parameterized in terms of coupling coefficients for different coupling paths, and their associated fall in strength as a function of feed-separation. An example of this model, fit to the source transit data and evaluated on meridian, is presented in \cref{fig:beam_ns}. Typical coupling strengths between adjacent feeds are found to be $\sim 15\%$ and $\sim 3\%$ for the direct-path and 1-bounce-path cases, respectively.  The coupling strength as a function of antenna separation falls differently for the two cases, and is estimated to be $\sim 1/|\Delta y_{ij}|^2$ and $\sim 1/|\Delta y_{ij}|^{1/2}$ for the direct and 1-bounce paths respectively.  Multi-bounce paths couple at less than one percent. Further details about the parameterization and performance of this model will be presented in a forthcoming paper.

There are at least two known limitations of the coupling model described above: 1) to date, it has not been able to fully account for the frequency dependence we observe in the source transit data (\cref{fig:beam_ns}), especially in the lower half of the frequency band, and 2) it predicts a N-S response modulation that is independent of E-W direction on the sky, which is inconsistent with the solar data (\cref{fig:beam_solar}).  There are at least two possible explanations for this: 1) the coupled feeds, $j$, (re)radiate a different base beam, $A_j({\bf n})$, than does the source feed, $i$, and/or 2) in addition to coupled feeds re-radiating the source signal, there is also a reflected signal that bounces directly off the ground plane and back to the cylinder before reaching the sky.  This reflected signal could have a slightly different delay parameter than the 1-bounce coupled signal, and is expected to have a different E-W profile than the coupled signal.

We are in the process of developing a richer model that incorporates these effects, parameterized by the electric field distribution in the cylinder aperture, as informed by the commercial software packages \sw{CST} \citep{CST} and \sw{GRASP} \citep{GRASP}.  From preliminary studies, it appears that the aperture field can be parameterized relatively compactly, and that the resulting model is qualitatively successful at fitting the features seen in the solar data.  Specifically, with $\sim$20 parameters to describe the aperture field, single-frequency fits to the solar data in \cref{fig:beam_solar} produce a model with residual errors of $\sim$10$^{-3}$ in the solar data, but which can be evaluated over the full sky.  Future work will involve using the holography and celestial source data in the model fits, so a detailed discussion of this effort will be deferred to a forthcoming paper.  Note that the coupling model described above is a special case of this more general multi-path model.

\subsection{Beam Model Usage}

In this section we summarize how various  beam models developed for CHIME have been used in scientific analyses to date.

\begin{itemize}
\item The celestial source model depicted in \cref{fig:beam_ptsrc} was developed for the stacking analysis presented in \citet{stacking2021}.

\item The detection of an exceptionally bright radio burst from a Galactic magnetar \citep{frbmagnetar-2020Natur.587...54C} occurred when the object was 22 degrees off of CHIME's meridian. Characterization of this rare event requires knowledge of the instrumental beam well off axis.  We use the solar data \citep{solar2022} and Taurus A (Tau A) holography data to measure CHIME's beam response there, enabling a measurement of the burst flux/fluence.

\item The  first CHIME FRB catalog \citep{frbcatalog-2021arXiv210604352T} gives an estimate of flux/fluence of each FRB. A beam model which gives the beam solid angle as a function of forward gain and frequency is required to model the statistical distribution of their brightness.  An early version of the \SI{2\pi}{\steradian} model depicted in \cref{fig:beam_SVD} is used for this work. This enables a measurement of the FRB sky rate, one of the main results from the paper.

\item CHIME/FRB is able to perform polarimetry on some events \citep{frbpolarimetry-2021ApJ...920..138M}.  While polarized beam models are not yet used in these measurements, CHIME's beam data have informed which systematic effects need to be included in the polarization fits as nuisance parameters. The most important of these is the differential response of the $X$ and $Y$-polarized beams near their half-power points, seen clearly in all  CHIME beam measurements.

\item The FRB team is building \textit{outrigger} cylindrical telescopes to provide a steady stream of sub-arcsecond localizations of FRBs. Data from CHIME holography show a lack of significant beam phase variation within a few degrees of meridian (\cref{fig:beam_holo}).  This result is crucial input to the design of the CHIME Outriggers, meaning the optical design of the Outriggers could differ somewhat from CHIME's design and not require beam phase re-calibration.
\end{itemize}

\section{Performance}
\label{sec:Performance}

In this section, we evaluate the performance of the instrument using data acquired over the first two years of operation and a number of dedicated measurements.  This performance evaluation includes an examination of the main sources of data loss, an assessment of the stability of the complex receiver gains, a characterization of the system temperature, an investigation into the effectiveness of the real-time RFI excision algorithm, and finally, a presentation of maps of the radio sky created from the CHIME \texttt{stack} dataset.

\subsection{Data Loss}
\label{sec:operations}

\begin{deluxetable}{l p{4cm} r}
    \tabletypesize{\footnotesize}
    \tablecolumns{3}
    \tablecaption{Primary sources of data loss.  \label{tab:observing_efficiency}}
    \tablehead{\multicolumn{1}{l}{Axis} & \multicolumn{1}{l}{Source of Data Loss} & \colhead{Fraction Lost}}
    \startdata
        Time  & Downtime & \SI{17}{\percent} \\
              & Daytime & \SI{50}{\percent} \\
              & Rain/Snowmelt & \SI{20}{\percent} \\
              & Phase mis-calibration \newline due to correlator restart & \SI{6}{\percent} \\
              & \textbf{Total} & \textbf{69\%} \\
        \hline
        Frequency & Non-operational GPU nodes & \SI{10}{\percent} \\
                  & RFI & \SI{42}{\percent} \\
                  & \textbf{Total} & \textbf{48\%} \\
        \hline
        Feed & Non-operational or \newline malfunctioning feed& \SI{3}{\percent} \\
              & Feed at ends of cylinder & \SI{6.25}{\percent} \\
              & \textbf{Total} & \textbf{9\%} \\
    \enddata
\tablecomments{Since each source of data loss is largely independent of all other sources, the total fraction of data lost is given by \break $f_{\mbox{\tiny total}} = 1 - \Pi_{i} (1 - f_{i})$ where $i$ runs over all the sources that are listed.}
\end{deluxetable}

CHIME has been operating continuously since its first-light ceremony on 7 September 2017.  The first year of operations was dedicated to commissioning the instrument, developing the real-time pipeline, and developing the calibration and flagging strategies.  Acquisition of data for the cosmological analysis began on October 7, 2018.  Since then, the daily data acquisition rate has averaged $\approx$ \SI{1}{\tera\byte\per\day}.  Of this daily total, approximately \SI{600}{\giga\byte} is the {\tt stack} dataset containing the primary science data.  The remainder is calibration, beam holography, housekeeping, and other engineering datasets.

\Cref{tab:observing_efficiency} summarizes the main sources of data loss between 7 October 2018 and 7 October 2020.  During this two-year period, the instrument was down for a total of   \SI{127}{\day} (\SI{17}{\percent}).  The majority of this time (\SI{102}{\day}) was due to planned hardware maintenance and software upgrades, which occurred approximately five times per year.  A further  25 days were unintended interruptions due to power failures, cooling failures and other accidental outages. 

The radio signal from the Sun dominates over the signal from the rest of the sky, even when the Sun is in the far sidelobes.  The signal from the Sun can be modelled and subtracted to a large extent; however, feed-to-feed variations in the gain or beam and inaccuracies in the model for the extended emission yield residuals that are significant compared to the noise and signal from the rest of the sky.  As a result,  data acquired when the Sun is above the horizon are currently excluded from the cosmological analysis.

Precipitation at the telescope site causes deterioration of the detected signal as a result of water pooling around the focal line electronics.  This signal deterioration is broadband and characterized by a reduction in gain, an increase in noise, and, occasionally, gain oscillations with periods ranging from seconds to minutes.  Accumulation of dry snow does not cause analogue signal deterioration, but snow-melt, which is more difficult to detect using weather data alone, does produce the same signal deterioration \reviewercomment{as rain}. Signals from the   2048 feeds are monitored for broadband, differential increases in their autocorrelations. This signature is used to identify and flag wet feeds before collating redundant baselines.  After each rain or snow-melt, roughly \SIrange{4}{12}{\percent} (inter-quartile range (IQR)) of the inputs are flagged for \SIrange{3}{21}{\hour} (IQR), effectively until they dry.  It is not yet clear if data acquired during these wet periods can be used in the science analysis.  Excluding it results in a $20\%$ reduction in observing time\reviewercomment{, preferentially occurring in months when nights are longest and therefore when we have the most useful data}.  Steps are being taken to improve focal line waterproofing.

The synchronization procedure implemented by the FPGAs does not guarantee that the phase of a common signal measured by two inputs on different ADC chips will remain constant through an FPGA restart.  This change in phase after FPGA re-synchronization is observed in the noise-source data, and the size of the phase change can be large compared to the requirements on instrument stability outlined in \secref{sec:phase}.  As a result, we mask the interval between each FPGA restart and the following point source calibration.  This results in an approximately $6\%$ reduction in observing time.

\Cref{tab:observing_efficiency} also lists the average fraction of the \SIrange{400}{800}{\mega\hertz} band that is masked due to RFI (as detailed in \secref{sec:noise_rfi}) and lost due to non-operational GPU nodes.  Note that in June 2020 the correlator software was upgraded to allow for much greater flexibility in the mapping between frequency channels and GPU nodes.  This gave us the capability to send frequency channels already contaminated by persistent RFI to the set of GPU nodes that are offline at any given time, which recovers a large fraction of the \SI{10}{\percent} of the band that was previously lost due to non-operational GPU nodes.

Finally, \Cref{tab:observing_efficiency} provides estimates of the fraction of the 2048 correlator inputs that are masked prior to collating redundant baselines.  The flagging broker masks approximately $3\%$ of inputs because they fail one or more of the tests described in \secref{sec:ReceiverSystem}.  In addition, in December 2019 we began applying a static mask that consists of the 8 feeds at the edge of each cylinder because it was determined that these feeds exhibit a highly non-redundant beam pattern.

\subsection{Stability}
\label{sec:stability}

\begin{figure}
    \centering
    \includegraphics[width=0.95\linewidth,keepaspectratio]{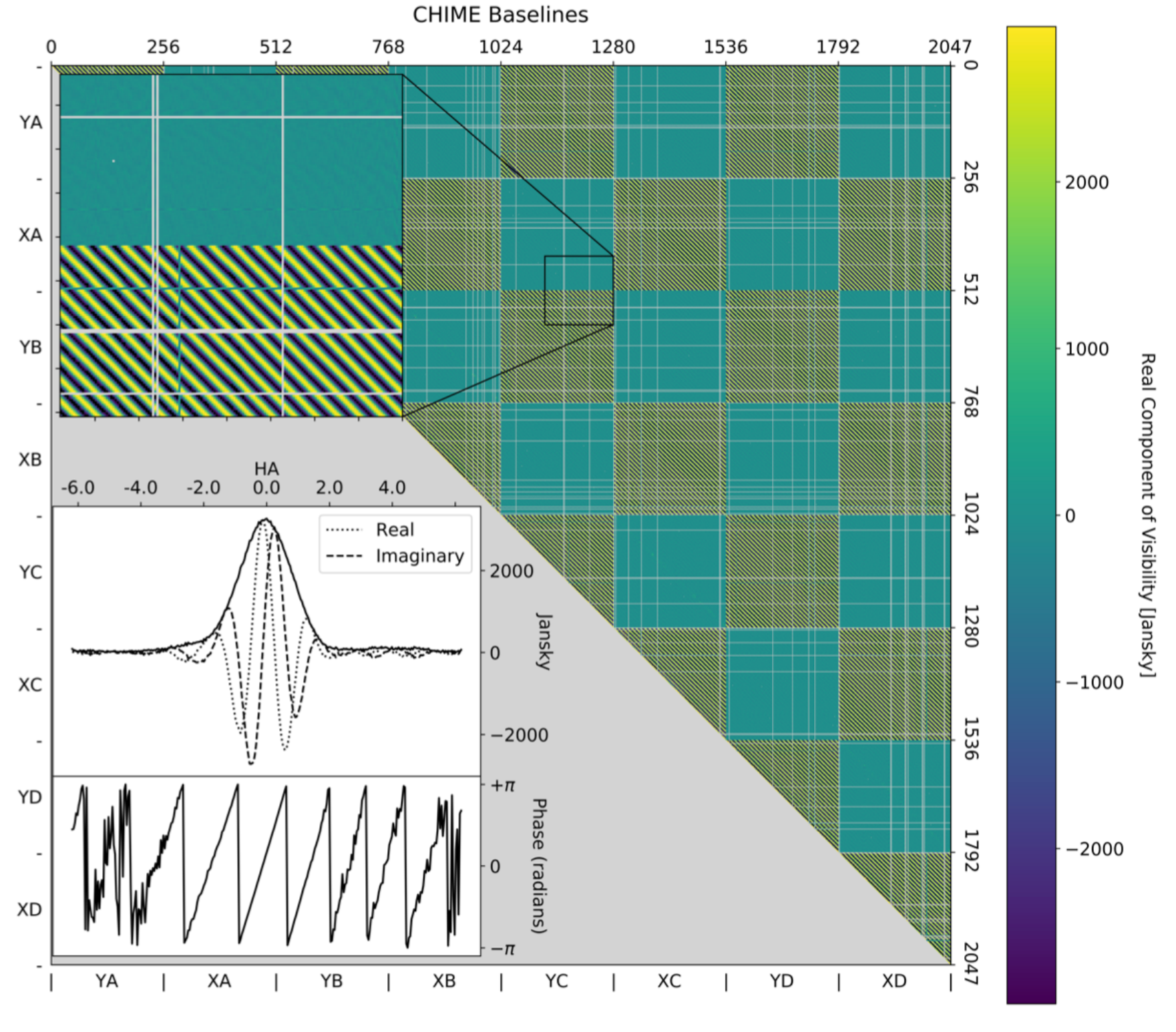}
    \caption{
    Real component of the calibrated visibilities at 758.203~MHz during the transit of Cyg A. Inset is the visibility for a single 22~m pure E-W baseline as a function of the hour angle of Cyg A, with the amplitude (solid), real (dotted), and imaginary (dashed) components shown in the top panel, and the phase shown in the bottom panel.
    }
    \label{fig:n2_visibility}
\end{figure}

The stability of the instrument is assessed using the complex gains measured by the \textit{calibration broker} (described in more detail in \secref{sec:ReceiverSystem}). The broker computes and stores gains using data from four bright source transits every day: Cassiopeia A, Cygnus A, Taurus A, and Virgo A, henceforth referred to as Cas A, Cyg A, Tau A, and Vir A respectively.  \cref{fig:n2_visibility} shows an example of $\Nfeed^{2}$ visibility data acquired during a Cyg A transit after applying the complex gains derived from the transit.  On any given day, one source is chosen as the primary calibrator (typically the brightest source to transit at night), but all of the transits are analyzed offline to assess stability. To help assess and maintain phase stability, a broadband noise source system is also employed, as described below.

We use end-to-end simulations to determine our stability requirements. Simulation of a CHIME-sized telescope is challenging due to computer resource limitations; therefore, we have performed simulations of a scaled-down instrument (with roughly 1/4 of CHIME's collecting area) to investigate these requirements, examining the anticipated accuracy of the \tcm power spectrum measured after the application of the Karhunen-Lo\`{eve} foreground filter described in \cite{2015shaw}. This work found the requirement for fractional variations in the complex gain to be less than \SI{1}{\percent}, which translates into phase errors smaller than \SI{0.007}{\radian} and amplitude errors smaller than \SI{0.7}{\percent}.  However, these requirements are derived from a simulation whose gain variations are constant across the band and un-correlated from input to input. Furthermore, it is unclear how these requirements scale with the size of the telescope, and neither of these conditions hold in the observed gain variations presented below; thus, these requirements serve as a rough guide only.  More realistic simulations designed to better reflect some of the observed complex gain variations have since been performed, and the resulting requirements are noted below where applicable.

\subsubsection{Amplitude}

\label{sec:amplitude}
\begin{figure}
   \centering \includegraphics[width=0.98\linewidth,keepaspectratio]{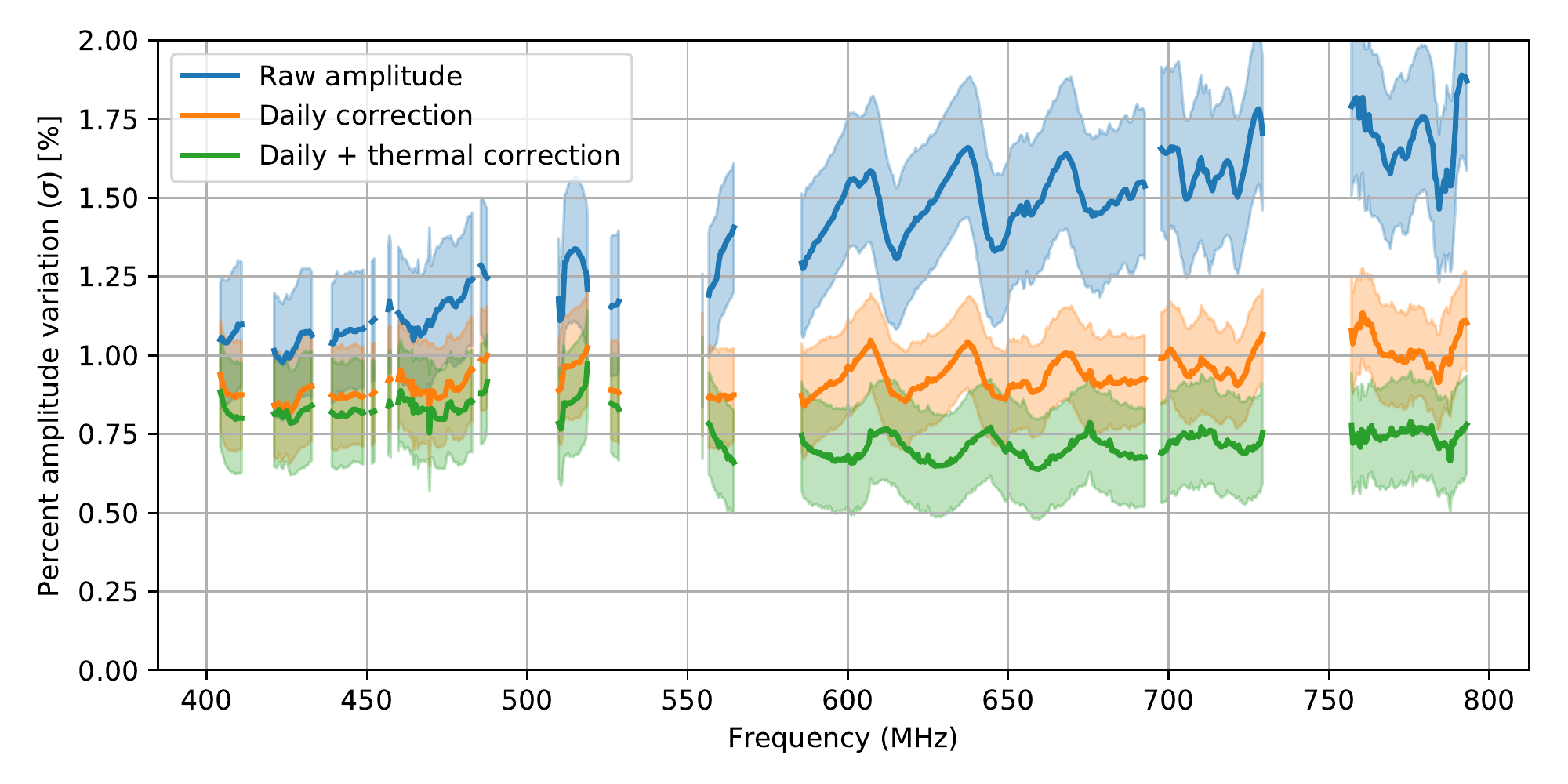}
    \caption{Gain amplitude stability. The blue band shows the standard deviation of the fractional gain amplitude variations as determined from 259 transits of Cas A, Cyg A and Tau A from June 2018 to July 2019. The {\em rms} stability is evaluated input by input; the central curve shows the mean stability across inputs while the band indicates the 1-sigma spread across inputs. The orange band shows the corresponding information with the gains corrected once per day using a previous transit. The green band shows the result of applying an additional correction to the orange data based on a linear regression to the ambient temperature change since the previous transit.  Gaps in the data correspond to known RFI-dominated bands.}
    \label{fig:stability_amplitude}
\end{figure}

\cref{fig:stability_amplitude} shows the fractional gain amplitude variations (standard deviation) for all correlator inputs as derived from the calibration broker gains over a full year from June 2018 to July 2019. These data include 259 gain solutions (\num{94} from Cas A, \num{89} from Cyg A, and \num{76} from Tau A), which have been scrubbed of RFI contaminated transits and anomalous gains mostly related to wetness of the instrument during rainy periods.

The blue curve indicates the intrinsic gain variations after outliers are removed but prior to applying any calibration corrections. It shows a pronounced slope with frequency which ranges from \SI{1}{\percent} at \SI{400}{\mega\hertz} to \SI{1.8}{\percent} (standard deviation) at \SI{800}{\mega\hertz}. A substantial portion of this variation is due to the thermal susceptibility of the instrument.

The orange curve shows the residual gain amplitude variations for the same data, but after applying a daily correction similar to that which is applied to the archived visibility data. This gives an indication of the gain variations present in the stored data prior to applying any subsequent corrections (see below).  To produce this curve, we take the difference between each transit's gain and a solution from the previous \SI{48}{\hour} (if available) and compute the standard deviation of the difference. This procedure brings the fluctuations down to a nearly flat \SIrange{0.9}{1}{\percent} level.

The green curve shows the residuals after correcting the orange data using the measured thermal susceptibilities and the ambient temperature change since the previous transit. This brings the variations down to $\sim$\SI{0.7}{\percent} (standard deviation).  The thermal correction flattens the residuals considerably, a consequence of the fact that the temperature susceptibility of the system gain rises with frequency from \SI{0.06}{\percent\per\kelvin} at \SI{400}{\mega\hertz} to \SI{0.2}{\percent\per\kelvin} at \SI{800}{\mega\hertz}.  This measured stability achieves the preliminary requirement described above, but with no margin.

By construction, the data tracked by the orange curve remove gain variations slower than $\sim$one day due to {\em any} source, while the data tracked by the green curve removes variations correlated with ambient temperature on all time scales.  We find that thermal regression applied to raw data (blue curve) and the daily-corrected data (orange curve) produced similar residuals.  This suggests that most of the variation on time scales longer than a day is thermal in origin, and that variations on shorter time scales are not well correlated with ambient temperature.

The analysis discussed above is carried out input by input, assuming nothing about how correlated the gain variations are across inputs. A singular value decomposition (SVD) analysis of the raw gain variations over input and time reveals a single dominant mode 
followed by a closely packed mode spectrum. The dominant mode accounts for about 60\% of the data variance at the lower end of the band and grows to over 80\% of the variance halfway to the high end of the band. The singular vector of the dominant mode is highly correlated with the ambient temperature, implying that an ambient-temperature-based correction largely accounts for the common-mode portion of the variance.  Thus, the residual variability after thermal regression (the green band in \cref{fig:stability_amplitude}) gives a good estimate of the non-common-mode variations in the system. 
These residual gain variations show some degree of correlation across inputs.  Making use of this to further improve the correction is under study.

The frequency structure of the gain stability depends on the declination of the source used to derive the gains. This appears to be due to a time and/or thermal dependence of the primary beam response of the instrument. This would be expected if feed-to-feed cross-talk depended on time and/or temperature, which, in turn, could result from thermal expansion and contraction of the CHIME structure. (See \secref{sec:phase} and \secref{sec:beams} for further discussion of these effects.)  Efforts to model this dependence are ongoing. The results shown in \cref{fig:stability_amplitude} are computed from the gains derived from the three brightest sources, so the frequency structure shown there is a weighted average of the response to these three sources.

\subsubsection{Phase}
\label{sec:phase}

\begin{figure}
   \centering \includegraphics[width=0.98\linewidth,keepaspectratio]{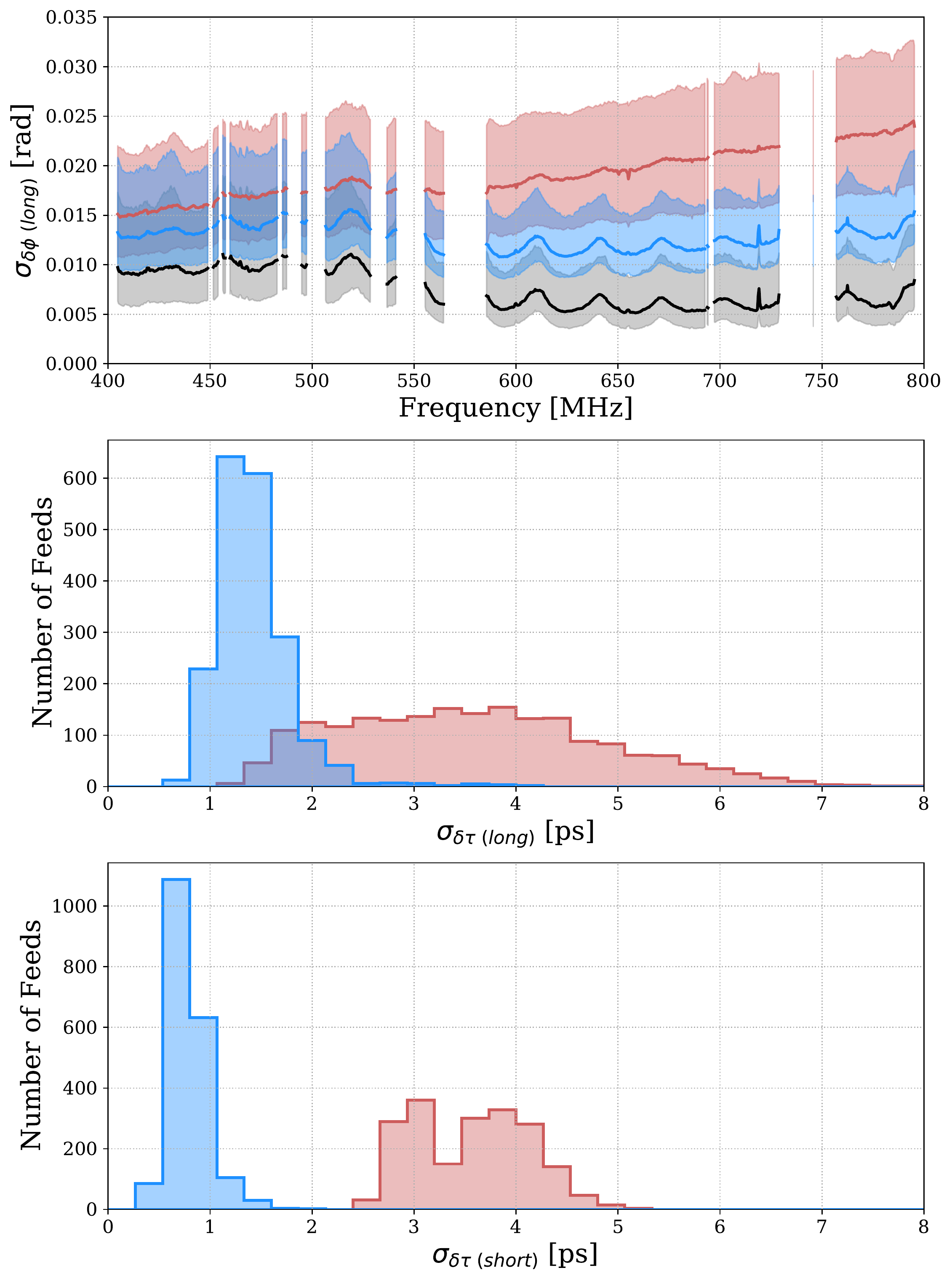}
    \caption{Gain phase stability.  Top: the standard deviation over \num{74} days of CHIME's phase response to Cas A at transit after applying a daily calibration derived from Cyg A.  Lines denote the median and bands denote the central \SI{68}{\percent} over the \num{2048} CHIME feeds.  Red indicates the raw phase variations.  Blue indicates the residual phase variations after correcting for delay variations caused by drift between copies of the \SI{10}{\mega\hertz} clock, thermal expansion of the focal line, and thermal susceptibility of analog receiver chain as tracked by the ambient temperature (see the text for a discussion of each of these effects).   Black indicates the residual phase variations after removing all delay-type variations \reviewercomment{by fitting and subtracting a model for the phase variations that scales linearly with frequency from each transit}.  The phases are referenced to the average phase over feeds of a given polarisation.  Middle: the standard deviation of the delay variations, shown as a histogram over feeds.  The red histogram indicates the raw delay variations while the blue histogram shows the residual delay variations after applying the three corrections listed above.  Bottom: same as the middle panel, but showing delay variations on short time scales ($\lesssim$ \SI{20}{\minute}), obtained by examining a window around the transit of Cyg A or Cas A when these sources are in the primary beam.  \reviewercomment{The black curve in the top panel, which corresponds to the perfect removal of all delay-type variations, is by definition equal to zero for all feeds in the bottom two panels.}
    }
    \label{fig:stability_phase}
\end{figure}

\cref{fig:stability_phase} summarizes the phase stability of CHIME as inferred from the response of each correlator input to the two brightest calibration sources (Cyg-A and Cas-A).  The measured phase variations are highly correlated across frequencies and, to first order, can be described by delay-type variations of the form
\begin{align}
\delta \phi_{ij}(t, \nu) &= 2 \pi \nu \, \delta \tau_{ij}(t) \label{eqn:delay_variations}
\end{align}
where $\delta \phi_{ij}$ is the change in the relative phase between inputs $i$ and $j$ at time $t$ for radio frequency $\nu$ due to a change in the relative delay $\delta \tau_{ij}$.  If we perfectly corrected all delay-type variations, the phase stability of the instrument would improve from the red curve to the black curve in \cref{fig:stability_phase}.  The dominant sources of delay variations are: relative drifts between copies of the \SI{10}{\mega\hertz} clock that defines the sampling rate of the ADCs; expansion and contraction of the telescope with ambient temperature; and changes in the electrical length of the \SI{50}{\meter} coaxial cables with temperature.  We describe these three sources of delay variation in turn, and outline the methods used to partially correct for them, to stabilize the phase.  After applying the corrections, the resulting stability is given by the blue curves in \cref{fig:stability_phase}.

The dominant source of phase instability on time scales $\lesssim$ \SI{20}{\minute} is relative drifts between the eight copies of the \SI{10}{\mega\hertz} clock that are separately distributed to each of the eight FPGA crates.  Each clock defines the sampling rate of the 256 ADCs within a crate.  The drifts are measured and corrected  using a single broadband noise source that is distributed to one input on each of the eight FPGA crates through a passive system of coaxial cables and power splitters.  The correlator computes the covariance of the noise source inputs over a \SI{10}{\second} integration for each of the \num{1024} native resolution frequency channels.  The largest eigenvector of this covariance matrix is used to estimate the response of the eight inputs to the signal from the noise source.  The phase of the response is referenced to the time of the last point-source calibration to remove static ripples caused by reflections in the distribution network.  Then, for each \SI{10}{\second} integration, the phase as a function of frequency is fit to \Cref{eqn:delay_variations} to extract the delay as a function of time, $\delta \tau_{ij}(t)$, for each FPGA crate $i$ relative to a reference $j$.  This is used as a proxy for the drift in the clock copy provided to that crate relative to the reference ADC input on the reference crate.

Examining the relative delay variations between the 4 crates within a single receiver hut, we find that the variations exhibit a sawtooth pattern with an \SI{8}{\minute} (east receiver hut) or \SI{6}{\minute} (west receiver hut) periodicity that mimics the temperature variations in that hut.  This periodicity tracks the cooling cycle of the chiller system in each hut.  This produces a relative delay variation of \SIrange{1}{2}{\pico\second} (standard deviation) between crates in the same hut.  Since the temperatures of the two huts cycle at different periods, the relative delay variations between crates in different huts are significantly larger: approximately \SIrange{6}{8}{\pico\second} (standard deviation).

A suite of simulations  is used to estimate the bias in the \tcm power spectrum due to realistic clock drifts.  We find that the clock drifts must have a standard deviation of $\lesssim$ \SI{1}{\pico\second} to ensure negligible bias in the power spectrum.  The bottom panel of \cref{fig:stability_phase} shows the improvement in the short-timescale delay noise that is achieved by regressing the delay variations obtained from point-source observations against the delay measured by the broadband noise source.  The residual delay variations have standard deviation $< \SI{1.5}{\pico\second}$ and are thus close to meeting our requirements.

Thermal expansion and contraction of the focal line introduce a temperature dependence to the north-south baseline distance that manifests as delay variations on timescales $\gtrsim$~\SI{20}{\minute}.  We can model this with the following expression
\begin{align}
&\delta \tau^{\mbox{\tiny focal-line}}_{ij}(t, \delta, \mbox{\textsc{ha}}) = \nonumber\\
&\left[ \cos{\ell} \sin{\delta} - \sin{\ell} \cos{\delta} \cos{\mbox{\textsc{ha}}} \right] \times \frac{\Delta y_{ij}}{c}\,\epsilon\,\delta T(t)  \label{eqn:focal_line_expansion}
\end{align}
where $\ell$ is the latitude of the telescope, $\delta$ is the declination of the source, \textsc{ha} is the hour angle of the source,  $\Delta y_{ij}$ is the nominal north-south baseline separation, $c$ is the speed of light, $\epsilon$ is the linear thermal expansion coefficient of the focal line, and $\delta T(t)$ is the difference between the ambient temperature and the nominal temperature.  Fitting the delay variations obtained from the point source transits to \Cref{eqn:focal_line_expansion} yields a thermal expansion coefficient of $\epsilon = \SI{21e-6}{\per\kelvin}$ for the focal line.  This is approximately equal to the coefficient for aluminum and roughly twice that of steel.  The focal line structure itself is made of steel while the cassettes that hold groups of 4 antennas to the focal line are made of aluminum, and are bolted to each of their neighbours.  The interplay of these components as the temperature changes is still under study, but our model fits the sky data well so we adopt the best-fit $\epsilon$ as a description of the instrument. 
The resulting delay error is the same for all redundant baselines, so the correction for this effect can be done offline, after collating these baselines.  However, the correction depends on sky position, so it needs to be implemented at the map-making stage.  This work is currently under development.

After controlling for drift in the clocks and thermal expansion of the focal line, the residual delay variations exhibit a correlation with ambient temperature.  Based on thermal chamber measurements of the components of the signal path, changes in the electrical length of the \SI{50}{\meter} coaxial cables are expected to be the dominant source of thermally-induced delay variations.  These changes in electrical length are the result of changes in the physical length of the cable from expansion of the centre conductor and changes in the dielectric constant due to a reduction in the dielectric density from expansion of the outer conductor. To first order, the observed delay variations can be modeled as
\begin{align}
\label{eq:coax_delay}
\delta \tau^{\mbox{\tiny coax}}_{ij}(t) = & (\alpha_{i} - \alpha_{j}) \,\bar{T}(t) + \bar{\alpha}\,[T_{i}(t) - T_{j}(t)]
\end{align}
where $\alpha$ is the thermal susceptibility of the coaxial cable, $T$ is the temperature of the coaxial cable, subscripts $i$ and $j$ refer to specific inputs, and a bar indicates the average over all inputs.  The first term is due to differences in the thermal susceptibility between cables while the second term is due to differences in the effective temperature of the cables.

In order to gauge the relative importance of the two terms in \Cref{eq:coax_delay}, we have installed three ``cable monitors'' that consist of two \SI{50}{\meter} coaxial cables that are routed to the focal line and then back along the same path, with one end connected to the noise source described above and the other end connected to the correlator.  There is one cable monitor routed to each of cylinders A, B, and C.  The cable monitor data are processed in the same manner as the noise source data described above.  The resulting delays are divided by 2 to account for the fact that the length of coaxial cable in the cable monitors is twice that of the CHIME on-sky inputs.  The measured delays are  regressed against the ambient temperature in order to measure the thermal susceptibility of the three cable monitors.  The average thermal susceptibility over cable monitors is $\bar{\alpha} = \SI{2.93}{\pico\second\per\kelvin}$.  The standard deviation over cable monitors is $\sigma_{\alpha} = \SI{0.04}{\pico\second\per\kelvin}$, which will result in relative delay variations with a standard deviation of $\sim$\SI{0.25}{\pico\second} given the temperature variations on a typical night.  Residual delay variations that are not explained by differences in thermal susceptibility are attributed to differences in the effective temperature of the cables.  These residuals have a standard deviation of $\sim$\SI{1.0}{\pico\second}, which implies effective temperature variations with a standard deviation of $\sim$\SI{0.3}{\kelvin} given the value of $\bar{\alpha}$ quoted above.

We characterize the difference in thermal susceptibility between CHIME correlator inputs by regressing the change in delay between point source transits against the change in ambient temperature.  The standard deviation of the thermal susceptibility over inputs is $\SI{0.3}{\pico\second\per\kelvin}$, which is much larger than we would expect from the scatter in the value of $\alpha$ measured for the three cable monitors.  If we randomly draw thermal susceptibilities for three inputs from the sample of 2048,  the probability they are all within $\SI{3}{\percent}$ like the cable monitors is $< \SI{2}{\percent}$.  This indicates that the analog receiver chain likely has some other source of susceptibility to the ambient temperature beyond the coaxial cables that is highly dependent on input.   Nevertheless, this thermal susceptibility is well characterized using the point source observations; we estimate that our uncertainty on the thermal susceptibility is $\sim \SI{0.05}{\pico\second\per\kelvin}$ using bootstrap resampling methods.

\cref{fig:stability_phase} shows in blue the residual delay variations after correcting for clock drift, expansion and contraction of the focal line, and thermal susceptibility of the analog receiver chain.  We find a standard deviation of $<\SI{1.5}{\pico\second}$ on $<\SI{20}{\minute}$ time scales and \SIrange{1}{2}{\pico\second} on \SI{3}{\hour} time scales.  The cable monitor data suggest that differences in the temperature of the coaxial cables are a significant contributor ($\sim \SI{1}{\pico\second}$) to the residual delay variation on long timescales.  We are actively investigating new techniques to measure and correct for the differences in coaxial cable temperature.

\reviewercomment{We characterize the phase stability of the instrument on longer timescales by examining changes in the phase between night-time transits of other pairs of bright point sources observed between February 2019 and March 2020.  The transit times of the four brightest point sources are spaced apart such that their various differences probe timescales ranging from \SIrange{0}{24}{\hour} with a roughly \SI{3}{\hour} sampling.  The worst performance occurs on \SI{18}{\hour} timescales where the post-correction delay variations have an RMS of $2.0\pm\SI{0.6}{\pico\second}$ (mean $\pm$ standard deviation over feeds).  This is a small degradation in the $1.6\pm\SI{0.6}{\pico\second}$ RMS delay stability observed on \SI{3}{\hour} timescales and shown in \cref{fig:stability_phase}.  If we expand the analysis to also include daytime transits, then we find a significant degradation in the delay stability, with the worst performance occurring on \SI{10}{\hour} timescales where the RMS is $3.4\pm\SI{1.1}{\pico\second}$.  This is a secondary reason to exclude the daytime data from the cosmology analysis, with the primary reason being contamination from solar radio emission.}

\subsection{Noise}
\label{sec:noise}

\begin{figure}
    \centering
    \includegraphics[width=0.98\linewidth,keepaspectratio]{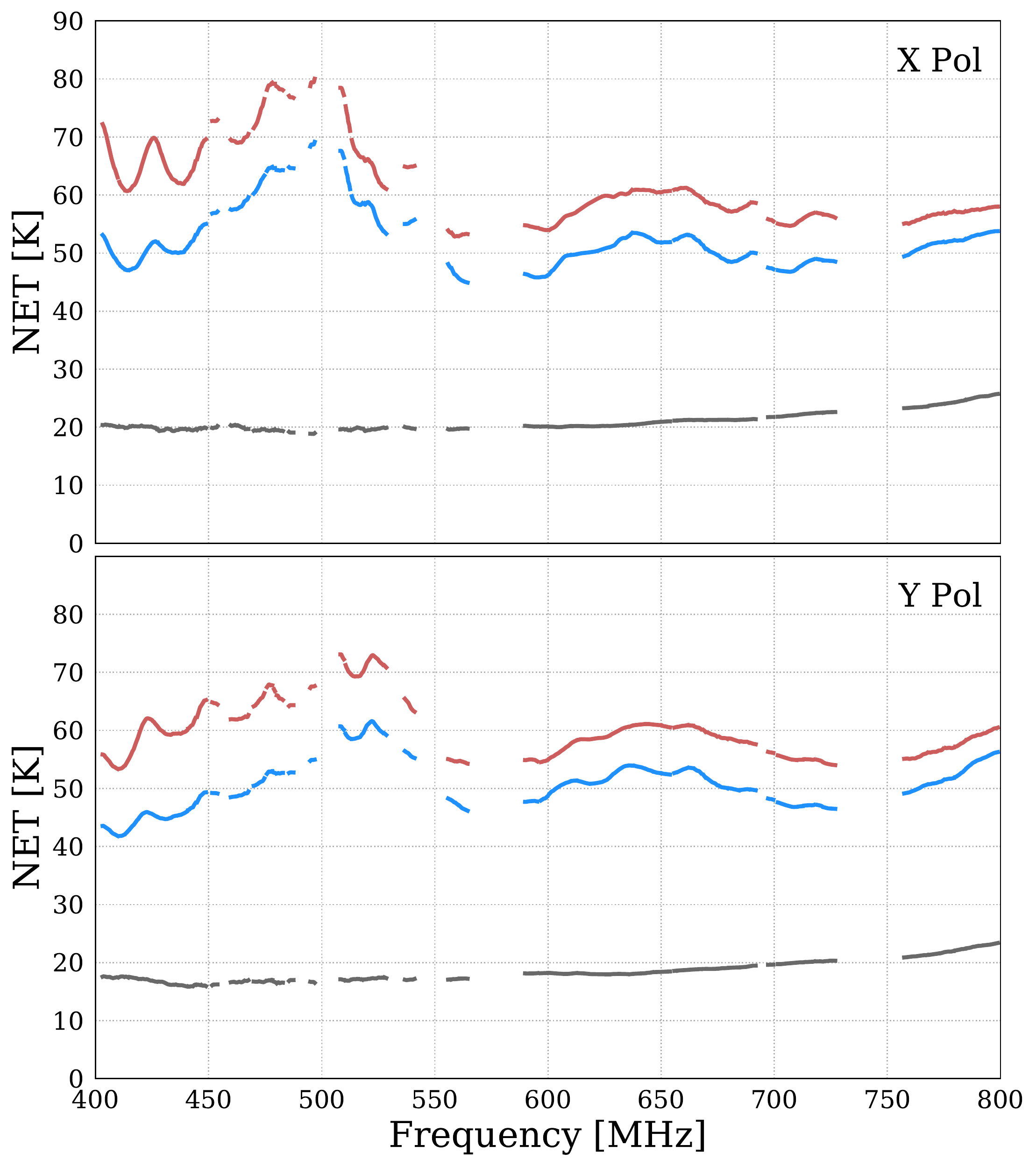}
    \caption{Noise Equivalent Temperature (NET) as a function of frequency for the two polarizations of a single CHIME antenna.  The antenna is located \SI{11}{\meter} south of the center of Cylinder B.  The receiver temperature (see text) is shown in black.  The system temperature when a dim part of the radio sky is transiting through the primary beam is shown in blue (corresponding to the median value between local sidereal times (LST) of 14:00 and 14:20).  The system temperature when a bright part of the radio sky is transiting through the primary beam is shown in red (corresponding to the median value between LST of 20:20 and 20:40).  All quantities are obtained from autocorrelation data collected on 2019-05-30 and calibrated to units of Kelvin using hot, cold, and ambient temperature loads as described in the text. 
    }
    \label{fig:net}
\end{figure}

Measuring the system temperature of the CHIME receivers using observations of the radio sky alone is challenging because it requires knowledge of the effective area of the antenna beam pattern.  Instead, we  perform an in-situ measurement of the system temperature referred to the LNA input 
of four CHIME receivers (two polarizations on each of two antennas)  by temporarily disconnecting the LNA from the antenna under test and connecting it to well-matched cold, ambient temperature, and hot  loads at 
\SI{80}{\kelvin},  $\sim$\SI{300}{\kelvin} and \SI{373}{\kelvin}.
We observe each regulated load for approximately 10 minutes,  re-connect the LNA  to the antenna, and resume normal observations.  

The autocorrelations recorded by the CHIME correlator during the measurement are corrected for bias due to quantization to \SI{4}{\bit} real + \SI{4}{\bit} imaginary, which is insignificant for sky measurements but is a significant correction for the hot and ambient temperature measurements.  
The autocorrelations are  converted to units of \si{\jansky} using the gains obtained from the visibility matrix at the transit of Cyg A occurring approximately 6 hours before the measurements, and regressed against the load temperature.  The slope of the regression is used to estimate the \si{\jansky\per\kelvin} factor that converts between flux density on the sky and temperature at the input to the LNA.  The intercept divided by the slope is used to estimate the \textit{receiver temperature}, by which we mean the noise temperature of the LNA, FLA, cables and ADC, referred to the LNA input.  The \si{\jansky\per\kelvin} calibration factor is applied to the autocorrelations collected the night following the measurement to estimate the system temperature. The resulting \textit{system temperature} measurements, referred to the LNA input  for the two polarizations of one of the antennas are shown in \cref{fig:net}. The results for both channels of the  other antenna are  consistent with these at the \SI{5}{\percent} level.

The receiver temperature increases from approximately \SI{20}{\kelvin} at \SI{400}{\mega\hertz} to \SI{25}{\kelvin} at \SI{800}{\mega\hertz}.  This is in good agreement with measurements of the LNA temperature made in the laboratory and described in \secref{sec:AnalogSystem}, indicating the LNA dominates the receiver noise, as expected from the design.  The system temperature when a dim part of the radio sky is transiting overhead is approximately \SI{50}{\kelvin}, but shows significant spectral structure that can be broadly separated into a \SI{150}{\mega\hertz} and \SI{30}{\mega\hertz} ripple.  The approximately \SI{30}{\kelvin} difference between the system temperature and the receiver temperature includes contributions from the 
radio sky, loss in the antenna balun,  ground spillover, transmission through the mesh, noise coupled from neighboring feeds, and antenna impedance mis-match in order of most significant to least significant contribution. 

The radiometer equation can be used to estimate the noise given the system temperature presented above and the number of baselines, integration time, and bandwidth.  In what follows, the variance of the data on different time scales is estimated directly and compared to our expectation based on the radiometer equation.  On short timescales, the variance of each visibility is estimated by differencing even and odd time samples at \SI{31}{\milli\second} cadence (see \secref{sec:ReceiverSystem}).  The radio sky does not change appreciably on these timescales and thus drops out of the difference.  This ``fast-cadence noise estimate'' shows good agreement with our expectation based on the radiometer equation after excluding events that are localized in time and frequency in a manner characteristic of RFI. On longer timescales, the variance can be estimated by differencing visibilities acquired at the same local sidereal time (LST) on different sidereal days. In general, these day-to-day variations are consistent with the fast-cadence noise estimate and integrate down with the number of redundant baselines that are stacked.  There are a few exceptions.  The residual complex gain instabilities described in \secref{sec:stability} dominate the day-to-day variations when the four brightest point sources are in the primary beam.  In addition, visibilities measured by the shortest intra-cylinder baselines, specifically those with a north-south distance less than \SI{10}{\meter}, are dominated by variations in sky brightness due to residual complex gain instabilities and also variations in the noise coupled between the feeds that form the baseline.

Similar results are obtained with beamformed data, differencing ``ringmaps'' of the sky (see \secref{sec:maps}) produced on different sidereal days. For maps constructed with inter-cylinder baselines only (thus excluding the shortest intra-cylinder baselines mentioned above), the day-to-day variation over most of the sky is consistent with the fast-cadence noise estimate after accounting for the number of baselines that are used to produce the maps.  The exception are pixels brighter than a few \si{\jansky\per\beam}, for which the noise is dominated by residual complex gain instabilities. The noise can be further reduced by stacking maps produced on multiple days.  In an analysis of 38 daily, inter-cylinder ringmaps spanning an interval of 73 days, the noise was observed to integrate down with the number of stacked days.

\subsection{RFI}
\label{sec:noise_rfi}

\begin{figure*}
    \centering
    \includegraphics[width=0.95\linewidth,keepaspectratio]{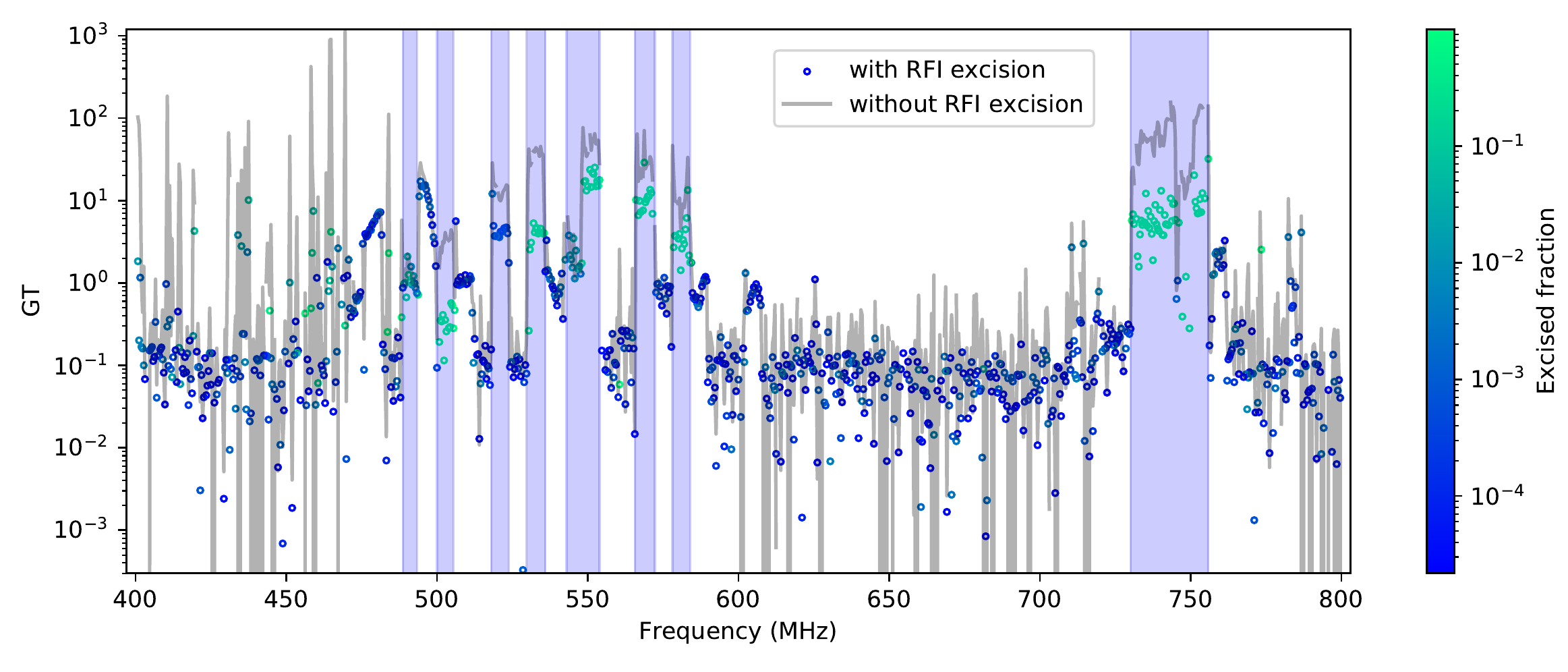}
    \caption{Result of the Gaussianity test (GT), \cref{eqn:gaussianity} for a single input on October 11, 2019 from 0:00-1:30 PDT before and after kurtosis based RFI excision.  The color of dots shows the average excised fraction over these  1.5 hours for each frequency channel. LTE and TV station bands are shown in purple. The Gaussianity of the data has improved in many frequency channels by excising less than \SI{1}{\percent} of the samples, i.e., their GT value is getting closer to zero after RFI excision.  Notice that almost all the data are automatically excised within the TV and LTE bands.  While this heavy excision improves the GT values for what remains, frequency channels in those bands still fail and are excised. 
    }
    \label{fig:rfi_excision}
\end{figure*}

\begin{figure}
    \centering
    \includegraphics[width=0.95\linewidth,keepaspectratio]{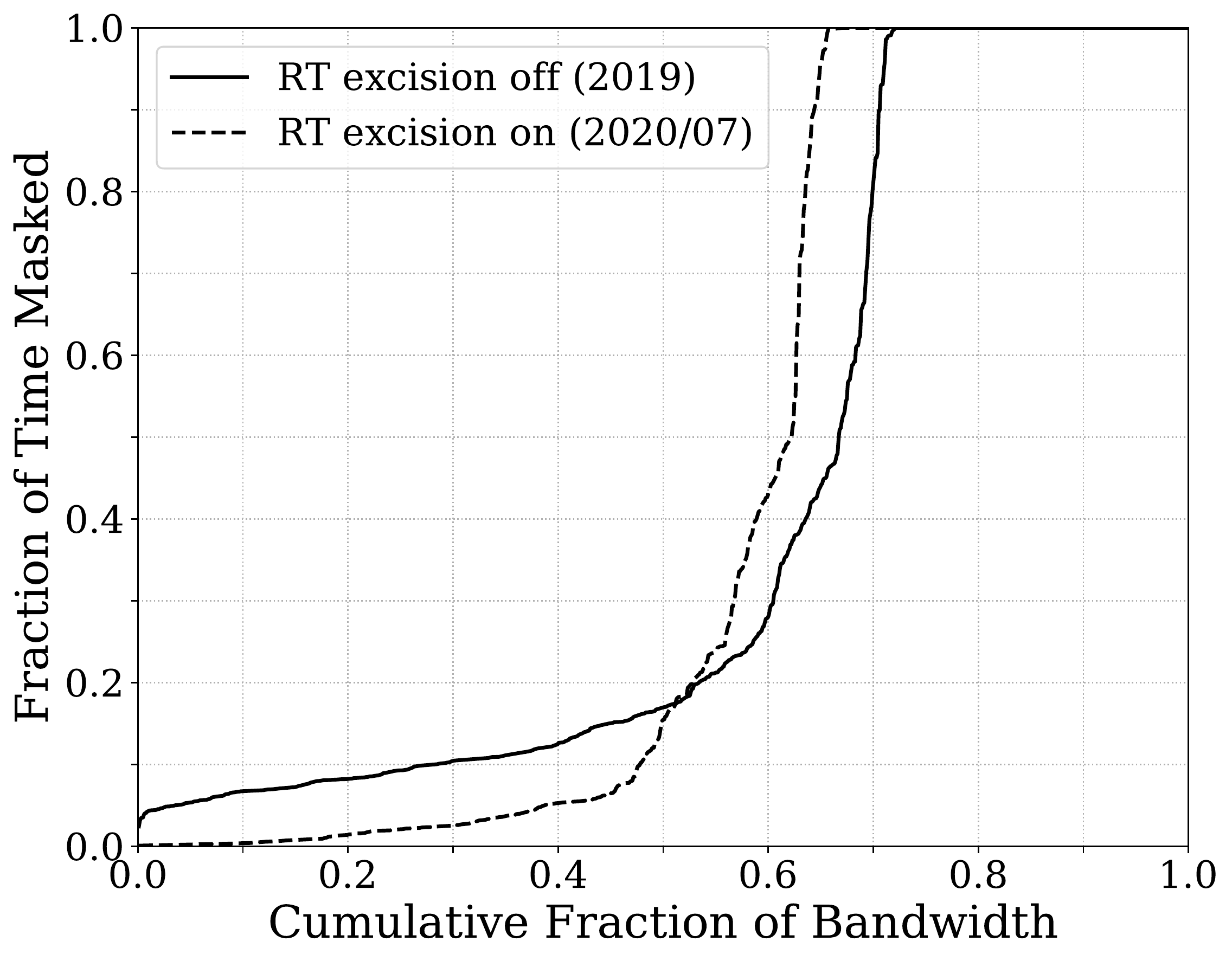}
    \caption{The cumulative fraction of the \SIrange{400}{800}{\mega\hertz} band which is masked  \textbf{less than} a particular fraction of time  by the off-line (second stage)  RFI excision algorithm.  The solid line indicates that all frequencies were masked at least 5\% of the time over 188 nights between 2018/12/22 and 2019/09/30 when the real-time (RT), kurtosis-based RFI excision was turned off.  
    The dashed line indicates that  nearly 40\% of the CHIME band was masked less than 5\%  of time  over 27 nights between 2020/06/26 and 2020/07/21 when the real-time RFI excision was turned on. The difference between the vertical asymptotes (30\% always masked in 2019, 40\% in 2020) is due to new radio transmitters nearby.}
    \label{fig:cdf_fraction_masked}
\end{figure}

The real-time RFI-excision algorithm described in \secref{sec:Xengine} was deployed in October 2019.  To evaluate its performance, the Gaussianity of the autocorrelations are compared before and after applying the RFI excision. The Gaussianity test value ($GT$) for signal $i$ is defined as
\begin{equation}
  GT_{i} = \sqrt{\frac{(1-f)\Delta t ~\Delta\nu}{2 ~(N - 2)} \sum_{j=1}^{N-1} \left(
  \frac{V_{ii}(t_{j+1}) - V_{ii}(t_{j})}{V_{ii}(t_{j+1}) + V_{ii}(t_{j})}\right)^{2}}-1 \ ,
  \label{eqn:gaussianity}
\end{equation}
where $\Delta \nu =\SI{390}{\kilo\hertz}$ is the channel bandwidth, $V_{ii}$ is the autocorrelation evaluated at $N$ times $t_j$,  $\Delta t (\sim \SI{10}{\second}$ ) is the integration time, with $(1-f)\Delta t$ remaining on average after high speed excision  and $f$ is the real-time excision fraction.  For a perfect Gaussian distribution the test will return $\sim 0$, and a large deviation from 0 indicates non-Gaussianity of the data.  The results of the test for a single input are shown in \cref{fig:rfi_excision}. Gaussianity of the data improves at all frequencies after applying the RFI excision, particularly in the \SIrange{600}{700}{\mega\hertz} band where excising less than \SI{1}{\percent} of the samples significantly improves the quality of the data. The algorithm excises \SI{15}{\percent} of the data on average.

The offline RFI excision algorithm described in \secref{sec:OfflinePipeline} masks frequencies and times where the measured sub-integration variance averaged over all cross-polar baselines deviates significantly from our expectation for radiometer noise.  Over 188 nights in 2019 the average fraction of the band that was masked was \SI{42}{\percent}, with little night-to-night variation.  During this interval the real-time RFI excision was turned off. \cref{fig:cdf_fraction_masked} shows as a solid line the cumulative distribution of frequency bins as a function of fraction of time masked over this interval.  About \SI{29}{\percent} of frequency bins are always masked, corresponding to the persistent sources of RFI discussed in \secref{sec:site}.  \cref{fig:cdf_fraction_masked} also shows as a dashed line this same quantity for 27 nights in mid-2020 when the real-time RFI excision was turned on.  The fraction of the band that is always masked increased to \SI{35}{\percent} because of a degradation in the RFI environment at DRAO, primarily due to (i) the appearance in early 2020 of the downlink for Rogers 600 MHz band, which introduced persistent RFI in a part of the spectrum that was previously clean (\SIrange{617}{627}{\mega\hertz}) (ii) the transition from partial to complete occupation of the LTE band at \SIrange{782}{788}{\mega\hertz}.  For the cleanest half  of the CHIME band the fraction of time that was masked decreased from almost \SI{15}{\percent}  to less than \SI{5}{\percent} with use of real-time RFI excision.

Since the real-time excision operates on the \SI{0.66}{\milli\second} and \SI{31}{\milli\second} frames, it is able to mask transient RFI events while discarding a much smaller fraction of the data than the offline algorithm that operates on the \SI{10}{\second} data frames does.   At present, the average fraction of the band that is masked is  roughly the same with either method, \SI{42}{\percent} but the fraction of the passband which is more than 95\% free of RFI is much higher with  rapid excision.

\subsection{Sky Maps}
\label{sec:maps}

\newcommand{\za}{\ensuremath{\mathrm{za}}}

\begin{figure*}
    \centering
    \includegraphics[width=0.98\textwidth,keepaspectratio]{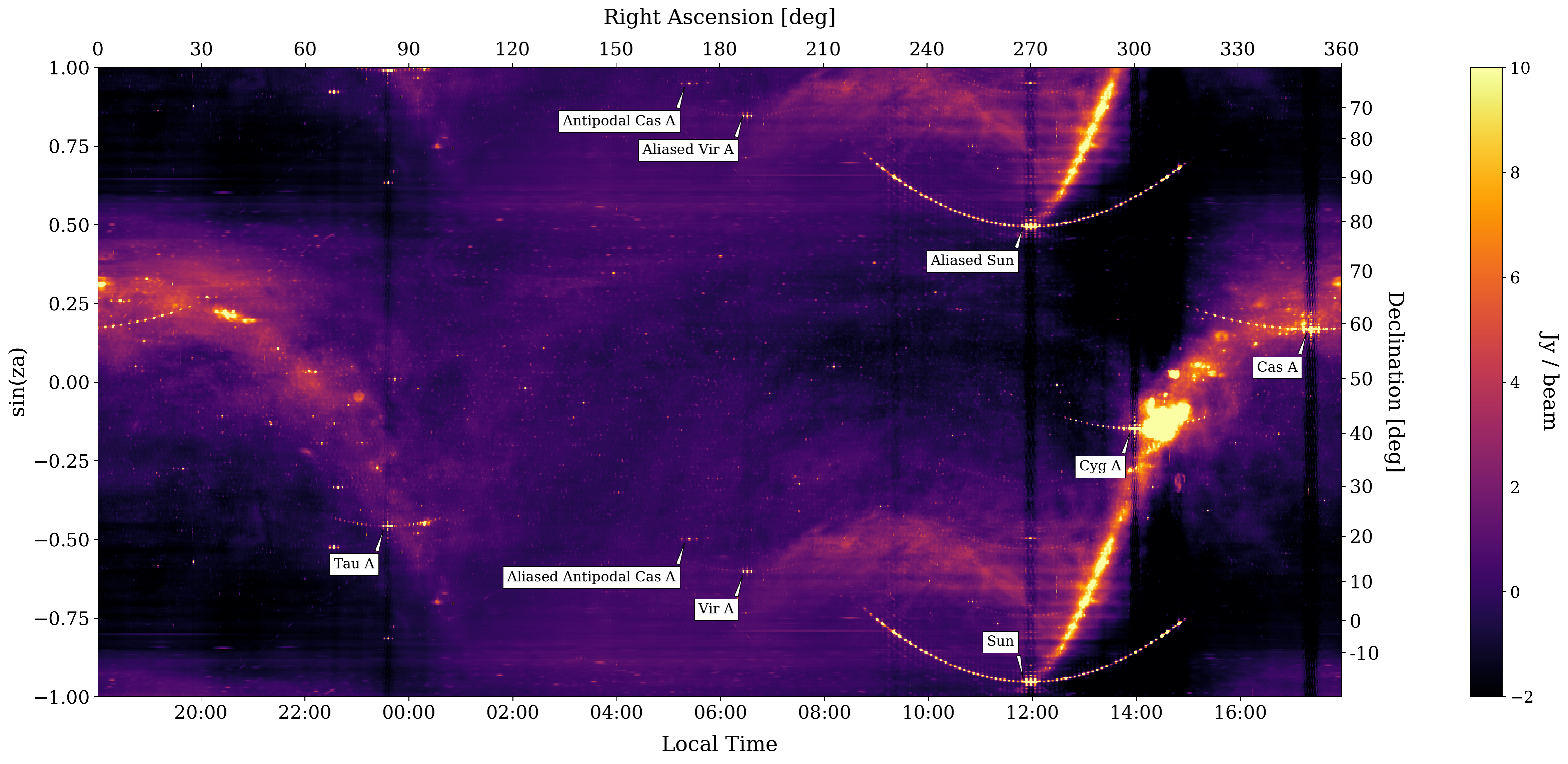}
    \caption{Map of the northern radio sky at \SI{679}{\mega\hertz} constructed from data collected by CHIME over a single sidereal day (2018-12-21/22), obtained by beamforming all YY visibilities (excluding autocorrelations) for each 10 second integration to a grid of 2048 declinations along the meridian, spanning from horizon to horizon and equally spaced in $\sin(\za)$. The map has been minimally processed, and no attempt has been made to deconvolve the transfer function of the instrument. The Sun and the four brightest point sources, and their aliases are identified. The map is shown with time (Pacific Standard Time, UTC$-8$) increasing from left to right to illustrate the CHIME observing strategy; therefore right ascension increases from left to right, opposite the astronomical convention. The image is plotted in the native time--$\sin(\za)$ coordinates; declination and right ascension (or, equivalently because all observations are at hour angle zero, local sidereal time) are labeled on the right and top.}
    \label{fig:ringmap}
\end{figure*}
We generate maps of the sky  for data quality assessment, for instrument characterization and as the starting point for Galactic science with CHIME; all have short-term and long-term goals. Our basic product is the ``ringmap''. We generate one-dimensional images along the meridian by Fourier transforming, one image for every ten-second time sample, and we assemble these into an all-sky image. These maps employ visibilities directly. The process is described in detail in \citep{stacking2021}. The cosmological stacking analysis is based on ringmaps with intra-cylinder baselines excluded in order to filter out diffuse Galactic emission and reduce the impact of noise crosstalk.

\subsubsection{Single-day Maps}

We show a ringmap produced from $YY$ visibilities using a single sidereal day of data in \cref{fig:ringmap}. The map is shown in the time-$\sin(\za)$ coordinate system, where $\za$ is the zenith angle, and with corresponding right ascension and declination labels on the top and right. We show 24 sidereal hours of data, with time increasing from left to right: right ascension also increases from left to right, opposite to the astronomical convention for sky images.

The ringmap of \cref{fig:ringmap} highlights a number of features of the Galaxy, our observing strategy, and instrumental features and artifacts. The large features of the radio sky dominate the map. The Galactic plane stretches across the sky, and the North Polar Spur rises from it. The Galactic plane and the North Polar Spur appear once at their true declinations and again at the top of the image, the result of aliasing. The spacing of feeds along the focal line is \SI{30}{\centi\meter}, more than half a wavelength for frequencies higher than \SI{500}{\mega\hertz}: the Fourier transform therefore produces an aliased response across much of the band. The bright sources --- the Sun, Cas A, Cyg A, Tau A and Vir A --- also have aliased versions; all except the Sun are unresolved by the CHIME beam and can be treated as point sources. Cas A is circumpolar, and a lower transit image, and its alias, are also seen. All bright sources are seen both at transit and in the sidelobes for several hours on either side of transit. Away from transit these sources appear to be at higher declination, producing the characteristic ``smile'' features on the ringmap. The point sources show a bright peak at the source right ascension and fainter peaks before and after transit, produced by the grating lobes; all the smile features have a dotted appearance. The shape of the smile is geometric and therefore frequency independent, but the positions of the grating lobe peaks along the smile are frequency dependent. Each time slice is an interferometric image lacking zero-spacing information, and therefore must average to zero. Consequently, the transit of each of the bright sources produces a vertical dark stripe of negative values across the map. Similar negative regions are evident at the right ascensions of particularly bright Galactic emission.

Crosstalk between adjacent feeds (see \cref{fig:beam_ns}) produces ripples in zenith angle which are evident as horizontal stripes in an uncorrected ringmap. We have reduced the striping by subtracting the median at each declination from the image. This process is quite effective at $|\za| \lesssim 25^{\circ}$, but some striping is still evident at larger zenith angles. 

\begin{figure*}
    \centering
\includegraphics[width=0.98\textwidth,keepaspectratio]{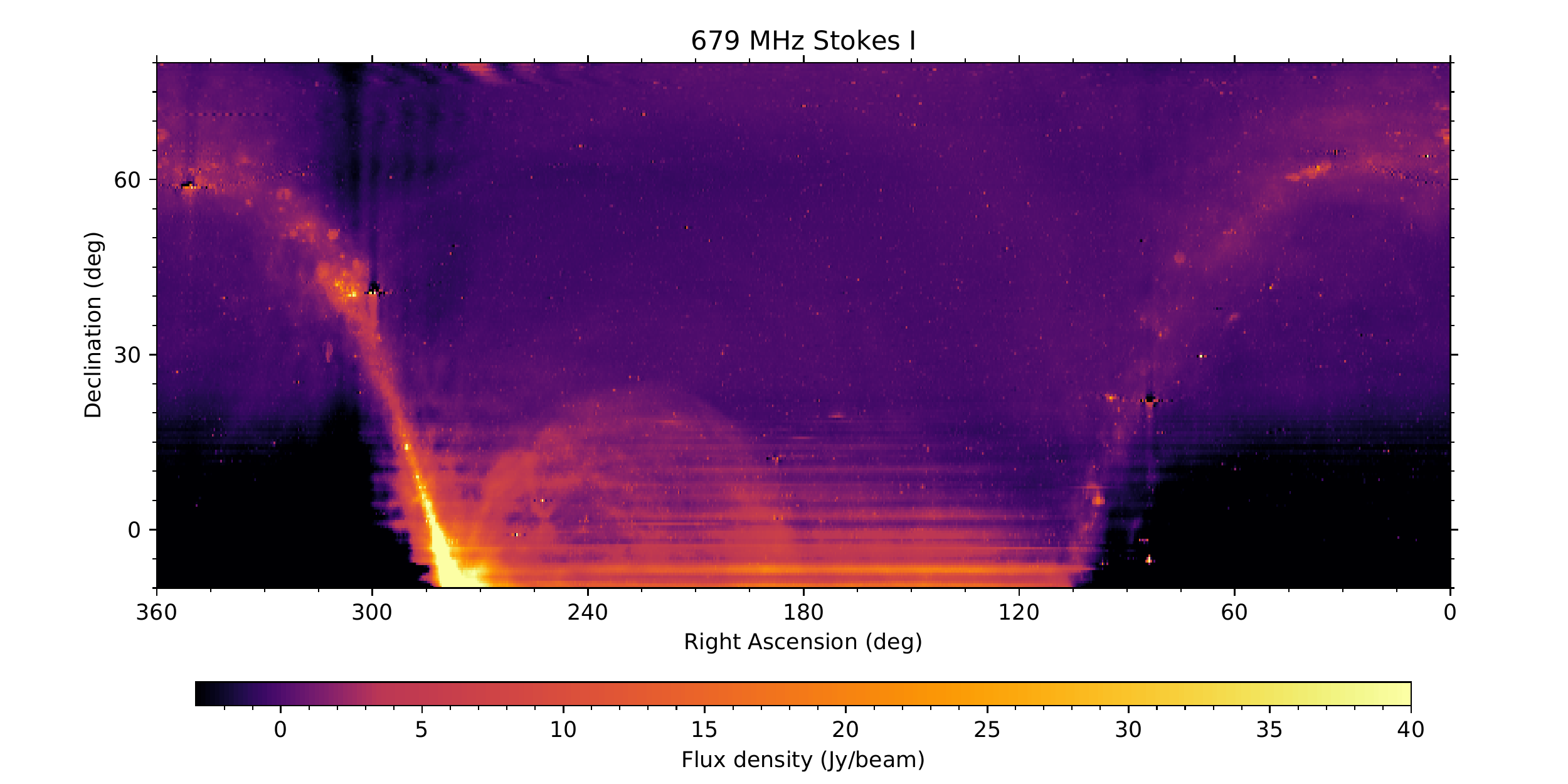}
    \caption{Map of the northern radio sky constructed from data collected by CHIME over 52 nights and stacked. This is a deconvolved Stokes $I = (XX + YY)/2$ ringmap obtained from all XX and YY visibilities using the stacked data, plotted in celestial coordinates in a plate carr\'ee  projection. The image shows most of the northern sky, oriented in the conventional way for astronomical images with right ascension increasing to the left (unlike \cref{fig:ringmap}).
    }
    \label{fig:bestmap}
\end{figure*}

\subsubsection{Stacked Maps}

In \cref{fig:bestmap} we show a stacked map, formed from data from nearly two months of observations. This too is a ringmap, but the data are combined as visibilities before the formation of the map. The stack uses night-time data from 52 24-sidereal hour periods (we call them ``days'' for brevity), divided into contiguous sets of days from different periods in the year chosen to provide complete coverage of the sidereal day (see \secref{sec:OfflinePipeline} for an overview of the daily processing pipeline). The stacking proceeds in two steps: first, days within a contiguous set are averaged together, and second, all these averages are combined.

In the first step, averaging over contiguous blocks, data deemed bad are masked (arising from the presence of the Sun or Moon, RFI, or data-quality flags) and any day with less than 70\% coverage after masking is discarded. Bias due to crosstalk is estimated by calculating the median visibility at each zenith angle in a one-hour region of right ascension where the sky signal is at low intensity. This value is subtracted from each individual day of data before stacking. Ideally, the same right ascension range would be used for all averages, but this is clearly impossible because the part of the sky transiting at night is changing with time of year. We compromised by choosing two right ascension ranges for the estimate of the crosstalk contribution. To ensure consistency we use a set where both these ranges transit at night; we derive an additive correction from that average and apply it to all averages. Daily calibration is based on either Cas~A or Cyg~A. To account for different beam responses at the locations of these two sources we derive a multiplicative amplitude correction at every frequency and apply it to all the averages prior to stacking. To remove the most prominent ``smile'' artifacts for display purposes, we subtracted Cas~A, Cyg~A, and Tau~A in visibility space.

\begin{figure*}
\centering
`\includegraphics[width=0.98\textwidth,keepaspectratio]{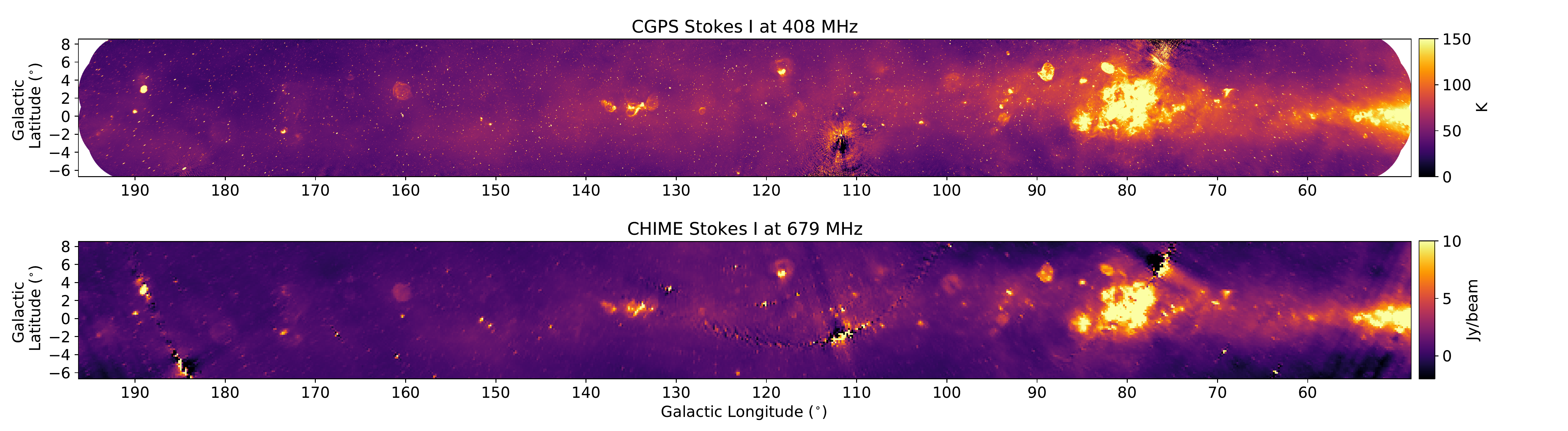}
\caption{Stokes $I$ maps of the Galactic plane from the Canadian Galactic Plane Survey at \SI{408}{\mega\hertz} \citep[CGPS, top panel;][]{Tung:2017} and CHIME at \SI{679}{\mega\hertz} (bottom panel; same data as in \cref{fig:bestmap}) in Galactic coordinates.}
\label{fig:cgps_chime}
\end{figure*}


The deconvolved ringmap at each frequency and declination is approximately given by the 1D convolution of the sky with the east-west profile of the primary beam at the corresponding frequency and declination, as described in detail in \citep{stacking2021}. By this method, we attain an estimate of the true sky at each declination by deconvolving the beam profile from each row of the ringmap.

In the 52-day map of \cref{fig:bestmap} we see all the features that are evident in the one-day map of \cref{fig:ringmap}, illustrating the fact that CHIME achieves a high signal-to-noise ratio even in one day. Both the  single-day and stacked maps are confusion-limited; a major benefit of stacked maps for Galactic science is the full sky coverage even with the elimination of daytime data. Within the envelope of the diffuse emission along the Galactic plane we can identify many well-known supernova remnants and \ion{H}{2} regions; these are evident in more detail in \cref{fig:cgps_chime}. The combination of the visibility-space subtraction of the brightest three point sources and the deconvolution removes the grating lobe copies of all point sources and the saturation of the image at the right ascension of the brightest sources.

In \cref{fig:cgps_chime}, we show the map from \cref{fig:bestmap} zoomed in on the Galactic plane and compared to a \SI{408}{\mega\hertz} Stokes $I$ map of the Galactic plane from the Canadian Galactic Plane Survey \citep[CGPS;][]{Tung:2017}. The CGPS \SI{408}{\mega\hertz}  data, obtained with the DRAO Synthesis Telescope, have an angular resolution of $\approx 3'$, and cover the area $52^{\circ}\leq l \leq 193^{\circ}$, $-6.5^{\circ}\leq b \leq 8.5^{\circ}$. Short spacings for the CGPS map are incorporated from the \cite{Haslam:1982} single-antenna data. There is good overall agreement between the CHIME and CGPS maps in the Galactic plane. Discrete objects such as supernova remnants, and more extended objects such as the W3/4/5 \ion{H}{2} region and the Cygnus X complex of \ion{H}{2} regions and stellar clusters, are distinctly visible in the CHIME data, and are well matched with the CGPS data in terms of structure and relative brightness. Although the CHIME data lack zero-spacings, and thus sensitivity to the largest scale structures, much of the diffuse emission visible in the CGPS map is also clearly discernible in the CHIME map. This is especially true of the bright extended emission at the low-longitude end of the CGPS coverage. The bright radio sources, Cyg A and Cas A, produce artifacts in both the CHIME and CGPS maps, although these are more easily mitigated in the CGPS through mosaicing of fields with a sufficiently dense sampling of pointings in those regions. While CHIME does not match the high angular resolution of the CGPS, its spectral coverage far exceeds that of the CGPS\footnote{The CGPS has a bandwidth of \SI{3.5}{\mega\hertz} at \SI{408}{\mega\hertz}, and a bandwidth of  \SI{35}{\mega\hertz}  at \SI{1420}{\mega\hertz}.}, allowing for more in-depth exploration of frequency-dependent phenomena in the Galaxy over a larger spatial extent.

Sky maps like these will be the main data product for science involving non-cosmological foregrounds. We will have all-sky images at hundreds of frequencies across an octave obtained with the same telescope, allowing analyses of spectral indices of point sources, extended objects, and diffuse emission. The Galactic signal is dominated by synchrotron emission, linearly polarized at its source, and Faraday rotated by the intervening magneto-ionic medium along virtually every line of sight.  A major scientific goal is to apply Faraday synthesis \citep{brentjens:2005} to the polarization data. We will derive Stokes $Q$ and $U$ maps, which will provide a valuable dataset for Faraday synthesis across the whole sky; the wavelength-squared range and resolution of the CHIME data provide the Faraday depth resolution to isolate discrete magnetic features, with Faraday depth resolution $\delta \phi \approx 3.8 / \Delta(\lambda^2) \approx \SI{9}{rad.m^{-2}}$ while retaining sensitivity to extended Faraday depth features, with $\phi_\mathrm{max-scale} \approx \pi \lambda_\mathrm{min}^{-2} \approx \SI{22}{rad.m^{-2}}$, in the Galaxy \citep{schnitzeler:2009}. Therefore it will be possible to distinguish between extended structures and multiple narrow features in Faraday depth space. Exploration of this parameter space is only beginning \citep{Dickey:2019,Thomson:2019}. The \SIrange[range-units = single, range-phrase=--]{400}{800}{\mega\hertz} polarization maps with $\approx 40'$ angular resolution from CHIME will form a component of the GMIMS survey, which includes a southern sky dataset obtained with the CSIRO Parkes Telescope \citep{Wolleben:2019} and a \SIrange[range-units = single]{1280}{1750}{\mega\hertz} northern sky dataset observed with the Galt Telescope \citep{Wolleben:2021}. If we are able to combine data across the \SIrange[range-units = single]{400}{1800}{\mega\hertz} range, we will achieve $\delta \phi \approx \SI{7}{rad.m^{-2}}$ and $\phi_\mathrm{max-scale} \approx \SI{110}{rad.m^{-2}}$, providing sensitivity to an unprecedented range of Faraday depth scales.

CHIME is an interferometer: it has coverage of the $(u,v)$ plane down to \SI{30}{\centi\metre} baselines, but not to zero baseline because autocorrelations of the signal from each feed are excluded from the analysis. To provide information on Galactic structure at the largest angular scales, a companion polarization survey will be made with a \SI{15}{\meter} radio telescope at DRAO, covering \SIrange[range-units = single]{350}{1050}{MHz}. These data, calibrated to an absolute scale of brightness temperature, will also provide the calibration of CHIME polarization data.

In addition, by observing the entire sky every day, we are sensitive to slow transients. We are cataloguing daily fluxes of 2723 point sources, primarily quasars, to characterize variability.

\section{Conclusions and Outlook}
 \label{sec:Summary}

We have built and are operating an extremely high mapping-speed instrument  designed to measure the three dimensional distribution of neutral hydrogen over the full Northern Hemisphere and the redshift range $0.8\leq z\leq 2.5$ with enough accuracy to provide useful constraints of the expansion history of the Universe.  

The instrument has been 
\reviewercomment{collecting data for cosmological analysis}
since \reviewercomment{late}  2018.  First results measuring the distribution of neutral hydrogen in three dimensional correlation with redshift catalogs of quasars and galaxies\reviewercomment{, using data from 2019, are presented in a companion paper \citep{stacking2021}.}  
CHIME is also monitoring the variability of \reviewercomment{2723} sources with daily cadence, and has produced confusion limited maps of polarized Galactic emission \reviewercomment{across the 400-800 MHz band}.

\begin{figure}[t]
    \centering
    \includegraphics[width=1.0\linewidth, trim = {8pt 8pt 8pt 0}]{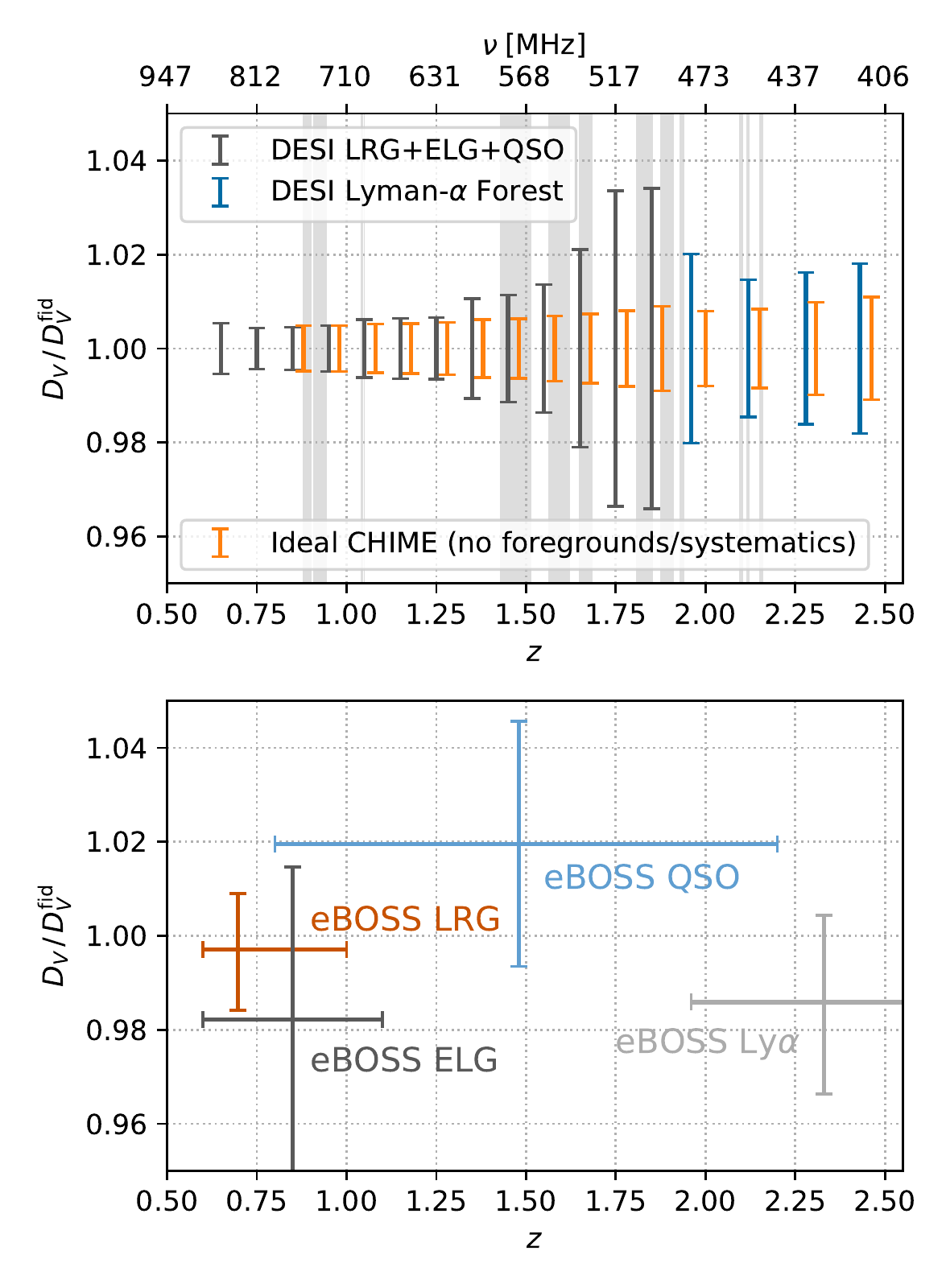}
    \caption{
    {\em Upper panel:} Projected constraints on the cosmic expansion history, parameterized using the spherically-averaged distance measure $D_V$ as a function of redshift, shown relative to a fiducial $\Lambda$CDM cosmology. For CHIME, the forecast error bars ({\em orange}) were calculated for 1~year of integration time using the Fisher matrix approach of \cite{bull2015}, assuming perfect foreground subtraction and no systematics. Each error bar is statistically independent.  We also show projections for the DESI clustering measurements ({\em black}), computed using the same formalism and based on combined constraints from the three clustering samples that overlap with CHIME's redshift coverage, and DESI Lyman-$\alpha$ forest measurements ({\em blue}), which we take from \cite{DESI:2016}. (See Appendix~\ref{sec:appforecast} for the details of these forecasts.) Shaded {\em grey} bands denote regions inaccessible to CHIME due to persistent sources of RFI. 
    {\em Lower panel:} Expansion history measurements from the final eBOSS survey, taken from the compilation in \cite{zhao2021}. Comparison to the CHIME  forecasts in the upper panel indicates  that the 
    \reviewercomment{intrinsic}
    \textit{statistical}  precision of CHIME is highly competitive with that of existing and near-future expansion history measurements.  The challenge is to understand systematic effects well enough that statistical errors dominate.
    }
    \label{fig:chime_forecast}
\end{figure}

\reviewercomment{To quantify the cosmological constraining power of CHIME under ideal conditions, in \cref{fig:chime_forecast} we show an updated forecast for the statistical precision of CHIME in measuring the cosmic expansion history using the BAO feature in the \tcm power spectrum. }
Compared to previous forecasts in the literature, these results use a more accurate version of CHIME's feed layout (\secref{sec:holo}), updated models for the mean \tcm brightness temperature and linear HI bias, and an empirically-derived estimate of CHIME's total system temperature, based on measurements presented in \secref{sec:noise}. We describe the methodology of these forecasts, which  mostly follows \cite{bull2015}, in Appendix~\ref{sec:appforecast}. In particular, Table~\ref{tab:forecastpars} lists the CHIME instrumental and survey characteristics used in these forecasts.  Note that these forecasts assume perfect foreground subtraction and the absence of systematic errors. (Persistent RFI bands are indicated in the figure.)

We also show the expected precision of a combined galaxy and quasar sample from the Dark Energy Spectroscopic Instrument (DESI; \citealt{DESI:2016}), computed within the same forecasting formalism; Lyman-$\alpha$ forest measurements expected from DESI (which we do not re-compute, but take from \citealt{DESI:2016}); and state-of-the-art measurements by the extended Baryon Oscillation Spectroscopic Survey (\citealt{Alam:2021}; specific measurements taken from \citealt{bautista2021,demattia2021,hou2021,duMasdesBourboux:2020}, and summarized in \citealt{zhao2021}). \cref{fig:chime_forecast} shows that CHIME's 
\reviewercomment{intrinsic}
statistical precision is competitive with DESI, and that CHIME on its own is in principle capable of percent-level BAO measurements over most of its band. 

\reviewercomment{Efforts in the coming years will be focused on realizing this potential, but we emphasize that we will need to overcome several challenges to do so.}
Foreground subtraction remains the primary 
\reviewercomment{obstacle}
to producing measurements which exploit CHIME's statistical power, and it is the main focus of our analysis effort.  The path to seeing BAO through a haze of Galactic emission many orders of magnitude brighter is to filter out the spectrally smooth Galactic components and keep the spectrally chaotic BAO signal.  Any systematic error which produces a rough or poorly understood spectral response mixes the Galactic and BAO signals. Thus, great care in the design has been taken to build a stable instrument with smooth, well characterized response.   Very precise measurement of the angular response of CHIME will be necessary to  perform component separation at the level required to  characterize the BAO,  because poorly understood frequency dependence of the angular response would lead to frequency variation of the Galactic contribution along an inferred line of sight.  \reviewercomment{\cite{stacking2021} describes a set of beam measurements and analysis methods that have allowed an initial detection of the \tcm signal, and work is underway to improve upon these methods. Other areas requiring further attention include mitigation of noise crosstalk between nearby feeds, RFI mitigation in the lower half of the CHIME frequency band, and development of analysis methods that are robust to residual uncertainties in gain calibration and beam knowledge.}

\reviewercomment{Overcoming these challenges}
has the potential to unlock a rich array of science targets accessible to \tcm intensity mapping. Beyond BAO, there is potential to constrain the linear growth rate of structures as a way to test general relativity \citep{obuljen2018,chen2019,castorina2019}; constrain models of cosmic inflation through  signatures in the primordial power spectrum of fluctuations \citep{xu2016,beutler2019} or non-Gaussian statistics in large-scale structure \citep{xu2015,karagiannis2020}; and probe the nature of dark matter \citep{carucci2015,bauer2021}. In addition, ``tidal reconstruction" techniques, which reconstruct large-scale (foreground-obscured) modes from the correlations they induce between smaller-scale modes \citep{zhu2018,modi2019,darwish2021}, can greatly expand the opportunities for cross-correlations with surveys of the cosmic microwave background or photometric galaxy redshifts. Additionally,  lower-frequency observations of the \tcm line are well-suited to probing the era of reionization \citep{furlanetto2019rei}, or more ambitiously,  the cosmic ``dark ages" up to $z \sim \mathcal{O}(100)$ \citep{furlanetto2019darkages}.

The instrument described here also acts as the front-end for several other systems, providing calibrated data to an FRB  detector \citep{FRB2018}, a 10-beam system which monitors all pulsars visible from Canada with up to daily cadence \citep{CHIMEPulsar:2021}, a system to search for cold clouds acting as 21cm absorption-line systems and a VLBI station \citep{Cassanelli2021}.  Among the accomplishments these new instruments have made is the discovery of half a dozen Galactic pulsars,   detection of an exceptionally bright radio burst from a Galactic magnetar \citep{frbmagnetar-2020Natur.587...54C}, pointing to possible similarities of magnetars and FRBs, and publication of the first substantial catalog of FRB  \citep{frbcatalog-2021arXiv210604352T}.
This broad range of additional scientific impact comes directly from achieving the sensitivity, large fractional bandwidth and enormous field of view that hydrogen intensity mapping requires.  

We have
  shown that CHIME is capable of generating a multitude of scientific results, and 
have demonstrated that one can build a very powerful instrument for a comparatively small cost when a clear scientific goal drives the design. We expect a steady flow of further results in the years to come.

\acknowledgments

We thank the Dominion Radio Astrophysical Observatory, operated by the National Research Council
Canada, for gracious hospitality and expertise.  The help and guidance we received from the scientific and technical staff have been crucial to our progress.  Additionally, the NRC has provided  support by leasing the CHIME site on their radio-protected campus to us, building a power substation,  allowing substantial access to the Galt 26m Telescope, providing office space and on-site lodging, and more.

CHIME
is funded by  grants from the Canada Foundation for Innovation (CFI) 2012 Leading Edge Fund (Project 31170), the CFI 2015 Innovation Fund (Project 33213), and by contributions from
the provinces of British Columbia, Qu\'ebec, and Ontario. Long-term data storage and computational support for analysis is provided by WestGrid\footnote{\url{https://www.westgrid.ca}} and Compute Canada\footnote{\url{https://www.computecanada.ca}}. CMC Microsystems and Canada's National Design Network (CNDN) provided test equipment and services that facilitated this research.\footnote{\url{https://www.cmc.ca}}

Additional support was provided by the University  of British Columbia, McGill University, and the University of Toronto. CHIME also benefits from NSERC Discovery Grants to several researchers,  funding from the Canadian Institute for Advanced Research (CIFAR), and from the Dunlap Institute for Astronomy and Astrophysics at the University of Toronto, which is funded through an endowment established by the David Dunlap family.
This material is partly based on work supported by the  NSF through   grants (2008031)  (2006911) and  (2006548) and by the Perimeter Institute for Theoretical Physics, which in turn is supported by the Government of Canada through Industry Canada and by the Province of Ontario through the Ministry of Research and Innovation.

\software{
   NumPy \citep{NumPy},
   SciPy \citep{SciPy},
   HDF5 \citep{HDF5},
   h5py \citep{h5py},
   Matplotlib \citep{Matplotlib},
   scikit-rf \citep{9632487},
   Skyfield \citep{Skyfield},
   caput \citep{caput},
   ch\_pipeline \citep{ch_pipeline},
   draco \citep{draco},
   kotekan \citep{kotekan},
   CST \citep{CST},
   GRASP \citep{GRASP},
   Prometheus \citep{Prometheus}, and
   Grafana \citep{Grafana}.
   }

\vfill\eject
\appendix

\section{Details of BAO Forecast}
\label{sec:appforecast}

\subsection{Fisher matrix formalism}

We project the constraints on baryonic acoustic oscillations from CHIME, in comparison with the Dark Energy Spectroscopic Instrument (DESI), which observes in the optical and has overlapping sky and redshift coverage with CHIME. We mainly follow the Fisher matrix method of \cite{bull2015}, using a modified version of their publicly available forecast code\footnote{\url{https://github.com/philbull/RadioFisher}}.

The Fisher matrix can be written as \citep{Seo:2007}
\begin{equation}
  F_{ij} = \int \frac{d^3k}{(2\pi)^3}V_{\rm eff}\frac{\partial C^S(\mathbf{k}, z)}{\partial \theta_i}\frac{\partial C^S(\mathbf{k}, z)}{\partial \theta_j}\ ,
\end{equation}
where  the effective volume $V_{\rm eff}$ is given by
\begin{equation}\label{equ:veff}
  V_{\rm eff} = f_{\rm sky}\int_{z_{\rm min}}^{z_{\rm max}}\frac{dV}{dz}dz\left[\frac{C^S(\mathbf{k}, z)}{C^S(\mathbf{k}, z) + C^N(\mathbf{k}, z)}\right]^2\ .
 \end{equation}
In the above expression, $f_{\rm sky}$ is the fractional sky coverage, $\int_{z_{\rm min}}^{z_{\rm max}}\frac{dV}{dz}dz$ is the physical volume of the survey, $C^S$ and $C^N$ are the signal and noise covariance respectively, and ${\theta}$ is the set of cosmological parameters to be constrained. In our case, $\boldsymbol{\theta}$ is the following set of parameters:
\begin{equation}
\boldsymbol{\theta} = \left\{
	D_A(z), H(z), A(z), \left[ b\sigma_8 \right](z), \left[ f\sigma_8 \right](z), \sigma_{\rm NL}
\right\} \ .
\end{equation}
The redshift-dependent parameters are constrained within redshift bins with width $\Delta z = 0.1$ for $z\leq 1.8$ and $\Delta z = 0.16$ for $z>1.8$, in order to match the binning of the DESI forecasts in \cite{DESI:2016}. The Hubble rate $H(z)$ and angular diameter distance $D_A(z)$ are transformed into the volume distance $D_{V}$ and Alcock-Pacynski term $F$ through
\begin{equation}
D_V(z) = \left[(1+z)^2D_A(z)^2\frac{cz}{H(z)}\right]^{\frac{1}{3}}\ ,
\quad
F(z) = (1+z) D_A(z) \frac{H(z)}{c}\ ,
\end{equation}
and we forecast the fractional errorbars on measurements of $D_V$ in each redshift bin. The amplitude $A(z)$ is defined by decomposing the matter power spectrum $P_{\rm m}$ into a smooth template $P_{\rm smooth}$ and oscillatory BAO factor $f_{\rm BAO}$,
\begin{equation}
P_{\rm m}(k, z) = \left[ 1 + A(z) f_{\rm BAO}(k) \right] P_{\rm smooth}(k, z)\ ,
\end{equation}
implemented using the method from \cite{bull2015}. The linear bias $b$, linear growth rate $f$, and fluctuation amplitude $\sigma_8$ have their usual meanings, while $\sigma_{\rm NL}$ is the redshift-space damping scale defined in the next section. In each redshift bin, the $D_V$ forecasts marginalize over the other parameters ($F$, $A$, $b\sigma_8$, $f\sigma_8$, and $\sigma_{\rm NL}$) with no priors.

\subsection{Signal models}

For CHIME, we take the HI signal covariance to be
\begin{equation}
  C^S(\mathbf{k}, z) = T_b(z)^2\left(b_{\rm HI}(z) + f(z)\mu^2\right)^2e^{-k^2\mu^2\sigma^2_{\rm NL}}P_{\rm m}(k,z)\ ,
  \label{eq:CS-HI}
\end{equation}
where $T_b(z)$ is the HI brightness temperature and $b_{\rm HI}(z)$ is the linear bias of HI. For $T_b(z)$, we use the expression from \cite{hall2013}, with the fitting formula for the mean HI density $\Omega_{\rm HI}(z)$ from \cite{crighton2015}, and for $b_{\rm HI}(z)$, we use the model from \cite{ansari2018}, that smoothly interpolates between measurements from the IllustrisTNG simulation at $z<2$ \citep{villaescusa-navarro2018} and the analytical approximation from \cite{castorina2017} at $z>2$. The large-scale effect of redshift-space distortions is accounted for in the $f(z)\mu^2$ term in Eq.~\eqref{eq:CS-HI}, where $f(z)$ is the linear growth rate  and $\mu$ is the angle of the wavevector to the line of sight. At smaller scales, the exponential factor roughly accounts for the ``Finger of God" effect that suppresses the observed clustering power beyond the cutoff scale $\sigma_{\rm NL}$.
The linear matter power spectrum $P_{\rm m}(k)$ is calculated using \sw{CAMB} \citep{Lewis:2000}.

For DESI, we use
\begin{equation}
  C^S(\mathbf{k}, z) = \left(b_{\rm g}(z) + f(z, k)\mu^2\right)^2e^{-k^2\mu^2\sigma^2_{\rm NL}}P(k)\ ,
  \label{eq:CS-gal}
\end{equation}
for the galaxy signal covariance, where $b_{\rm g}(z)$ is the linear galaxy bias and the other components are the same as Eq.~\eqref{eq:CS-HI}. For these forecasts, we combine the luminous red galaxy (LRG), emission line galaxy (ELG), and quasar (QSO) samples, by summing their expected number densities in each redshift bin, as given for the DESI baseline survey in Sec.~2.4.2 of \cite{DESI:2016}, and using a number-density-weighted mean of the corresponding linear bias factors, also taken from \cite{DESI:2016}. (We do not consider the bright galaxy sample, because its redshift range does not overlap with that of CHIME.)

For these forecasts, we adopt fiducial cosmological parameters from the \emph{Planck} CMB-only best-fit $\Lambda$CDM model \citep{PlanckParam:2018},
\begin{equation}
 h = 0.6732,\; \Omega_{\rm m} = 0.3158,\; \Omega_K = 0,\;  \Omega_b = 0.0494,\;  w = -1,\;  n_s = 0.966,\;  \sigma_8 = 0.812,\;  N_{\rm eff} = 3.046.
\end{equation}
Following \cite{bull2015}, we choose the nonlinear dispersion scale to be $\sigma_{\rm NL} = 7$\,Mpc, corresponding to power being significantly damped at $k \gtrsim 0.14$\,Mpc$^{-1}$. This value is higher than recent values from the literature, both for HI and DESI-like galaxies (e.g.~\citealt{stacking2021}), but is justified here because it limits the sensitivity of our forecasts to nonlinear scales where the assumptions in Eqs.~\eqref{eq:CS-HI} and~\eqref{eq:CS-gal} break down. Also, we make use of the BAO information only, instead of the full shape of the HI or galaxy power spectrum (e.g.~\citealt{sailer2021}). While a full-shape analysis would provide increased constraining power, it it also more likely to be affected by foregrounds and systematics, so we aim to be conservative in that respect by restricting to BAO only.

\begin{table}
\centering
\begin{tabular}{l|l}
\hline
\hline
Sky coverage, $S_{\rm sky}$ (deg$^2$) & $31000$ \\
Redshift range, $[z_{\rm min}, z_{\rm max}]$ & [$0.8$, $2.5$]  \\
Channel width, (kHz) & $390$ \\
Number of redshift bins, $n_{\rm bin}$ & $15$ \\
System temperature$^*$, $T_{\rm sys}$ (K) &  $55$ \\
Integration time, $t_{\rm tot} $ (yr) & $1$ \\
Number of antennas per cylinder, $N_{\rm ant}$ & $256$ \\
Number of polarizations per antenna, $n_{\rm pol}$ & $2$ \\
Number of cylinders, $N_{\rm cyl}$ & $4$ \\
Cylinder width, $w_{\rm cyl}$ (m) & $20$ \\
Cylinder spacing (edge-to-edge) (m) & $2$ \\
Physical cylinder length (m) & $100$ \\
Illuminated cylinder length$^\dagger$, $l_{\rm cyl}$ (m) & $78$ \\
Antenna spacing, $d_{\rm ant}$ (m) & $0.3048$ \\ \hline
Minimum baseline, $b_{\rm min}$ (m) & $0.3048$ \\
Maximum baseline, $b_{\rm max}$ (m) & $102$ \\
\hline
\hline
\end{tabular}
\caption{The instrumental and observing parameters of CHIME used in our forecasts. $^*$Note that the quoted system temperature includes both instrumental and sky contributions, and is based on the noise measurements in \secref{sec:noise}. $^\dagger$In our chosen forecasting formalism \citep{bull2015}, the length of the cylinder that is instrumented with feeds is relevant; for CHIME, this length is $l_{\rm cyl} = 256 \times d_{\rm ant} \approx 78\,{\rm m}$.}
\label{tab:forecastpars}
\end{table}

\subsection{Noise models}

We mainly follow \cite{bull2015} in approximating the noise covariance for CHIME as
\begin{equation}
C^N (\mathbf{k}, z) =
	\frac{T_{\rm sys}(z)^2}{\nu_{21}n_{\rm pol}t_{\rm tot}}
	\frac{\lambda^4S_{\rm sky}}{A_e^2 S_{\rm FOV}}
	\frac{1}{n(\mathbf{u})}\ ,
\end{equation}
where $\nu_{21}$ is the HI line emission rest frequency, $n_{\rm pol}$ is the number of polarizations per antenna,  and $\lambda(z)$ is the observing wavelength corresponding to emission from redshift $z$.
For the system temperature $T_{\rm sys}(z)$, we use a constant 55\,K, based on the observations in \secref{sec:noise}; note that this includes both instrumental and sky contributions, which are usually modelled separately in forecasts.
We take the total integration time $t_{\rm tot}$ to be 1 year. The sky coverage of CHIME, $S_{\rm sky}$, is 31000\,deg$^2$, corresponding to $f_{\rm sky}\approx0.75$. We approximate the instantaneous field of view of a single cylinder as $S_{\rm FOV} \approx 90^{\circ} \times \lambda / w_{\rm cyl}$ \citep{Newburgh:2014}, where $w_{\rm cyl}$ is the cylinder width. The effective collecting area per antenna is denoted by $A_e$, and in this formalism, it takes the following form for a cylinder telescope:
\begin{equation}
A_e = \eta \frac{l_{\rm cyl}}{N_{\rm ant}} w_{\rm cyl}  \ ,
\end{equation}
where $\eta$ is the aperture efficiency assumed to be 0.7 in our case (following \citealt{bull2015}), $l_{\rm cyl}$ is the length along the cylinder axis that is instrumented with feeds, and $N_{\rm ant}$ is the number of antennas per cylinder.  $n(\mathbf{u})$ is the $(u,v)$-plane baseline number density of CHIME, calculated using the code from \cite{bull2015} with details given in their Appendix C, accounting for the fact that adjacent CHIME cylinders are separated by 2$\,$m (implying that the shortest nonzero East-West baseline is 22$\,$m). We are only interested in the ideal statistical constraining power of CHIME, so we assume perfect foreground cleaning and no systematics, so that the noise covariance  includes only the instrumental thermal noise. In addition, we assume Gaussian beams with equal response across the sky, and neglect the intrinsic HI shot noise, which is far smaller than the thermal noise~\citep{villaescusa-navarro2018}. Table~\ref{tab:forecastpars} summarizes the instrumental parameters used in our forecast for CHIME.

The noise covariance for a galaxy survey is dominated by the shot noise due to the limited number of galaxies detected in the observed region at a particular redshift. We assume that this noise covariance is Poissonian for DESI, and is thus
\begin{equation}
C^N (z) = \frac{1}{n(z)}\ ,
\end{equation}
where $n(z)$ is the comoving galaxy number density at redshift $z$. We convert the quoted values for $dN / (dz\,d{\rm deg}^2)$ from Sec.~2.4.2 of \cite{DESI:2016} into $n(z)$ values within each redshift bin, and sum over the LRG, ELG, and QSO samples. We adopt the sky coverage of the DESI baseline survey at 14000\,deg$^2$.

\bibliography{paper}{}
\bibliographystyle{aasjournal}

\end{document}